\documentclass[a4paper,11pt]{article}
\pdfoutput=1
\usepackage{jheppub}
\usepackage{amsthm}
\usepackage{amsmath}
\usepackage{float}
\usepackage{graphicx}
\usepackage{subcaption}
\usepackage{slashed}
\usepackage{amssymb}
\usepackage{xcolor}
\usepackage[normalem]{ulem}

\newcommand*\diff{\mathop{}\!d}

\newcommand{\nn}{\nonumber}

\newcommand{\be}{\begin{eqnarray}}
\newcommand{\ee}{\end{eqnarray}}
\newcommand{\ma}{\mathrm}

\newcommand{\bs}{\boldsymbol}

\begin{document}
\title{Computationally Efficient Description of Medium Response to Jets in Heavy Ion Collisions}

\author[1]{Jorge Casalderrey-Solana}
\affiliation[1]{Departament de F\'\i sica Qu\`antica i Astrof\'\i sica \& Institut de Ci\`encies del Cosmos (ICC),\\ 
Universitat de Barcelona, 
Barcelona, Spain}

\author[2,3]{Jos\'e Guilherme Milhano}
\affiliation[2]{LIP - Laborat\'orio de Instrumenta\c c\~ao e F\'isica Experimental de Part\'iculas,\\
Avenida Prof. Gama Pinto, 2, 
Lisboa, Portugal}
\affiliation[3]{Departamento de F\'isica, Instituto Superior T\'ecnico (IST), Universidade de Lisboa,
\\ 
Avenida Rovisco Pais 1, 
Lisbon, Portugal}

\author[4,5]{Daniel Pablos}
\affiliation[4]{Departamento de F\'isica, Universidad de Oviedo,
Oviedo, Spain}
\affiliation[5]{Instituto Universitario de Ciencias y Tecnolog\'ias Espaciales de Asturias (ICTEA),\\ 
Calle de la Independencia 13, 33004 Oviedo, Spain}

\author[6]{Krishna Rajagopal}
\affiliation[6]{Center for Theoretical Physics --- A Leinweber Institute,\\ Massachusetts Institute of Technology, Cambridge, MA 02139 USA}

\author[7]{Xiaojun Yao}
\affiliation[7]{InQubator for Quantum Simulation, Department of Physics,\\ University of Washington, Seattle WA 98195 USA}

\emailAdd{jorge.casalderrey@ub.edu}
\emailAdd{gmilhano@lip.pt}
\emailAdd{pablosdaniel@uniovi.es}
\emailAdd{krishna@mit.edu}
\emailAdd{xjyao@uw.edu}

\preprint{MIT-CTP/6083, 
IQuS@UW-21-131}
\abstract{We develop an Efficient Wake procedure for computing the distribution of hadrons originating from jet wakes in heavy ion collisions --- the hydrodynamic response of a droplet of quark-gluon plasma to the energy and momentum deposited in it by high-energy partons propagating through it. The procedure employs the linearity of linearized hydrodynamics
and takes account of the effects of both longitudinal expansion and transverse radial flow on the hydrodynamic evolution of the wakes and on the resulting particle production at the freezeout hypersurface.
It makes repeated use of template solutions to linearized hydrodynamics in a Bjorken flow background with no transverse flow, templates that need only be computed once, and uses suitable rotations and boosts to map fluctuations from these templates to fluctuations at a point on the freezeout hypersurface in a way that 
incorporates the effects of the radial flow.
We benchmark this procedure by comparing its results to results obtained from full $(3+1)$-dimensional nonlinear hydrodynamics calculations, find reasonable 
agreement, and find that our Efficient Wake procedure yields a {\it much} better description of the distribution of hadrons originating from jet wakes than does the older oversimplified procedure employed in the Hybrid Model.
And, the Efficient Wake procedure {\it is} computationally efficient: it is at least tens of thousands of times faster than full nonlinear hydrodynamics calculations.
Hence, we anticipate that when our new procedure is implemented in Monte Carlo analyses of jets in heavy ion collisions, for example in the Hybrid Model, it will greatly improve the description of the soft component of many jet and jet substructure observables as compared to experimental data.
}
\maketitle

\section{Introduction}

When jets, which are collimated sprays of hadrons produced in high energy collisions, are produced in a heavy ion collision, they serve as good probes of the quark-gluon plasma (QGP) produced in the same collision. Two aspects of jet physics in proton-proton collisions are keys to the utility of jet measurements in heavy ion collisions for this purpose. First, each jet originates from a hard parton produced in a high momentum transfer scattering of partons from the incident protons (or nuclei) and the cross-sections for these hard scattering processes are reliably calculable. The initial hard parton then showers, and the development of the parton shower is also well-studied via from-first-principles QCD methods, often formulated as in effective field theory. The processes via which the shower of partons hadronizes are also well-measured and well-described. Second, the many jet and jet substructure observables that can be measured in heavy ion collisions have been widely studied in proton-proton collisions, both theoretically and experimentally. These two aspects together mean that when experimentalists look at how measured attributes of  jets and their substructure  in a sample of jets selected and reconstructed from heavy ion collisions differ from those in proton-proton collisions, the baseline upon which they are standing is well-established and well-understood. 

The modifications to jets and jet substructure in heavy ion collisions originate, broadly speaking, in two ways. First, as the energetic partons in a jet shower plow through the droplet of strongly coupled quark-gluon plasma produced in the same heavy ion collision, the jet shower is modified. Shower partons lose energy, and splittings in the shower can be modified in various ways, including via medium-induced radiation and via the scattering of jet partons off partons in the medium. All of these effects modify the energy and structure of the parton shower, and hence of the jet that results. Second, the jet modifies the medium through which it passes. In particular, the energy and momentum lost by the jet partons is deposited into the medium, creating a hydrodynamic wake in the droplet of quark-gluon plasma. Since this wake carries momentum in the jet direction, when it later hadronizes the soft particles that result must contribute to what an experimentalist reconstructs and calls a jet. That is, in heavy ion collisions jets as reconstructed by an experimentalist include not only the hadrons formed via the hadronization of the (modified) parton shower but also some hadrons originating from the wake in the droplet of QGP that the jet excites.

Theoretical studies of jet and jet substructure observables in proton-proton collisions are under good control, thanks to the separation of perturbative and nonperturbative physics, which enables the construction of factorization formulas for various observables. This approach is well established for the study of jets in proton-proton collisions, but it faces difficulties and challenges in heavy ion collisions due to the jet-medium interaction, which has two effects. The first effect is often called ``jet quenching": when the highly energetic partons (whose energy is much higher than the temperature of the QGP) in a jet travel through the droplet of QGP, they interact with the QGP  constituents and, as a result, lose energy to the medium. Many studies have been devoted to understanding the parton energy loss mechanism~\cite{Gyulassy:1993hr,Wang:1994fx,Baier:1994bd,Baier:1996kr,Zakharov:1996fv,Baier:1996sk,Gyulassy:1999zd,Gyulassy:2000fs,Wiedemann:2000za,Guo:2000nz,Gyulassy:2000er,Wang:2001ifa,Arnold:2002ja,Jeon:2003gi,Majumder:2007zh,CasalderreySolana:2011rz,Ovanesyan:2011xy,MehtarTani:2010ma,Blaizot:2013hx,Blaizot:2013vha,Chesler:2014jva,Casalderrey-Solana:2014bpa,Casalderrey-Solana:2015vaa,Chesler:2015nqz,He:2015pra,Ghiglieri:2015zma,Ghiglieri:2015ala,Casalderrey-Solana:2016jvj,Cao:2017zih,Hulcher:2017cpt,Caucal:2018dla,Casalderrey-Solana:2018wrw,He:2018xjv,Casalderrey-Solana:2019ubu,Isaksen:2020npj,Huss:2020whe,Barata:2021byj,Barata:2021wuf,Isaksen:2022pkj,Isaksen:2023nlr,Pablos:2024muu,Kudinoor:2025gao,Beraudo:2025nvq,Andres:2026qrt,Leitao:2026fgh,Hulcher:2026dht,Kudinoor:2026wcs}, which is a complicated task since the problem involves multiple scales: from the hard splitting of partons in the shower, to the soft, strongly coupled dynamics of the QGP medium. As a consequence of the partons in the jet shower losing energy, the jet itself loses energy and in addition the shape and substructure of the jet can be modified. 

It is well known that jet energy loss is sensitive to the history of splittings in the parton shower such that the amount of energy lost by the jet and the modification of the jet substructure vary enormously from one jet to another, even for jets originating from hard partons with the same initial energy. Modeling these event-by-event (here, jet-by-jet) fluctuations in how jets interact with the QGP medium and lose energy is crucial as for many jet observables they make a dominant contribution to how measured quantities differ in heavy ion collisions relative to proton-proton collisions. For example, because the jet production rate drops rapidly with jet energy, selecting a sample of jets in heavy ion collisions whose energy is above some cut selects those jets whose particular fragmentation history results
in selecting a sample of jets that are narrower, harder, and have very different jet-by-jet fluctuations than when the same selection criterion is applied to jets in proton-proton collisions~\cite{Milhano:2015mng,Rajagopal:2016uip,Casalderrey-Solana:2016jvj,Brewer:2017fqy,Hulcher:2017cpt,
Mehtar-Tani:2017web,Casalderrey-Solana:2018wrw,Casalderrey-Solana:2019ubu,Caucal:2019uvr,
Du:2020pmp,
Caucal:2021cfb,
Brewer:2021hmh,Pablos:2022mrx,Hulcher:2022kmn,Kudinoor:2025gao,Hulcher:2026dht}.
In addition, jet energy loss depends on the position in the transverse plane at which the initial hard scattering from which the jet originates occurred, as well as the direction in which the jet travels.  A jet that originates near the edge of the nuclear collision and travels outward cannot lose as much energy as a jet with the same fragmentation pattern that travels inward.  We note that the local temperature within the droplet of QGP produced in a heavy ion collision varies rapidly in time as well as in space, as it expands and cools rapidly after its formation. This too must be taken into account in modeling any measured jet observable. For all of these reasons and more, a significant amount of theoretical work has focused on the development of Monte Carlo studies of jet production in heavy ion collisions~\cite{Gyulassy:1994ew,Lokhtin:2005px,Renk:2008pp,Lokhtin:2008xi,Armesto:2009fj,Schenke:2009gb,Zapp:2011ya,Zapp:2012ak,Majumder:2013re,Zapp:2013vla,Renk:2013pua,Wang:2013cia,Casalderrey-Solana:2014bpa,Casalderrey-Solana:2015vaa,Casalderrey-Solana:2016jvj,Cao:2017hhk,Cao:2017zih,Hulcher:2017cpt,Cao:2017qpx,Casalderrey-Solana:2018wrw,He:2018xjv,Ke:2018jem,Casalderrey-Solana:2019ubu,Putschke:2019yrg,Dai:2019ddc,Caucal:2019uvr,Caucal:2020xad,Ke:2020clc,JETSCAPE:2020mzn,Brewer:2021hmh,Hulcher:2022kmn,Kurkela:2022qhn,Luo:2023nsi,Kudinoor:2025gao,Beraudo:2025nvq,Hulcher:2026dht,Kudinoor:2026wcs}. 
Recently, it has also been observed that future calculations performed on quantum computers could more directly study quantum interference effects in jet quenching~\cite{Barata:2021yri,Yao:2022eqm,Barata:2022wim,Barata:2023clv,Castro:2025ocx,Barata:2026icn}.

The second consequence of the jet-medium interaction is the medium response to the energy and momentum lost by the jet --- which must be deposited in the medium. 
The energy and momentum lost by the jet will evolve in the soft medium and, once it has hydrodynamized, will form a wake consisting of sound waves (regions in the droplet of QGP that are compressed and heated) as well as a region behind the jet in which the QGP fluid is moving in the jet direction. As the period of time during which the droplet of QGP and the jet wake within it evolve hydrodynamically comes to an end,  the jet wake will turn into hadrons as the medium hadronizes. Due to momentum conservation, the jet wake carries a net momentum in the jet direction corresponding to the momentum lost by the jet.
This means that the hadrons produced when the jet wake hadronizes have a net momentum  that is correlated with the momentum of the original jet. 
What an experimentalist reconstructs as a jet using a jet finding algorithm will then inevitably include both particles from the jet wake and particles originating from the parton shower of the initial energetic particles. No background subtraction procedure can remove all of the particles from the jet wake. Therefore, it is mandatory to develop a good theoretical description of the evolution of, and particle production from, the jet wake in order to calculate almost any jet or jet substructure observables, the only exceptions being observables whose definition includes sufficiently aggressive grooming so as to remove all soft particles entirely. 

Since the bulk dynamics of the droplet of QGP produced in a heavy ion collision is well described by hydrodynamics, the jet wake can also be described hydrodynamically as long as it hydrodynamizes quickly after the energy and momentum deposition by the jet. We shall make this assumption throughout this paper. One of the most interesting longer term goals of work in this direction is to compare theoretical predictions based upon assuming that the evolution of the wake is governed by hydrodynamics (followed by hadronization) with data with the hope of identifying deviations that can teach us about pre-hydrodynamic evolution of jet wakes. 

Early studies treated the energy and momentum lost by the jets as external currents and derived linearized hydrodynamic equations, assuming that the perturbation caused by the jet energy and momentum deposition into the medium is small~\cite{CasalderreySolana:2004qm,Ruppert:2005uz,CasalderreySolana:2006sq,Neufeld:2008fi,Neufeld:2008dx,Qin:2009uh,Neufeld:2009ep,Ayala:2012bv,Ayala:2014sua,Yan:2017rku}. More recent studies relaxed the assumption that the perturbation is linear and carried out more complete analyses using full hydrodynamics~\cite{Chaudhuri:2005vc,Betz:2010qh,Floerchinger:2014yqa,Tachibana:2014lja,Tachibana:2017syd,Tachibana:2020mtb,Cao:2020wlm,Cao:2022odi}.  A combination of hydrodynamics and weak-coupling partonic transport has also been 
developed~\cite{Chen:2017zte,Zhang:2018urd,Yang:2021qtl,Zhao:2021vmu,Yang:2022nei};  however, these full hydrodynamic analyses are computationally very expensive. The key assumption in these studies is that the jet wake can be described hydrodynamically, which means the energy-momentum tensor of the disturbance caused by the jet can be approximated by long-wavelength modes in a gradient expansion. However, close to the jet, the gradients are expected to be larger and the gradient expansion may break down. Nevertheless, various numerical studies of the out-of-equilibrium dynamics of gauge theories in both strongly-coupled and weakly-coupled scenarios~\cite{Chesler:2010bi, Heller:2011ju, Kurkela:2015qoa, Chesler:2015bba, Chesler:2016ceu} (reviews can be found in Refs.~\cite{ CasalderreySolana:2011us, Chesler:2015lsa, Busza:2018rrf, Florkowski:2017olj, Romatschke:2017ejr}) as well as  studies of hydrodynamic and pre-hydrodynamic 
attractors~\cite{Heller:2015dha,Kurkela:2015qoa,Strickland:2017kux,Blaizot:2019scw,Jaiswal:2019cju,Kurkela:2019set,Brewer:2019oha,Brewer:2022vkq} indicate that hydrodynamics may be applicable even when gradients are large and the system is far from local thermal equilibrium, as in a
pre-hydrodynamic epoch during which the dynamics is dominated by one or a few low-lying modes in an effective Hamiltonian~\cite{Brewer:2019oha,Brewer:2022vkq,Rajagopal:2024lou,Rajagopal:2025nca,DuPlessis:2026qjy}. 
Furthermore, explicit analysis of the medium response to an energetic colored particle in strongly-coupled gauge theories~\cite{Chesler:2007an,Gubser:2007ga,Chesler:2007sv,Chesler:2008wd,Chesler:2008uy,Gubser:2009sn,Chesler:2011nc,CasalderreySolana:2011us} demonstrates that, at least in the strongly coupled fluid of ${\cal N}=4$ supersymmetric Yang-Mills theory, the perturbation of the stress energy tensor caused by these energetic particles behave hydrodynamically even at distances as small as $1/\pi T$ away from the jet.

The authors of Ref.~\cite{Casalderrey-Solana:2016jvj} introduced an oversimplified calculation that yields a simple analytical expression for the momentum distribution of the soft hadrons that results as a jet wake freezes out.  
They then implemented this crude description of jet wakes in the Hybrid Model.
This calculation yields a distribution in the azimuthal angle of the momenta of the hadrons originating from a jet's wake relative to that of the jet itself that includes only a constant term and a cosine term,  proportional respectively to the 
energy and momentum that the jet loses to the hydrodynamic medium~\cite{Casalderrey-Solana:2016jvj}. This feature has been confirmed more recently via a different calculation that begins 
from kinetic theory~\cite{Kurkela:2026fiu}.
In both these contexts, this particular feature relies both upon assuming that the wake is hydrodynamic and upon neglecting transverse radial flow; the additional simplifying assumptions behind the  full analytic expression derived in Ref.~\cite{Casalderrey-Solana:2016jvj} are described there.  
Comparisons to then-available experimental data~\cite{Casalderrey-Solana:2016jvj} already indicated that 
the analytic expression from Ref.~\cite{Casalderrey-Solana:2016jvj}
yields a
$p_T$-distribution for the hadrons from the wake that is somewhat softer than in data
and indicates that a constant+cosine angular dependence
corresponds to a distribution of hadrons around the jet that is broader in angle
than in data.
More recent comparisons to experimental data~\cite{CMS:2025dua,CMS:2026mur}
indicate that 
the depletion of soft hadrons in the direction opposite to the jet caused by a jet wake can reasonably be characterized by a constant+cosine angular dependence.

In our previous work~\cite{Casalderrey-Solana:2020rsj}, we treated the wake as a hydrodynamical perturbation using the framework of linearized hydrodynamics on top of a Bjorken flow, which is a $(1+1)$-dimensional boost-invariant solution to the equations of viscous hydrodynamics with longitudinal expansion  but  no transverse expansion. We followed the evolution of the wake using linearized hydrodymamics until freezeout, when the wake becomes soft hadrons. By comparing our results to experimental data, we found that the particles produced from the wake left behind by a high-energy parton traversing the Bjorken fluid 
contain too many very soft particles ($p_T\lesssim 1$~GeV), just as is the case in the much more crude description of the jet wake that we have implemented in the Hybrid Model beginning with Ref.~\cite{Casalderrey-Solana:2016jvj}. 
Since the Bjorken fluid has no transverse flow, in Ref.~\cite{Casalderrey-Solana:2020rsj} we attempted to add effects of transverse flow by hand by boosting the momenta of the particles produced from the jet wake, using the position-dependent transverse flow at freezeout extracted from a $(2+1)$-dimensional viscous hydrodynamics simulator ``VISHNU''~\cite{Song:2007ux,Shen:2014vra}. We found that the transverse flow effect hardens the $p_T$ spectrum, meaning in this case that more particles with transverse momenta above $2$ GeV are produced, as the experimental data seem to require. We also noticed that the sound wave in the spacetime rapidity direction leads to a double-peak structure in the spacetime rapidity distribution of the wake-induced perturbations to quantities such as the energy and momentum densities, if and only if the deposition 
of energy and momentum into the fluid by the jet
ends long before freezeout. That is, this structure develops only if the wake has a long time to evolve hydrodynamically in the expanding background fluid. The double-peak structure in the spacetime rapidity distribution of the perturbed energy and momentum densities may lead to a double-peak structure in the observed momentum-rapidity distribution of the particles produced from the wake. Whether or not this happens depends on the magnitude and direction of the transverse flow velocity of the background fluid experienced by the perturbed fluid. Although the procedure of adding the transverse flow effect by hand that we introduced in Ref.~\cite{Casalderrey-Solana:2020rsj} improves the description of the particle production from the jet wake relative to the crude treatment in the Hybrid Model~\cite{Casalderrey-Solana:2016jvj} in the direction that a comparison with experimental data demands, in Ref.~\cite{Casalderrey-Solana:2020rsj} we did not perform any detailed comparisons between our simple but improved-by-hand description of the jet wake and a full nonlinear hydrodynamic analysis.

In this work, we will further improve upon our linearized hydrodynamics description of jet wakes by taking into account the transverse expansion of the droplet of QGP, as well as by accounting for the effects of transverse flow  at freezeout, in both cases doing so much more systematically. 
We will begin in Section~\ref{sect:linear} by reprising our previous work~\cite{Casalderrey-Solana:2020rsj} because we shall base the approach that we develop here on the use of template solutions to the $(1+1)$-dimensional linear hydrodynamics equations that describe the wake created by the deposition of a unit of energy and momentum in a Bjorken flow background, with no transverse flow. The full solutions of Ref.~\cite{Casalderrey-Solana:2020rsj} describing the wake of a high-energy parton in this simplified background fluid with no
transverse flow can then be composed from these template solutions via the method of Green's functions.
In Section~\ref{sect:lin_flow} we turn to the problem of interest, namely describing the wake of a high-energy parton traversing a droplet
of hydrodynamic QGP that is expanding radially in the transverse plane, as well as undergoing boost-invariant longitudinal expansion.  We develop a procedure that allows us to use the same template solutions constructed in Section~\ref{sect:linear} to obtain an approximate, but computationally efficient, description of the wakes of high-energy partons in this background. To use the method of Green's functions in full would require constructing orders of magnitude more different template solutions in this background itself. Our procedure does rely upon the linearity of linearized hydrodynamics, as in the method of Green's functions: we compute the hydrodynamic perturbation (the wake) at a specified point on the freezeout surface by superposing contributions sourced from each segment of the high-energy parton trajectory.
For each such source, we introduce a procedure via which we identify from which spacetime point, in which of the templates that we have tabulated
in Section~\ref{sect:linear}, we must map the hydrodynamic perturbations in the Bjorken-flow-template solution --- rotated and boosted suitably so as to take account of the transverse radial flow in the background fluid --- to hydrodynamic perturbations at each specified point on the freezeout surface.  The ``Efficient Wake'' procedure that we introduce here is efficient precisely because it takes advantage of the Bjorken flow template solutions, which need only be computed once.

Because the Efficient Wake procedure takes account of the effects of the strong transverse radial flow in the expanding droplet of QGP on the wake, it yields a very substantially better description of the hadrons produced when the droplet of fluid, including the wake therein, freezes out as described in Section~\ref{sect:cf} than is obtained via the oversimplified treatment introduced in Ref.~\cite{Casalderrey-Solana:2016jvj} and used since then in Hybrid Model studies. We benchmark our results for a  representative sample of 50 illustrative events, each containing a high-energy parton with a random point of origin in the transverse plane sampled using a Glauber model that describes the overlap between colliding nuclei, each traveling in a random direction 
through the droplet of QGP until they reach the freezeout hypersurface
against results obtained in Section~\ref{sect:music} by solving full nonlinear $(3+1)$-dimensional hydrodynamics for the jet wakes using the MUSIC 
code~\cite{Schenke:2010nt,Schenke:2010rr,Ryu:2015vwa,Paquet:2015lta}. Via this comparison,
we show that our approach is computationally less expensive than solving the full nonlinear hydrodynamics problem by at least a factor of many tens of thousands and can be faster by a factor of $200,000$ --- an Efficient Wake procedure indeed.  The approximations that we make in the procedure that we introduce in Section~\ref{sect:lin_flow} are not controlled by the smallness of the hydrodynamic perturbations, but we quantify their validity by comparing our results for the sample of 50 high-energy partons to the results from full nonlinear hydrodynamics calculations in Section~\ref{sect:results}. We conclude and look ahead in Section~\ref{sect:conclusions}.

We shall see that the momentum distribution of the particles produced from the wakes of a high-energy parton at freezeout depends sensitively and enormously on ``details'' of their point of origin, direction, and of the background fluid flow. This means that there is no way to obtain realistic predictions via calculating the wake only for some average configuration.  Quantitative predictions require sampling appropriately over many configurations in each one of which the dynamics of the wake and of the resulting particle production varies considerably.  Looking ahead, when in future we implement the calculation that we develop here into a jet Monte Carlo code, we will want to sum over $10^5$-$10^6$ jets, each with its own branching parton shower, and with each parton in each shower creating a unique wake which yields varying soft particle production at freezeout.
This makes computational efficiency imperative.

Our procedure allows us to investigate the effects of transverse radial flow effects on the wakes of high-energy partons and the distribution of hadrons that these wakes yield at freezeout case-by-case. We shall find that the spectra and angular and rapidity distributions of the soft particles originating from the wakes of high-energy partons are very different for distinct  configurations, depending on the value of the local transverse flow and the relative angle between the flow and the direction of motion of the high energy parton. As noted above, the very substantial dependence of the final state particle production from a jet wake on these ``details'' makes it imperative to sample over all configurations, with reliable knowledge of the wake from each, if one is to have a chance of describing experimental data.  Our approach makes this computationally efficient, which is to say possible.

Deploying the framework for treating particle production from jet wakes that we present in this paper in a jet Monte Carlo code is future work. What we do here is to gain confidence in our approach by comparing 
the particle production in our linearized description of jet wakes, built up from superposing templates, with that from a full nonlinear $(3+1)$-dimensional viscous hydrodynamics calculation for our sample of 50 high-energy partons with each of two different initial energies, finding good agreement with a more computationally efficient procedure that is faster by more than four orders of magnitude.

\section{Linearized Hydrodynamics on Bjorken Flow, and Template Solutions}
\label{sect:linear}

In this Section, 
we will begin by reviewing the framework of linearized hydrodynamics in a background Bjorken flow, which we developed in our 
previous work~\cite{Casalderrey-Solana:2020rsj}.
We will then describe how we construct
template solutions to the
$(1+1)$-dimensional 
equations that describe the linearized hydrodynamic response
of the Bjorken flow to the deposition of a unit of energy 
and momentum in the $x$-direction
at a specified point.
We will then show how to recover the full solutions of Ref.~\cite{Casalderrey-Solana:2020rsj}
that describe the wake that a high-energy parton excites in the 
Bjorken hydrodynamic background
by superposing template solutions.
The template solutions can be thought of as analogous to Green's functions,
with the analogue of the delta function source in the standard Green's function method being the 
deposition of one unit of energy and momentum at one point. The composition of the full solutions of Ref.~\cite{Casalderrey-Solana:2020rsj} from templates
is then the direct analogue of solving an inhomogeneous differential equation via the method of Green's functions.  
In Sections~\ref{sect:lin_flow} 
and \ref{sect:cf}
we shall deploy the template solutions from this Section in the service of describing how the wake that a high-energy parton deposits in a droplet of fluid that is expanding radially in the transverse plane (in addition to experiencing boost-invariant longitudinal Bjorken expansion) modifies the 
distribution of soft particles produced at freezeout. Our approach will be similar in spirit to the Green's function method, although not a direct analogue.

\subsection{First-order Hydrodynamics}

The system that we shall treat in subsequent Sections consists of a high-energy parton from a jet and the expanding cooling droplet of QGP. Without the jet-medium interaction, the bulk dynamics of the QGP is governed by conservation of energy and momentum, which can be written in terms of the stress-energy tensor $T^{\mu\nu}$
\be
\label{eqn:hydro0}
\nabla_\mu T^{\mu\nu} = 0 \,,
\ee
where $\nabla_\mu$ denotes the covariant derivative.
In general, the stress-energy tensor is constructed by using symmetry properties of the system and a systematic gradient expansion. The building blocks include the local energy density $\varepsilon$ and the fluid velocity field $u^\mu$ satisfying the normalization condition $u^2=1$. By using these building blocks, one can write down the most general expression for $T^{\mu\nu}$ that satisfies the symmetry properties and organize it in terms of the order of derivatives in the gradient expansion. To linear order in the gradient expansion, we have
\be
\label{eqn:Tmunu}
T^{\mu\nu} = (\varepsilon + P)u^\mu u^\nu - P g^{\mu\nu} + 2\eta \nabla^{\langle\mu} u^{\nu\rangle} \,,
\ee
where we have neglected the bulk viscosity for simplicity. Here $P$ denotes the pressure, $g^{\mu\nu}$ is the mostly minus metric of the spacetime under consideration, $\eta$ stands for the shear viscosity and 
\be
2\nabla^{\langle\mu} u^{\nu\rangle} = \Delta^{\mu\rho} \nabla_\rho u^\nu + 
\Delta^{\nu\rho} \nabla_\rho u^\mu -
\frac{2}{3} \Delta^{\mu\nu} \nabla_\rho u^\rho\,,
\ee
is the shear term, where we have defined $\Delta^{\mu\nu} \equiv g^{\mu\nu} - u^\mu u^\nu$.

When the jet-medium interaction is accounted for, jets lose energy and momentum during their evolution inside the QGP and the hydrodynamic equation (\ref{eqn:hydro0}) governing the bulk dynamics is modified. The total energy and momentum of the whole system are still conserved. But, if we focus on the bulk dynamics and assume the energy and momentum lost by the jets thermalize quickly and behave hydrodynamically, we can treat the energy and momentum lost by the jets as source terms in the conservation equations:
\be
\label{eqn:full_hydro}
\nabla_\mu T^{\mu\nu} = J^{\nu} \,,
\ee
where the external current $J^{\nu}$ depends on the distribution and in-medium dynamics of the jets. If the energy and momentum perturbation induced by the energy and momentum from the jet that is deposited into the background fluid is small, we can expand the stress-energy tensor to linear order in the perturbation and decompose the total stress-energy tensor into the background part and the perturbation, $T^{\mu\nu} = T^{\mu\nu}_0 + \delta T^{\mu\nu}$, which satisfy the following hydrodynamic equations, respectively:
\be
\label{eqn:hydro1}
\nabla_\mu T_0^{\mu\nu} &=& 0 \\
\nabla_\mu \delta T^{\mu\nu} &=& J^\nu \,.
\ee

In this Section, we will consider the case where the energy and momentum lost by a high-energy parton is deposited into a longitudinally boost invariant fluid that has no transverse expansion, also known as a Bjorken flow, which we will review now.

\subsection{Bjorken Flow}
To define the Bjorken flow, it is convenient to use the Milne coordinate system $(\tau,x,y,\eta_s)$\footnote{The coordinate transformation from Minkowski to Milne coordinates is given by $\tau=\sqrt{t^2-z^2}$ and $\eta_s=\frac{1}{2}\log \big( \frac{t+z}{t-z}\big)$.} where the subscript $s$ is inserted to distinguish the spacetime rapidity coordinate $\eta_s$ from the shear viscosity $\eta$. The metric of this coordinate system is given by
\be
g_{\mu\nu} = \rm{diag} (1, -1, -1, -\tau^2) \,.
\ee
In this coordinate system, the fluid velocity field of the unperturbed Bjorken flow is given by
\be
\label{eqn:u0}
u^\mu_0 \equiv (u^\tau_0, u^x_0, u^y_0, u^{\eta_s}_0) = (1,0,0,0) \,,
\ee
where the subscript $0$ is included to distinguish the quantities of the background fluid from those of the perturbed fluid. In Minkowski coordinates, the fluid velocity field is written as
\be
\label{eqn:u0_cartesian}
u^\mu_0 = (u^t_0, u^x_0, u^y_0, u^z_0) = (\cosh\eta_s,0,0,\sinh\eta_s) \,. 
\ee
Then, the hydrodynamic equation that describes the background Bjorken flow can be shown to take the form
\be
\label{eqn:bjorken}
\frac{\partial \varepsilon_0}{\partial\tau}  + \frac{\varepsilon_0 + P_0 }{\tau} - \frac{4\eta_0}{3\tau^2} = 0\,.
\ee
In practice, to solve the hydrodynamic equation (\ref{eqn:bjorken}), we also need an equation of state that connects the local pressure and energy density.
In this work, 
for simplicity we will use the conformal equation of state
\be
\label{eqn:eos1}
\frac{P_0}{\varepsilon_0} = \frac{1}{3}
\ee
for which the speed of sound is $c_s^2=1/3$.
We also need an ansatz for the shear viscosity and, again for simplicity, we will use the result for a strongly coupled conformal fluid, 
$\eta_0=s_0/(4\pi)$~\cite{Kovtun:2004de},  where the entropy density of the unperturbed fluid is related to its energy density and  pressure via
\be
\label{eqn:eos2}
\varepsilon_0+P_0 = Ts_0 \,.
\ee
To obtain the time dependence of the temperature in the Bjorken background flow, we will further assume $\varepsilon_0 \propto T^4$ which, together with Eq.~\eqref{eqn:eos1}, implies $P_0 \propto T^4$.

\subsection{\label{linbj}Linear 
Perturbations to Bjorken Flow}

The perturbed fluid velocity field can be written as
\be
\label{eqn:u}
u^\mu = u^\mu_0 + \delta u^\mu = (1, \delta u^x, \delta u^y, \delta u^{\eta_s}) \,,
\ee
which has no perturbation in $u^\tau_0$ at linear order because of the normalization condition $u^\mu u_\mu=1$. 
At the same time, the energy density, pressure and viscosity are also perturbed ($\varepsilon=\varepsilon+\delta\varepsilon$, $P=P_0+\delta P$, $\eta=\eta_0+\delta \eta$) with the perturbations written as $\delta\varepsilon$, $\delta P = c_s^2 \delta \varepsilon$ and $\delta\eta =\gamma_\eta \delta\varepsilon$, respectively, where $\gamma_\eta$ is defined as
\be
\label{eqn:gamma-eta}
\gamma_\eta \equiv \frac{\eta_0}{\varepsilon_0+P_0} = \frac{\eta_0}{Ts_0}\,.
\ee
The total stress-energy tensor $T^{\mu\nu}$ with the perturbation can be obtained by plugging all the perturbed quantities into Eq.~(\ref{eqn:Tmunu}). Subtracting the background stress-energy tensor $T^{\mu\nu}_0$ from $T^{\mu\nu}$ and expanding to  linear order in perturbation  leads to the stress-energy tensor $\delta T^{\mu\nu}$. After some algebra, we find the equations describing the linearized hydrodynamics on Bjorken flow: 
\be
\partial_\tau \delta\varepsilon + \Big(1-\frac{\gamma_\eta}{\tau}\Big)\frac{\delta\varepsilon + \delta P}{\tau}
+ \partial_x \Big( (\varepsilon_0 + P_0)\delta u^x + \frac{4\eta_0}{3\tau}\delta u^x \Big) && \nn\\
+ \partial_y \Big( (\varepsilon_0 + P_0)\delta u^y + \frac{4\eta_0}{3\tau}\delta u^y \Big) + \partial_{\eta_s} \Big( (\varepsilon_0 + P_0)\delta u^{\eta_s} - \frac{8\eta_0}{3\tau}\delta u^{\eta_s} \Big)&=& J^\tau \label{FirstLinearizedHydroEq}\\
\Big( \partial_\tau + \frac{1}{\tau} \Big) \Big( (\varepsilon_0 + P_0) \delta {\bs u}^\perp + \frac{2\eta_0}{3\tau} \delta {\bs u}^\perp \Big) + {\bs \partial}^\perp \delta P  + \frac{2\gamma_\eta}{3\tau}{\bs \partial}^\perp  \delta\varepsilon   && \nn\\
- \eta_0 \Big( \partial^{\perp2} + \frac{\partial_{\eta_s}^2}{\tau^2} \Big) \delta {\bs u}^\perp - \frac{1}{3}\eta_0  {\bs \partial}^\perp  \Big({\bs \partial}^\perp \cdot \delta {\bs u}^\perp + \partial_{\eta_s} \delta u^{\eta_s} \Big) &=& {\bs J}^\perp \label{SecondLinearizedHydroEq} \\
\Big( \partial_\tau + \frac{3}{\tau} \Big) \Big( (\varepsilon_0 + P_0) \delta u^{\eta_s} - \frac{4\eta_0}{3\tau} \delta u^{\eta_s} \Big) + \frac{1}{\tau^2}\partial_{\eta_s} \delta P - \frac{4\gamma_\eta}{3\tau^3}\partial_{\eta_s} \delta\varepsilon  &&\nn\\
- \eta_0 \Big( \partial^{\perp2} + \frac{\partial_{\eta_s}^2}{\tau^2} \Big) \delta u^{\eta_s} - \frac{1}{3\tau^2} \eta_0 \partial_{\eta_s} \Big( {\bs \partial}^\perp \cdot \delta {\bs u}^\perp + \partial_{\eta_s} \delta u^{\eta_s} \Big) &=& J^{\eta_s} \,,\label{ThirdLinearizedHydroEq}
\ee
where the bold symbols denote Euclidean 2-vectors in the transverse plane and the dot product between two Euclidean 2-vectors is given by ${\bs a}\cdot{\bs b} = a^xb^x + a^yb^y$. All the perturbed quantities and the external source terms are functions of $(\tau,x,y,\eta_s)$.

We will solve the linearized hydrodynamics equations in momentum space.  To this end, we define the momentum perturbation in terms of the velocity perturbation
\be
\label{eqn:define_gperp}
{\bs g}^\perp(\tau, {\bs x}^\perp, \eta_s) &\equiv & (\varepsilon_0 + P_0) \delta {\bs u}^\perp \\
\label{eqn:define_geta}
g^{\eta_s} (\tau, {\bs x}^\perp, \eta_s) &\equiv & (\varepsilon_0 + P_0) \delta u^{\eta_s} \,.
\ee
The Fourier transform that connects position and momentum space is given by
\be
X(\tau, {\bs x}^\perp, {\eta_s}) & = & \int \frac{\diff k^{\eta_s} \diff^2k^\perp }{(2\pi)^3} e^{i{\bs k}^\perp \cdot {\bs x}^\perp + i k^{\eta_s} {\eta_s}} \tilde{X}(\tau, {\bs k}^\perp, k^{\eta_s}) \,,
\ee
where the symbol $X$ denotes $\delta\varepsilon$, $g^x$, $g^y$, $g^{\eta_s}$ or $J^\mu$. The equations of linearized hydrodynamics ---  (\ref{FirstLinearizedHydroEq}), (\ref{SecondLinearizedHydroEq}) and (\ref{ThirdLinearizedHydroEq}) ---  can now be written in momentum space as
\be
\label{eqn:li_hydro1}
\Big(\partial_\tau + \frac{1+c_s^2}{\tau} \Big) \delta \tilde{\varepsilon} + i{\bs k}^\perp \cdot \tilde{\bs g}^\perp  + i k^{\eta_s} \tilde{g}^{\eta_s} &=& \tilde{J}^\tau\\
\label{eqn:li_hydro2}
\Big(\partial_\tau + \frac{1}{\tau} \Big) \tilde{\bs g}^\perp + ic_s^2 {\bs k}^\perp \delta\tilde{\varepsilon} + \gamma_\eta \Big( {k^\perp}^2 + \frac{{k^{\eta_s}}^2}{\tau^2} \Big) \tilde{\bs g}^\perp + \frac{1}{3}\gamma_\eta {\bs k}^\perp ({\bs k}^\perp \cdot \tilde{\bs g}^\perp + k^{\eta_s} \tilde{g}^{\eta_s}) &=& \tilde{\bs J}^\perp \\
\label{eqn:li_hydro3}
\Big(\partial_\tau + \frac{3}{\tau} \Big) \tilde{g}^{\eta_s} + \frac{ic_s^2 k^{\eta_s} }{\tau^2}  \delta\tilde{\varepsilon} + \gamma_\eta \Big( {k^\perp}^2 + \frac{{k^{\eta_s}}^2}{\tau^2} \Big) \tilde{g}^\eta + \frac{1}{3\tau^2}\gamma_\eta k^{\eta_s} ({\bs k}^\perp \cdot \tilde{\bs g}^\perp + k^{\eta_s} \tilde{g}^{\eta_s}) &=& \tilde{J}^{\eta_s} \,.\ \ \ \ \ \ \ 
\ee
In these equations, we have dropped terms that are suppressed by a factor $\gamma_\eta / \tau$ relative to terms that have no such factors. This approximation is well motivated by the gradient expansion used in the construction of hydrodynamics. (The gradient of fluid velocity fields scales as $1/\tau$ in the Bjorken flow.)

To solve the linearized hydrodynamic equations, we also need explicit expressions for the external source terms $\tilde{J}^\mu$. To make our considerations general, we will write the external currents in terms of the energy loss rate ${\diff E}/{\diff\tau}$. Various studies of jet energy loss rates have been carried out in both the weak-coupling and strong-coupling scenarios, but the hydrodynamization process via which the energy lost by the jet partons and deposited in the medium becomes a perturbation to the hydrodynamic fluid is not 
fully understood. For both simplicity and concreteness, in the following construction we will assume that the energy and momentum deposited into the hydrodynamic fluid by the jets thermalize locally in the QGP fluid immediately after the deposition.

While depositing energy into the fluid, the jets evolving inside the QGP also lose momentum, which also modifies the hydrodynamic fluid. The momentum loss rate can be related to the energy loss rate under the assumption that the parton losing energy moves at the speed of light with a fixed spacetime rapidity $\eta_{s\,\rm{parton}}$ and transverse direction $\hat{n}^\perp_{\rm parton}$, meaning that
\be
\label{eqn:dP}
\diff P^i &=& v^i_{\rm parton} \diff E  \\
\label{eqn:vperp}
v^\perp_{\rm parton} &=& \frac{\hat{n}^\perp_{\rm parton}}{\cosh(\eta_{s\, {\rm parton}})} \\[4pt]
\label{eqn:vz}
v^z_{\rm parton} &=& \tanh{(\eta_{s\, {\rm parton}})} \,.
\ee
We can then relate the external source terms $J^\mu$ to the energy and momentum loss rates by using Stokes' theorem. (A detailed explanation of this application of Stokes' theorem can be found in the Appendix of Ref.~\cite{Casalderrey-Solana:2020rsj}.) Assuming the perturbation induced by the jet energy loss drops sufficiently fast at large spacetime rapidity, we can integrate the external currents over a spacelike hypersurface at a fixed $\tau$ to obtain
\be
\label{eqn:conserve_ef}
\int \tau \diff x \diff y \diff \eta_s  \,J^t &=& \frac{\diff}{\diff\tau} \int \tau \diff x \diff y \diff \eta_s\, \delta T^{t \tau}  =\frac{\diff E}{\diff\tau}
\\
\label{eqn:conserve_pf}
\int \tau \diff x \diff y \diff \eta_s  \,J^i &=& \frac{\diff}{\diff\tau} \int \tau \diff x \diff y \diff \eta_s\, \delta T^{i \tau}  =\frac{\diff P^i}{\diff\tau}\,,
\ee
where the superscripts $t, \, i=x,\, y,\, z$ denote Minkowski spacetime 
components of $J^\mu$ and $\delta T^{\mu\tau}$.  The physical meaning of the above equations \eqref{eqn:conserve_ef} and \eqref{eqn:conserve_pf} is just conservation of energy and momentum: the energy and momentum lost by the parton become part of the energy and momentum stored in the hydrodynamic fluid and, hence,  the total energy and momentum are conserved. These equations only constrain the normalization of the external source terms $J^\mu(\tau,x,y,\eta_s)$ but do not determine their dependence on spacetime coordinates, which is necessary to carry out the Fourier transform and obtain the external currents in spatial momentum space $\tilde{J}^\mu$ that show up in the linearized hydrodynamic equations \eqref{eqn:li_hydro1}, \eqref{eqn:li_hydro2} and \eqref{eqn:li_hydro3}. We will assume that the energy and momentum perturbations are initially distributed in a small region around the local spacetime point where the parton lost energy and momentum, energy and momentum that immediately thermalized.  More specifically, we assume the source terms that specify the initial energy and momentum perturbation distributions can be parametrized by Gaussian functions
\be
\label{eqn:Jmu}
&&J^{\mu} (\tau,x,y,\eta_s) \nn\\
&=& \frac{C^\mu(\tau)}{(2\pi)^{3/2}\sigma_x^2 \sigma_{\eta_s}} \exp\bigg(-\frac{(x-x_{\rm parton}(\tau))^2+(y-y_{\rm parton}(\tau))^2}{2\sigma_x^2} - \frac{(\eta_s -\eta_{s\, {\rm parton}})^2}{2\sigma_{\eta_s}^{\ 2}}\bigg)  \,, \ \ \ \ \ \ \ \ 
\ee
where $(x_{\rm parton}(\tau),y_{\rm parton}(\tau),\eta_{s\, {\rm parton}})$ denotes the transverse position and the spacetime rapidity at time $\tau$,  with the spacetime rapidity of the parton being constant during the evolution.
Here $\sigma_x$ and $\sigma_{\eta_s}$ represent the Gaussian widths in the transverse and spacetime rapidity directions respectively. In practice, we will choose the Gaussian widths in the source terms (\ref{eqn:Jmu}) to be $\sigma_x=\frac{1}{\pi T}$ and $\sigma_{\eta_s} = \frac{1}{\pi}$ respectively. The $\tau$-dependent normalization factors $C^\mu(\tau)$ are fixed by the energy and momentum loss rates through the conservation laws, i.e. Eqs.~\eqref{eqn:conserve_ef} and \eqref{eqn:conserve_pf}. Using the relation between contravariant vectors in the Minkowski coordinate system and those in the Milne coordinate system, we find
\be
\label{eqn:C_tau}
C^\tau(\tau) &=& \frac{1}{\cosh(\eta_{s\, {\rm parton}})} \frac{1}{\tau}\frac{\diff E}{\diff \tau}\exp\Big(-\frac{\sigma_{\eta_s}^2}{2}\Big) \\
\label{eqn:C_perp}
C^\perp(\tau) &=& \frac{\hat{n}^\perp_{\rm parton}}{\cosh(\eta_{s\, {\rm parton}})} \frac{1}{\tau}\frac{\diff E}{\diff \tau} \\[4pt]
\label{eqn:C_eta}
C^{\eta_s}(\tau) &=& 0 \ .
\ee
The fact that $C^{\eta_s}(\tau) = 0$ just reflects our assumption of the constant spacetime rapidity of the parton, i.e., the parton does not deposit momentum along the spacetime rapidity direction. Combining Eqs.~\eqref{eqn:Jmu}, \eqref{eqn:C_tau}, \eqref{eqn:C_perp} and \eqref{eqn:C_eta}, we find that the external currents in momentum space take the form
\be
\label{eqn:Jtilde}
\widetilde{J}^\mu(\tau,{\bs k}^\perp,k^{\eta_s}) &=& C^\mu(\tau) \exp\bigg( -\frac{({\bs k}^\perp\sigma_x)^2 + (k^{\eta_s} \sigma_{\eta_s})^2}{2} \bigg) \nn\\
&\times& \exp\Big(-i \big( k^x x_{\rm{parton}}(\tau) + k^y y_{\rm{parton}}(\tau) +  k^{\eta_s} \eta_{s\,\rm{parton}} \big) \Big) \,,
\ee
where ${\bs k}^\perp=(k^x, k^y)$ is a Euclidean 2-vector. The linearized hydrodynamic equations \eqref{eqn:li_hydro1}, \eqref{eqn:li_hydro2} and \eqref{eqn:li_hydro3}, together with the expression \eqref{eqn:Jtilde} for the external currents, complete our construction of the linearized hydrodynamics on top of a Bjorken flow. 

This concludes our review. 
In the next subsection, we will discuss how to efficiently solve the linearized hydrodynamic equations via numerical methods that take advantage of their linearity by first building a library of templates and then building solutions of interest by superposition.

\subsection{Numerical Construction of Templates}
\label{sec:NumericalSolution}

In the following, we will solve the linearized hydrodynamic equations efficiently by employing their linearity. 
In Section~\ref{sec:GFBj}, we will consider
jet wakes
on top of a Bjorken flow background.
For this simplified example,
we will use the method of Green's functions.  But later, 
when we consider a background that includes radial flow in Section~\ref{sect:lin_flow}, we will use the Green's functions from this Section, but will not directly follow the
method of Green's functions.  For this reason, henceforth we will refer to the Green's functions as template solutions.
We will begin by solving the linearized hydrodynamic equations for point-like source terms with one unit of energy and momentum deposition along a chosen axis, which we will choose to be the $x$-axis.
In Section~\ref{sec:GFBj}, we will 
then take suitable linear combinations of these template solutions, in which the weight of the linear combination is determined by the amount of energy and momentum lost by the jet partons under consideration. Proper rotations and shifts of the template solutions are needed to account for the partons moving in varying directions, at varying locations.
In Section~\ref{sect:lin_flow},
when we consider a droplet of QGP with transverse (in fact radial) flow, we will develop a method 
for using these template solutions motivated by, but different from, the method of Green's functions.
The advantage of this method, where we begin by computing a library of template solutions, is computational efficiency. The template solutions only need to be solved for once and saved into files. To obtain the solutions of wakes left behind by jet partons in real physically relevant studies and the resulting particle production, all we will need to do is take weighted linear combinations and perform proper rotations, boosts and shifts, which has a low computational cost.

The first thing we need to do is obtain the template solutions, which are solutions to the linearized hydrodynamic equations in response to a point-like deposition of one unit of energy and momentum along the $x$-axis in a Bjorken flow background. We take the one unit to be $1$ GeV. The deposition time for each template is specified as $\tau^d_{\rm temp}$, which we will vary so that template solutions with different deposition times can be superposed later. The spatial deposition point is arbitrary and we choose it to be the origin of the transverse plane and $\eta_s=0$. That is, we set $x_{\rm parton}(\tau^d_{\rm temp})=0$, $y_{\rm parton}(\tau^d_{\rm temp})=0$ and $\eta_{s\,{\rm parton}}=0$ in Eq.~\eqref{eqn:Jtilde}. Later, when we take the linear combinations needed to describe realistic situations in which the jet parton keeps losing energy along a straight path, we will take the direction- and position-dependence of the deposition into account by properly rotating each template solution and shifting the origin of each template solution.

We denote the template solutions 
at a time $\tau>\tau^d_{\rm temp}$ by
$\delta\varepsilon_{1x}(\tau^d_{\rm temp}; \tau-\tau^d_{\rm temp},x,y,\eta_s)$, ${\bs g}^\perp_{1x}(\tau^d_{\rm temp}; \tau-\tau^d_{\rm temp},x,y,\eta_s)$ and $g^{\eta_s}_{1x}(\tau^d_{\rm temp}; \tau-\tau^d_{\rm temp},x,y,\eta_s)$, 
with the subscript $1x$ indicating that this solution describes the response of the fluid to the deposition of 1 GeV of energy and 1 GeV$/c$ of momentum in the $x$-direction.
The first argument reminds us that the template solutions also depend directly on the deposition time $\tau^d_{\rm temp}$, since the temperature of the background fluid at this deposition time determines both the width of the Gaussian source terms in the transverse plane  
and the time it takes for the fluid to evolve from this deposition time 
until freezeout.

As the discussion above makes apparent, in order to obtain the template solutions, we shall need the time dependence of the temperature of the background Bjorken fluid. The temperature of the Bjorken fluid at an arbitrary time $\tau$ can be obtained from Eqs.~\eqref{eqn:bjorken}, \eqref{eqn:eos1} and \eqref{eqn:eos2}, which leads
in the case of a conformal fluid to
\be
\label{eqn:Tvstau}
\frac{\diff T(\tau)}{\diff\tau} = -\frac{T(\tau)}{3\tau} + \frac{1}{9\pi\tau^2}\,,
\ee
where we have taken the ratio of the shear viscosity and the entropy density of the unperturbed fluid to be $\eta_0/s_0=1/(4\pi)$ and have used the fact that for a conformal fluid the energy density of the unperturbed fluid satisfies $\varepsilon_0\propto T^4$ and the speed of sound satisfies $c_s^2=1/3$.\footnote{
For a generic equation of state and a generic temperature-dependent specific shear viscosity, 
instead of Eq.~\eqref{eqn:Tvstau} we would have
$$
\frac{dT(\tau)}{d\tau} 
= -\frac{c_s^2(T)\,T}{\tau}\left(1-\frac{4
\frac{\eta_0}{s_0}(T)}{3 \tau T}\right)
$$
with $c_s$ the speed of sound.
}
The solution to the temperature evolution equation (\ref{eqn:Tvstau}) is given by
\be
\label{eqn:sol_Tvstau}
T(\tau) = \Big(T(\tau_0)+\frac{1}{6\pi\tau_0} \Big) \Big( \frac{\tau}{\tau_0} \Big)^{-1/3} - \frac{1}{6\pi\tau} \,.
\ee
We fix the initial temperature by first fixing the starting time of the Bjorken flow to be $\tau_0=0.4$~fm$/c$ and taking the freezeout temperature to 
be $T^f=145$~MeV. For the center of the droplet of QGP produced in a central PbPb collision with center of mass energy $\sqrt{s_{NN}}=5.02$~TeV, 
a full hydrodynamic simulation such as the one described in Section~\ref{sect:music} gives a freezeout time of $\tau^f=13.6$~fm$/c$ for the values of $\tau_0$
and $T^f$ that we have specified. 
With this as motivation, we will use $\tau^f=13.6$ fm$/c$ for the Bjorken flow background, meaning that in all of our templates we take $\tau^f_{\rm temp}=13.6$~fm$/c$.
With the freezeout time chosen as above, the initial temperature of the Bjorken flow is then $T_0=446$ MeV for the center of the QGP in central collisions, which is obtained from Eq.~(\ref{eqn:sol_Tvstau}) by using $T(\tau^f)= T^f=145$ MeV. In real heavy ion collisions, the freezeout temperature is independent of the centrality of the collision and the location but the freezeout time depends on them. In future studies of collisions with a nonzero impact parameter, the values of $T_0$ and $\tau^f$ for the Bjorken background flow will be different.
And, in the central collisions with radial flow that we shall consider in Section~\ref{sect:lin_flow},
different points in the expanding cooling droplet of QGP will have different freezeout times. We will discuss how to handle this in 
Sections~\ref{sect:lin_flow} and~\ref{sect:cf}. 
Note also that in these later Sections, when we include the effects of radial transverse flow, the time dependence of the temperature of the background fluid, $T$, will no longer be given by Eq.~\eqref{eqn:sol_Tvstau} and will be different at different points in the transverse plane. 
This is a concrete indication
that when we employ the template solutions from this Section, which are solutions to the equations of linearized hydrodynamics in
a Bjorken flow background with no radial flow, in Section~\ref{sect:lin_flow},
we will not be able to follow the method of Green's functions directly there as we shall be able to do in this Section.

We carry out the calculation of the template solutions by first solving the Fourier transformed version of the linearized hydrodynamic equations in  momentum space, namely Eqs.~\eqref{eqn:li_hydro1}, \eqref{eqn:li_hydro2}, \eqref{eqn:li_hydro3}, and then taking the inverse Fourier transform. The momenta are discretized on a grid of size $N_x\times N_y\times N_{\eta}$, where $N_i$ is the number of grid points in the $i$-direction in momentum space. In obtaining numerical solutions to the linearized hydrodynamics equations, we have employed $N_x=N_y=333$ and $N_\eta=751$. 
The momentum ranges are chosen to be $k^x\in[-9.41,9.41]$ GeV, $k^y\in[-9.41,9.41]$ GeV, $k^\eta\in[-60,60]$. For each momentum grid point, we solve the linearized hydrodynamics equations
in momentum space by using the fourth-order Runge-Kutta method. 
In practice, we speed up the calculation by employing symmetries. The solutions of the wake are invariant under both  reflections $y\to -y$ and $\eta_s\to -\eta_s$, since we are taking the $x$-axis to be the direction of the momentum deposited in the fluid 
and hence $\tilde{J}^y= \tilde{J}^{\eta_s} =0$. By examining Eqs.~\eqref{eqn:li_hydro1}, \eqref{eqn:li_hydro2} and \eqref{eqn:li_hydro3} and noticing that the source terms in momentum space are also invariant under the reflections $y\to -y$ 
and $\eta_s\to -\eta_s$ 
since 
we are taking the origin 
of our spatial coordinates to be at
the deposition point,
we find that the solutions satisfy the following symmetry properties:
\be
\delta\tilde{\varepsilon}(\tau, k^x, k^y, k^{\eta_s}) &=& \delta\tilde{\varepsilon}(\tau, k^x, -k^y, k^{\eta_s}) = \delta\tilde{\varepsilon}(\tau, k^x, k^y, - k^{\eta_s}) \\[4pt]
\tilde{g}^x(\tau, k^x, k^y, k^{\eta_s}) &=& \tilde{g}^x(\tau, k^x, -k^y, k^{\eta_s}) = \tilde{g}^x(\tau, k^x, k^y, - k^{\eta_s}) \\[4pt]
\tilde{g}^y(\tau, k^x, k^y, k^{\eta_s}) &=& -\tilde{g}^y(\tau, k^x, -k^y, k^{\eta_s}) = \tilde{g}^y(\tau, k^x, k^y, - k^{\eta_s}) \\[4pt]
\tilde{g}^{\eta_s}(\tau, k^x, k^y, k^{\eta_s}) &=& \tilde{g}^{\eta_s}(\tau, k^x, -k^y, k^{\eta_s}) = - \tilde{g}^{\eta_s}(\tau, k^x, k^y, - k^{\eta_s}) \,.
\ee
Therefore, we only need to solve the linearized hydrodynamic equations in momentum space for $k^y\geq0$ and $k^{\eta_s}\geq0$. Solutions with $k^y<0$ or $k^{\eta_s}<0$ can be obtained by using the above symmetry relations.

After solving the linearized hydrodynamics equations at each point in the momentum-space grid,
we perform the inverse Fourier transform to obtain a template solution 
\begin{eqnarray}
\delta\varepsilon_{1{x}}(\tau^d_{\rm temp};\tau-\tau^d_{\rm temp},x,y,\eta_s)\, ,\nonumber\\ {\bs g}^\perp_{1{x}}(\tau^d_{\rm temp};\tau-\tau^d_{\rm temp},x,y,\eta_s)\, , \nonumber\\
g^{\eta_s}_{1{x}}(\tau^d_{\rm temp};\tau-\tau^d_{\rm temp},x,y,\eta_s)\, ,
\label{eq:TemplateSolution}
\end{eqnarray}
in position space.  
Recall that the subscript {$1{x}$} reminds us that this is a template solution describing the response of the Bjorken flow to a {\it unit} deposition of energy and momentum, with the momentum along the $x$-axis.
With the choices of the momentum ranges that we have made above, the spatial resolution is $\Delta x=\Delta y=\frac{\pi}{(k^x)_{\ma{max}}} \approx 0.0666$ fm in the transverse plane and $\Delta \eta_s = \frac{\pi}{60}\approx 0.052$ in spacetime rapidity. The range of  $(x, y, \eta_s)$-space included in our calculation is a box of size $\frac{\pi N_x}{(k^x)_{\ma{max}}} \times \frac{\pi N_y}{(k^y)_{\ma{max}}}  \times \frac{\pi N_\eta}{(k^\eta)_{\ma{max}}} $ centered at the origin, which corresponds to the deposition point. The upper limits of the momentum ranges specified above are chosen such that contributions outside the ranges are suppressed exponentially since the Gaussian source terms are suppressed exponentially. 

We compute template solutions \eqref{eq:TemplateSolution}
for a set of values of $\tau^d_{\rm temp}$ in the range between $\tau_0=0.4$~fm$/c$ and $\tau^f_{\rm temp}=13.6$~fm$/c$ at intervals $\Delta \tau$, which we take to be $\Delta \tau=0.4$~fm$/c$ meaning that we compute and tabulate 34 template solutions
\eqref{eq:TemplateSolution}.
We choose this range of deposition times because later we will need to superpose templates describing the response of the (Bjorken) flow
to unit-depositions along the entire trajectory of a high-energy parton starting from $\tau_0$ until either it loses all of its energy or (in Section \ref{sect:lin_flow}) it exits the
droplet of QGP or until $\tau^f_{\rm temp}$, whichever comes first.
We shall refer to the value of $\tau^{d,k}_{\rm temp}$ for the $k$'th template as $\tau^{d,k}_{\rm temp}\equiv k \Delta\tau$, where $k$ runs from 1 to 34.
Since later we will only be interested in 
the hydrodynamic perturbations at the freezeout time, for each of the 34 template solutions \eqref{eq:TemplateSolution} we shall actually tabulate its value at a grid of points in $(x,y,\eta_s)$ only at $\tau=\tau_{\rm temp}^f$. We shall choose to record the values of the hydrodynamic perturbations at the freezeout time on a grid with spacings $\Delta x=\Delta y=0.395$~fm and $\Delta\eta_s=\frac{\pi}{20}\approx 0.157$,
which is a coarser grid than the grid on which we have done the calculation: we record the values of the hydrodynamic perturbations only at $1/6$ of all the $x$ and $y$ values and $1/3$ of all the $\eta_s$ values at which we have calculated them. We make this choice for computational efficiency 
in our subsequent calculations.
We have checked that the hydrodynamic perturbations at the freezeout time hardly vary on length scales shorter than the spacings in this coarsened grid. 
We also choose to tabulate the perturbations only over the range $-3.456<\eta_s<3.456$, corresponding to 45 points in the coarsened $\eta_s$-grid, because the wake caused by a unit deposition of energy and momentum at the earliest possible deposition time has not reached larger values of $|\eta_s|$ at the freezeout time.
The grid of points at which we tabulate the hydrodynamic perturbations for each of the 34 templates has 55 grid points in $x$ and $y$, extending over a range $-10.77~{\rm fm}<x<10.57~{\rm fm}$ and  $-10.77~{\rm fm}<y<10.57~{\rm fm}$.

With the tabulated
template solutions \eqref{eq:TemplateSolution}
in hand, recalling that these are solutions to the linearized hydrodynamic equations in response to a point-like $1$ GeV deposition of energy and momentum along the $x$-axis at time $\tau^d_{\rm temp}$, we are now ready to take weighted linear combinations 
and apply suitable rotations and shifts 
in order to obtain linearized hydrodynamics solutions describing the wake of a high-energy parton losing energy
in the Bjorken flow background. We shall complete this simplified example in the next subsection.

\subsection{\label{sec:GFBj}Linearized Hydrodynamics on Bjorken Flow}

As a simple, but illustrative, example of the use of template solutions, 
in this subsection we show how to use the templates that we have computed and tabulated in the previous subsection to obtain the solutions
to linearized hydrodynamics
that describe the wake that high-energy parton traveling along a straight line excites in a Bjorken flow background, reproducing the
results of Ref.~\cite{Casalderrey-Solana:2020rsj}.

We consider a high-energy parton that starts at $(\tau_i,x_i,y_i,\eta_s=0)$, moves at the speed of light in an arbitrary direction in the transverse plane specified by the polar angle $\theta$, and keeps losing energy and momentum from time $\tau_i$ until $\tau_e$. It is important that in our notation we maintain a distinction between the time $\tau_e$ when
energy loss stops and the freezeout time $\tau^f$, since not every jet parton keeps losing energy until freezeout: some partons lose all of their energy at a time $\tau_e$ before reaching the freezeout hypersurface. (We also note that in the next Section, the freezeout hypersurface
will anyway not be at a constant $\tau^f$.)
We shall also treat $\tau_i$ as 
 an arbitrary parameter; in the Hybrid Model it is usually taken as either 0.4 or 0.6 fm$/c$ for the parton that initiates the jet shower, with partons that originate at splittings within the shower having later initial times. 
 Between the initial time $\tau_i$ and the time $\tau_e$ when energy loss ends, the total amount of energy and momentum deposited by the high-energy parton into the background fluid is given by
\be
\Delta E_{\rm{tot}} &=& \int_{\tau_i}^{\tau_e} \diff\tau \frac{\diff E}{\diff\tau} \\
\Delta P^i_{\rm{tot}} &=& \int_{\tau_i}^{\tau_e} \diff\tau \frac{\diff P^i}{\diff\tau} = v^i_{\rm parton} \int_{\tau_i}^{\tau_e} \diff\tau \frac{\diff E}{\diff\tau} \,,
\ee
where the direction of motion of the parton is the initial direction of the spatial velocity ${\bs v}_{\rm parton}$ 
of the parton.
We assume the parton moves at the speed of light so $v^i_{\rm parton} v^i_{\rm parton} = 1$ ($i=x,y,z$ is summed and should be distinguished from the $i$ in $\tau_i$ which stands for ``initial''). Since $\eta_{s\,{\rm parton}}=0$ throughout, we have $v^x_{\rm parton} = \cos\theta$, $v^y_{\rm parton} = \sin\theta$ and $v^z_{\rm parton}=0$. To use the template solutions, we first rotate and then shift the parton under consideration such that it starts at the spatial origin and moves along the positive $x$-axis. Then the full linearized hydrodynamics solution for the 
wake in the fluid created by this high-energy parton moving along its entire trajectory,
starting at the origin and moving along the $x$-axis, is obtained via superposition of the template 
solutions \eqref{eq:TemplateSolution} that we have computed and tabulated in the previous subsection, and takes the form
\be
\delta\varepsilon(\tau, x,y,\eta_s) &=& \int_{\tau_i}^{\tau_e} \diff\tau' \frac{\diff E(\tau')}{\diff\tau'} \delta\varepsilon_{1x} \big(\tau'; \tau-\tau',x-\tau'+\tau_i,y,\eta_s \big)\nonumber \\
g^x(\tau, x,y,\eta_s) &=& \int_{\tau_i}^{\tau_e} \diff\tau' \frac{\diff E(\tau')}{\diff\tau'} g^x_{1x} \big(\tau'; \tau-\tau',x-\tau'+\tau_i,y,\eta_s \big)\nonumber \\[4pt]
g^y(\tau, x,y,\eta_s) &=& \int_{\tau_i}^{\tau_e} \diff\tau' \frac{\diff E(\tau')}{\diff\tau'} g^y_{1x} \big(\tau'; \tau-\tau',x-\tau'+\tau_i,y,\eta_s \big)\nonumber \\[4pt]
g^{\eta_s}(\tau, x,y,\eta_s) &=& \int_{\tau_i}^{\tau_e} \diff\tau' \frac{\diff E(\tau')}{\diff\tau'} g^{\eta_s}_{1x} \big(\tau'; \tau-\tau',x-\tau'+\tau_i,y,\eta_s \big) \,,
\label{eq:Bjorken-flow-wake-solution}\ee
in which the $x-\tau'+\tau_i$ term indicates the proper shifts of 
the template solutions that are needed to account for the fact that, 
inside the integrals on the right-hand sides of these equations, the unit deposition that occurs 
at $\tau^d_{\rm temp}\equiv \tau'$ occurs at a position $(x,y,\eta_s)=(\tau'-\tau_i,0,0)$
as the parton is moving along the $x$-axis. 
The solution for the wake of a high-energy parton that begins
at
$(\tau_i,x_i,y_i,\eta_s=0)$ and moves 
in a direction in the transverse plane that makes an angle $\theta$ with the $x$-axis can be obtained by translating and rotating the deposition point, the trajectory of the high energy parton, and the template solution
\eqref{eq:Bjorken-flow-wake-solution} in the transverse plane. In practice, the integrals on the right-hand sides of Eqs.~\eqref{eq:Bjorken-flow-wake-solution} are evaluated as Riemann sums with $\Delta\tau'=0.4$~fm$/c$, which corresponds to the times at which we have evaluated and tabulated the template solutions.

This completes our discussion of the efficient numerical procedure to solve the linearized hydrodynamics equations describing the evolution of the wake left in a Bjorken fluid by a high-energy parton. To apply this result, we would need to specify the rate at which the high-energy parton loses energy, $dE/d\tau$, as a function of the initial parton energy,  $\tau$, and the QGP temperature $T$ at the spacetime point where the high-energy parton is at time $\tau$.
If we use the same parameters specifying the background Bjorken flow as used in Ref.~\cite{Casalderrey-Solana:2020rsj} and consider a high-energy parton losing energy from $\tau_i=0.6$ fm$/c$ to $\tau_e=4.6$ fm$/c$ with the strong-coupling rate of energy loss as used in the Hybrid Model and in Ref.~\cite{Casalderrey-Solana:2020rsj}, integrating Eqs.~\eqref{eq:Bjorken-flow-wake-solution} will exactly reproduce the jet wake perturbations to the Bjorken flow solution depicted via plots of the wake energy $\delta\varepsilon$
and wake momentum $(g^x,g^y,g^{\eta_s})$ 
as functions of $x$ and $y$ at different values of $\tau$
in 
Figs.~1, 2, and 4 of Ref.~\cite{Casalderrey-Solana:2020rsj} for the viscous case. 
We have confirmed that evaluating these integrals using the template solutions that we have tabulated in the previous subsection, which is to say
evaluating the integrals as Riemann sums with $\Delta \tau=0.4$ fm$/c$, reproduces the profiles to a good precision.

In the next Section, we shall employ the 
template solutions \eqref{eq:TemplateSolution} from 
Section~\ref{sec:NumericalSolution} in the service of describing the wake produced by a high-energy parton traversing a radially expanding droplet of QGP. We shall not, however, use the full solutions~\eqref{eq:Bjorken-flow-wake-solution} to the equations of  linearized hydrodynamics in a Bjorken flow background without transverse flow that we have obtained here by the method of Green's functions and that we have presented here as a simplified example that reproduces the results of Ref.~\cite{Casalderrey-Solana:2020rsj}.

\section{Linearized Hydrodynamics with Transverse Flow}
\label{sect:lin_flow}
 
 As is well known, the QGP created in heavy ion collisions is expanding rapidly in the transverse plane in addition to along the longitudinal (beam) direction. This means that, unlike in a Bjorken flow (\ref{eqn:u0_cartesian}), the fluid velocity has non-vanishing components along the $x$- and $y$-axes. For an ultrarelativistic collision the longitudinal dynamics at each point in the transverse plane can reasonably be approximated as boost invariant over some wide range of rapidity, meaning that (absent any jet wakes) the 
fluid velocity field may be parameterized as
\be
\label{eqn:u0_transverse_cartesian}
(u^t_0, u^x_0, u^y_0, u^z_0) = {\gamma}(\cosh\eta_s, {v}_x, {v}_y, \sinh\eta_s)\,,
\ee
where ${\gamma}^{-1} \equiv \sqrt{1-{v}_x^2-{v}_y^2}$. The transverse velocity fields ${v}_x, {v}_y$ depend on the position in the transverse plane as well as on proper time, and so does the temperature $T$. This dependence is determined via hydrodynamic simulation of the collision dynamics.

Events in which one or more jets have excited wakes in the hydrodynamic fluid will in general
have flow patterns that cannot be described by the parametrization in Eq.~\eqref{eqn:u0_transverse_cartesian}. Indeed, if there is a significant transfer of energy and momentum from the hard partons in a jet shower to the bulk QGP matter, the flow fields that describe the droplet of fluid in such an event are not boost invariant. This means that, as a matter of principle, the complete description of the collective dynamics of events in which hard partons are produced demands event-by-event, boost-noninvariant, $(3+1)$-dimensional hydrodynamic simulations~\cite{Tachibana:2014lja,Tachibana:2017syd,Chen:2017zte,Zhang:2018urd,Yang:2021qtl,Zhao:2021vmu,Yang:2022nei}.
Pursuing this approach is computationally costly. Any more tractable approach must involve some simplifying assumptions.  The approach in Ref.~\cite{Casalderrey-Solana:2016jvj} was to simplify the problem so brutally as to neglect the hydrodynamic evolution almost completely, leaping directly to assumptions about the soft particle production from the wake at freezeout whose only virtue was that they enforced energy and momentum conservation. We can now do much better.

Our most important working assumption throughout is simply that the amount of energy and momentum transferred by the jet partons to the hydrodynamic fluid is small, compared to the total energy of the event  and also as compared to the total energy and momentum of the volume of fluid through which the jet wake propagates.
Upon making this very reasonable assumption, 
the wake(s) created by hard parton(s) plowing through the droplet of fluid will be small
perturbations on top of the bulk fluid, with the bulk fluid velocity field parameterized as in Eq.~\eqref{eqn:u0_transverse_cartesian}.
Upon making only this assumption, the hydrodynamic evolution of the wake(s) that hard parton(s) excite in the fluid could then be determined by studying linearized hydrodynamic perturbations on top of the flow fields determined by standard numerical solution of the unperturbed bulk dynamics.  The challenge, however, is that since the bulk flow fields are different at different points in the transverse plane, and since each hard parton has its own point of origin and direction and therefore traverses a unique path in the transverse plane, the description of the perturbed fields would still require an event-by-event solution of the full linearized hydrodynamic equations, which 
would still be computationally 
expensive.

\subsection{Our Procedure, in Eight Steps}

In this paper, we develop an approach that: (i) takes advantage of the template solutions computed and 
tabulated in Section~\ref{sec:NumericalSolution};
(ii) yields a much better description of the soft hadrons originating from jet wakes than 
in the oversimplified treatment introduced a decade ago~\cite{Casalderrey-Solana:2016jvj}, as judged by comparing our results for 50 illustrative events to those obtained by solving full nonlinear hydrodynamics equations for the
jet wakes; and (iii) is computationally much less expensive 
than event-by-event solutions of the linearized hydrodynamic equations would be.

In order to describe the soft hadrons originating from jet wakes,
we need to know the perturbation to the fluid temperature and to the hydrodynamic velocity fields \eqref{eqn:u0_transverse_cartesian} 
at each specified point on the freezeout surface resulting from the wake
of a jet.  As these perturbations are small, it is a good approximation
to assume that they can be obtained by linear superposition of all such perturbations at the specified point on the freezeout surface resulting from
each of many individual depositions of a unit of energy and momentum (each with a specified deposition time, deposition point of origin, and deposition direction) that,
with proper weighting, add up to the energy and momentum deposited in the droplet of QGP by a jet.
Solving this linearized hydrodynamics problem fully via the method of Green's functions would require templates built upon
a background velocity field \eqref{eqn:u0_transverse_cartesian} that includes transverse flow.
What we shall do instead is
to make the further approximation that we can employ the
templates tabulated in Section~\ref{sec:NumericalSolution} that
are built upon the Bjorken flow background velocity field \eqref{eqn:u0_cartesian}.
This further approximation is
not controlled by the smallness
of the hydrodynamic perturbations.
We shall check its validity by comparing our results for a
small sample of 50 high-energy partons to results from a full nonlinear hydrodynamics calculation in Section~\ref{sect:results}.

Our procedure can be described as follows:
\begin{enumerate}
\item
We choose a specified point on the freezeout hypersurface $(\tau^f,x^f,y^f,\eta^f_s)$.
\item 
We shall assume in this paper that the jet parton whose wake we analyze travels along a constant $\eta_s$ trajectory,
and  without further loss of generality we shall assume that it has $\eta_s=0$.
We choose the time for an individual deposition, $\tau^d$, and the position of this deposition in  the transverse plane, $(x^d,y^d)$, and choose the direction in the transverse plane in which the jet parton is propagating.
We consider the deposition of a unit of energy and momentum into the fluid at this time and position, with the direction of the deposited momentum specified by the direction of the jet parton. 
\item 
We identify the trajectory that begins at this deposition point in spacetime and ends at the specified spacetime point on the freezeout hypersurface
that follows a path such that the trajectory has the
same velocity --- that we denote $\vec\beta$ ---
in the local fluid rest frame at every point on the trajectory.\footnote{Without further specification, there is an ambiguity in defining the axes of the local fluid rest frame: any rotation of the axes is as good as any other. We define the local fluid rest frame as the frame obtained by starting from the lab frame axes,  boosting along the beam direction to the rapidity of the fluid cell, and then making a second transverse boost with the transverse velocity of the fluid.}
We shall refer to this trajectory as the ``relative flow trajectory'' connecting the
deposition point $(\tau^d,x^d,y^d,0)$ to the
freezeout point $(\tau^f,x^f,y^f,\eta^f_s)$.
If these points are connected by a relative flow trajectory with $\vec\beta=0$, namely a flow line, then a perturbation that has no velocity relative to the fluid in which it finds itself is carried along by the fluid flow from $(\tau^d,x^d,y^d,0)$ to $(\tau^f,x^f,y^f,\eta^f_s)$, with $\eta_s^f=0$ for a flow line. 
For a generic point on the freezeout surface, this is not the case, and our challenge below will be to find the $\vec\beta$ such that the two points are connected by a relative flow trajectory with relative velocity $\vec\beta$.
Note that we shall
only consider central collisions in which the transverse flow velocity is in the radial direction.

\item 
We compute the proper time between the deposition point and the freezeout point along the relative flow 
trajectory identified in (3); we denote this proper time by $\Delta \tau_{\vec\beta}$.

\item 
We then select the template solution from Section~\ref{sec:NumericalSolution} in which the proper time between the deposition time and the freezeout time in the template is $\Delta\tau_{\vec\beta}$, meaning that we choose the template in which the deposition time in the template is $\tau^d_{\rm temp}=\tau^f_{\rm temp}-\Delta\tau_{\vec\beta}$.  
We then rotate this template solution
in the transverse plane by an angle 
that we shall denote by $\phi^d_{\vec\beta=0}$ defined in such a way
that, after this rotation, the direction of the unit momentum deposition in the template (which is the $x$-direction in the templates tabulated in Section~\ref{sec:NumericalSolution}) points in the direction in which the jet parton is propagating --- in the local fluid rest frame at the deposition point.
In this rotated template solution, we draw a straight-line trajectory that starts from the deposition point, has velocity $\vec\beta$
and extends for a proper time $\Delta\tau_{\vec\beta}$.
The end of this trajectory identifies a 
point in the template solution that 
we shall denote by 
$(\tau^f_{\rm temp},x^f_{\rm temp},y^f_{\rm temp},\eta^f_{s\,{\rm temp}})$.

\item 
We map the perturbations to the temperature and the Bjorken background flow at the point 
$(\tau^f_{\rm temp},x^f_{\rm temp},y^f_{\rm temp},\eta^f_{s\,{\rm temp}})$
in the template solution to the perturbations in 
the temperature and hydrodynamic velocity
at the point $(\tau^f,x^f,y^f,\eta_s^f)$ on
the freezeout surface chosen in (1) in the local fluid rest frame at that point. We shall describe the rotation of coordinate axes employed in mapping
the velocity perturbation from the point in the template solution to the point on the freezeout surface
below, in Section~\ref{sec:Utility}.

\item 
With all of this groundwork laid, we employ the linearity of linearized hydro: we repeat the procedure (2)-(6) above for
all the unit depositions of energy and momentum that, with proper weighting, 
add up to all of the energy and momentum deposited into the hydrodynamic fluid by the jet parton along its entire trajectory, and add up the resulting perturbations with proper weights at the 
point on the freezeout surface 
$(\tau^f,x^f,y^f,\eta_s^f)$
chosen in (1).

\item 
We carry out the entire procedure for all points on the freezeout hypersurface, and then obtain the 
distribution of soft hadrons originating from the perturbations at the freezeout surface resulting from the jet wake via the conventional Cooper-Frye freezeout prescription.

\end{enumerate}
In the following subsections, we describe the implementation of this procedure in full detail.

{
First, though, we note a further simplifying assumption that we will 
employ in steps (3) and (4). Although throughout our procedure we will use 
a freezeout hypersurface taken from a realistic
hydrodynamical simulation for heavy ion collisions with zero impact
parameter obtained via the code 
MUSIC~\cite{Schenke:2010nt,Schenke:2010rr,Ryu:2015vwa,Paquet:2015lta} as described in Section~\ref{sect:cf}, in steps (3) and (4) of the procedure above (which allow us to determine from which point in which rotated template solution we will take the 
hydrodynamic perturbations that in step (5) we map to a point on the freezeout hypersurface) we shall make 
the further simplifying assumption 
that the transverse
velocity of the fluid is purely radial and takes the blast-wave 
form~\cite{Schnedermann:1993ws,Teaney:2001av,Huovinen:2001cy}
\begin{equation}
\label{blastwave_flow}
\gamma v= \sigma r \ .
\end{equation}
As we describe in Appendix~\ref{app:model},
this approximation is
motivated at a qualitative level by results from hydrodynamic simulations
for heavy ion collisions with zero
impact parameter.\footnote{We show in Appendix~\ref{app:model}
that the  simplified blast-wave profile
that we employ provides a reasonable description of a full azimuthally symmetric hydrodynamic simulation
for values of $r$ ranging from 0 to near the freezeout surface and for times starting from as early as 
$\tau=4~{\rm fm}/c$ until the freezeout time.
}
We fix the value of the constant $\sigma$ in Appendix~\ref{app:model} by 
fitting the blast-wave form to full results from hydrodynamic simulations of central heavy ion collisions with center of mass energy $\sqrt{s_{\rm NN}}=5.02$~TeV obtained with the MUSIC code, 
obtaining $\sigma=0.12$~fm$^{-1}$.
The time independence of this simplifying ansatz will allow us to complete 
the specification of steps (3) and (4) of our procedure in a semi-analytical fashion.

As should be apparent from its description, although our procedure is well specified and concrete
and has a rationale behind it, there is no sense in which it is a controlled approximation. Its virtue derives entirely from the very substantial computational speed-up 
(we quantify this in Section~\ref{sect:music})
that is obtained via employing the
templates from 
Section~\ref{sec:NumericalSolution} obtained via a straightforward calculation in a Bjorken background rather than solving for the Green's functions for  linearized hydrodynamics in
a background with transverse flow.

Steps (1) and (2) of the procedure set out above need no further explanation.
We shall accomplish the goals set out in steps (3) and (4) of the procedure above in stages in the next three subsections. These subsections 
serve in sum to accomplish the goals in steps (3) and (4).  
We shall describe how we accomplish steps (5) and (6) in 
Section~\ref{sec:Utility}, and in
Section~\ref{seclinsup} we shall complete the full description of how we implement this procedure with steps (7) and (8).

\subsection{\label{step1}Absorbing the Fluid 
Rapidity at the Freezeout Point}

The first stage of the 
construction is to absorb 
the longitudinal velocity of the background flow at the freezeout point $(\tau^f,x^f,y^f,\eta_s^f)$
by performing a ``rapidity-dependent Lorentz transformation''. By this, we mean a Lorentz transformation in which the  parameter ${\rm y}$ that specifies the Lorentz boost depends on 
$\eta_s^f$ according 
to $\rm y(\eta^f_s)=\eta^f_s$. This transformation does not affect the transverse coordinates. After the transformation, the time and beam direction coordinates of the freezeout point, which we denote by $t^f_\perp$ and $z^f_\perp$ respectively and refer to as transverse coordinates, become:
\begin{eqnarray}
t^f_\perp&=&\tau^f \cosh{\left(\eta^f_s-\rm y(\eta^f_s)\right)}=\tau^f \nonumber\\
z^f_\perp&=&\tau^f \sinh{\left(\eta^f_s-\rm y(\eta^f_s)\right)}=0 \,,
\label{eq_perpcoord}
\end{eqnarray}
where, as stated before, $\rm y(\eta^f_s)=\eta^f_s$. 
By construction, the velocity field of the fluid 
at the freezeout point is purely transverse when it is written in terms of the transverse coordinates $(t^f_\perp , x^f, y^f, z^f_\perp  )$
and in the absence of any perturbations 
can be denoted by
\be
\label{eq_peru}
u^f_{0\perp}= \gamma \left(1,{v}_x, {v}_y,0\right)\,.
\ee
To simplify notation in the following, henceforth we will drop the subscript ``$\perp$'' from the transverse coordinates. That is,  $t$ and $z$ below refer to the transverse coordinates obtained after this first stage of our construction.

\subsection{Identifying $\Delta\tau_{\vec\beta_\perp}$ and $\vec\beta_\perp$}\label{step2}

Our goal is to identify 
the $\vec\beta$ such that the relative flow trajectory with velocity $\vec\beta$,
see step (3) of our procedure, starting from the deposition point $(\tau^d,x^d,y^d,0)$ 
passes through the freezeout point $(\tau^f,x^f,y^f,\eta_s^f)$.
Since the deposition point and the freezeout point are at different $\eta_s$, the velocity $\vec\beta$ will have a longitudinal component $\beta_z$ 
as well as a transverse component $\vec\beta_\perp$.
After the transformation in Section~\ref{step1},
the transverse coordinates of the freezeout point are $(t^f,x^f,y^f,0)$, where as promised we have dropped the subscript $\perp$. 
The transformation in Section~\ref{step1} does not transform the coordinates of the deposition point, since  ${\rm y}(\eta_s^d) = \eta_s^d =0$, meaning that $t^d=\tau^d$.
This means that in these transverse coordinates, our principal goal is to identify
the $\beta_\perp$ 
such that the relative flow trajectory with velocity $\vec\beta_\perp$,
starting from the deposition point $(t^d,x^d,y^d,0)$ 
passes through the freezeout point $(t^f,x^f,y^f,0)$.
We shall also compute the proper 
time
$\Delta\tau_{\vec\beta_\perp}$ along the relative flow trajectory in these transverse coordinates; in Section~\ref{step3}, we shall compute
the proper time $\Delta\tau_{\vec\beta}$  along the relative flow trajectory specified in the original coordinates by $\vec\beta=(\vec\beta_\perp,\beta_z)$.

As a warm-up exercise, that will turn out to be of direct utility, we begin by computing the flow line (the relative flow trajectory 
with $\vec\beta=0$) from the deposition point --- which will almost certainly not cross the freezeout hypersurface at the point of interest to us.
We can determine the flow line originating from any deposition point, namely
the integral curve of the fluid velocity field that passes through the deposition point.  For any boost-invariant flow, with or without transverse flow, the flow line originating from any deposition point will be at a constant $
\eta_s$, in our case at $\eta_s=0$ since $\eta^d_s=0$. 
For the simplified case of a central collision, the transverse
fluid flow is radial which means that the flow
line from any deposition point is also radial, and can be obtained by
integrating the flow equation 
\begin{equation}
\label{eq_flowline}
\frac{dr}{dt} =v\, , \quad  \frac{d\phi }{dt}=0 \, ,
\end{equation} 
where $t$ is related to $\tau$ by the Lorentz transformation constructed in the previous subsection and where $r$ and $\phi$ are polar coordinates for the transverse $(x,y)$ plane.  Because the flow is radial, $\phi$ is constant along a flow line.

Recalling that flow lines are relative velocity trajectories with $\vec\beta=0$, we denote
the proper time in the local fluid rest frame at a point along the flow line by $t_{\vec\beta=0}$, with the convention that $t_{\vec\beta=0}=t^d$ at the deposition point.  The time $t_{\vec\beta=0}$ is then the time measured by an observer floating in the fluid carried along the flow line. And, $t_{\vec\beta=0}$ 
serves to label different points along the flow line.
For any infinitesimal step in the transverse lab frame time $t$ (defined in the previous subsection),
the corresponding infinitesimal step
in $t_{\vec\beta=0}$ 
is given by the projection of the interval along the fluid velocity direction:
\begin{equation}
\label{eq:dtcdif}
   \frac{dt_{\vec\beta=0}}{dt}\equiv u_\mu \frac{dx^\mu}{dt} =  \left(\gamma - \gamma v \dot r\right)  .
\end{equation}
We have written this expression in a form that 
will apply when we take $\vec\beta\neq 0$ below;
for the case of flow lines with $\vec\beta=0$,
$\dot r=v$ and 
$dt_{\vec\beta=0}=dt/\gamma$, confirming that $t_{\vec\beta=0}$ is indeed the proper time along the flow line.

The flow line originating from the deposition point will cross the freezeout hypersurface at some spacetime point,
since the temperature of the fluid decreases along the flow line and at some point reaches $T^f$.
We denote the radial position in the transverse plane of the spacetime point on the freezeout surface 
connected to the deposition point 
by its flow line by $r^f_{\vec\beta=0}$.
The transverse time of this spacetime point is then specified by the function 
$t^f(r^f)$ that defines the shape of the freezeout hypersurface.
With the blast-wave 
approximation~\eqref{blastwave_flow}, the flow equation that serves to specify $r^f_{\vec\beta=0}$
can be written as
\begin{equation}
\label{tofr_flow}
t^f(r^f_{\vec\beta=0})-t^d= \int_{r^d}^{r^f_{\vec\beta=0}} \frac{d\rho}{v(\rho)}\, .
\end{equation}
Recalling that the function $t^f(r^f)$ 
is known, this equation serves to specify
$r^f_{\vec\beta=0}$.  
Once $r^f_{\vec\beta=0}$ has been determined, the left-hand side of
Eq.~\eqref{tofr_flow}
can be evaluated, meaning that we have then 
determined the time difference between the transverse time at the deposition point, 
$t^d$, and the transverse time at 
the point where the
flow line from the deposition point crosses the freezeout hypersurface,  $t^f(r^f_{\vec\beta=0})$.

We are, however, interested in the time from deposition to freezeout in the local fluid rest frame along the flow line, which is to say as measured by an observer floating in the fluid carried along the flow line.  We shall denote this time by $\Delta t_{\vec\beta=0}\equiv t^f_{\vec\beta=0} - t^d$, recalling that at the deposition point $t_{\vec\beta=0}=t^d$.
In the blast-wave approximation, with  Eq~\eqref{blastwave_flow},
this takes the form
\begin{eqnarray}
\label{eq:Dtc}
\Delta t_{\vec\beta=0}&=&
\int_{t^d}^{t^f(r^f_{\vec\beta=0})} \frac{dt'}{\gamma}
=\int_{t^d}^{t^f(r^f_{\vec\beta=0})} dt' \left(-v^2\gamma +\gamma\right)
=\int_{t^d}^{t^f(r^f_{\vec\beta=0})} dt' \left(-\sigma r \frac{dr}{dt'} +\gamma\right)
\nonumber\\
&=&- \frac{1}{2} \sigma  \left(\left(r^f_{\vec\beta=0}\right)^2-(r^d)^2\right) + \int_{t^d}^{t^f(r^f_{\vec\beta=0})} dt' \sqrt{1+\sigma^2 r^2 (t')} \,,
\end{eqnarray}
where $r(t')$ is the solution to the flow 
equation~\eqref{eq_flowline}. Although this integral can be performed explicitly for the flow line originating from the deposition point, we will leave Eq.~\eqref{eq:Dtc} in this form as this is the form that we will need in our analysis of relative velocity trajectories with $\vec\beta\neq 0$, below.
Note that there are two ways in  which 
$\Delta t_{\vec\beta=0}$ is not (yet) the 
$\Delta \tau_{\vec\beta}$ that we are seeking. 
We need to repeat this analysis with $\vec{\beta}\neq 0$; and, we need to restore the fluid rapidity at the freezeout point. 
We shall take these on in turn.

We turn now to repeating the above analysis for a relative flow  trajectory with $\vec\beta\neq 0$ 
originating from the deposition point,
not a flow line.  Recall from step (3) in our procedure that a relative flow trajectory 
has the same velocity $\vec\beta$ in the local fluid rest frame (velocity relative to that of the fluid) at every point on the trajectory. 
We begin by analyzing the effects of $\vec{\beta}_\perp$.
Since the flow is radial, relativistic addition  of velocities gives us the equation for $dr/dt$ along
a relativistic flow trajectory in transverse coordinates:
\begin{equation}
\label{eq:relfl}
\frac{dr}{dt}=\frac{{\vec\beta}_\perp\cdot \hat r + v(r) }{1+v(r) \,{\vec \beta}_\perp \cdot \hat r } \,,
\end{equation}
where $\hat r$ is the unit vector in the radial direction in the transverse plane and
where ${\vec \beta}_\perp\cdot \hat r \equiv \beta_x \cos \phi + \beta_y \sin \phi$.

A relative velocity trajectory with some generic $\vec\beta_\perp$ originating from some generic deposition point 
with nonzero $r^d$
has a nonzero $\hat \phi$ component, meaning that it is not parallel to the local fluid velocity. This means
that in the local fluid rest frame,
$\vec\beta_\perp\cdot \hat\phi$ is in general nonzero, 
and (unlike for flow lines, which satisfy  Eq.~\eqref{eq_flowline}) 
the angle $\phi$ is not constant along a
relative velocity trajectory.
 The equation for $d\phi/dt$ can be obtained by requiring that distances in the direction in the transverse plane that is perpendicular
 to the fluid velocity are the same in the local fluid rest frame as they are in the transverse lab frame 
 coordinates  $(x,y)$,  
 as we expect for a Lorentz transformation. Infinitesimally, this condition is given by  
\begin{equation}
r d\phi= dy_{\vec\beta_\perp} \cos \phi  - dx_{\vec\beta_\perp} \sin \phi  \,,
\end{equation}
with $(dx_{\vec\beta_\perp}, dy_{\vec\beta_\perp})$ an infinitesimal displacement 
around a point on the relative velocity trajectory
in the fluid rest frame at that point. 
We can then identify
$\vec{\beta}_\perp=(dx_{\vec\beta_\perp}/dt_{\vec\beta_\perp}, dy_{\vec\beta_\perp}/dt_{\vec\beta_\perp})$, where $dt_{\vec\beta_\perp}$ is the time interval in the fluid rest frame along the relative flow trajectory, defined in Eq.~\eqref{eq:dtcdif}. We conclude that $\phi$
along a
relative velocity trajectory with relative velocity $\vec\beta_\perp$ satisfies 
\be
r\frac{d\phi}{dt_{\vec\beta_\perp}
}= \beta_y \cos\phi - \beta_x \sin \phi \,,
\ee
which can equivalently be rewritten as
\be
r\frac{d\phi}{dt}=\left(\gamma - \gamma v \dot r\right) \left(\beta_y \cos\phi - \beta_x \sin \phi\right) \,,
\label{eq:dthetadt}
\ee
on account of Eq.~\eqref{eq:dtcdif}.
The coupled equations~\eqref{eq:relfl} and \eqref{eq:dthetadt} together specify the relative velocity trajectory with a given $\vec\beta_\perp$ in the $(r,\phi)$ transverse plane.  Our task is to find the value of $\vec\beta_\perp$ such that the relative velocity trajectory with this $\vec\beta_\perp$ originating from the deposition point intersects the freezeout surface in transverse coordinates at the point $(t^f,x^f,y^f,0)$.

We now analyze the effect of the longitudinal relative velocity $\vec\beta_z$.
This must in general be nonzero because the relative velocity trajectory whose velocity relative to the local fluid rest frame is $\vec\beta$ begins at the deposition point where $\eta_s=0$ and ends at the freezeout point where $\eta_s=\eta_s^f$.  (Note that the deposition point has a nonzero $\tau^d$ and $z=0$; it is not at the origin.)
Since $\beta_z$ is defined in the fluid rest frame and since the fluid moves in the transverse plane, we can again use the relativistic composition of velocities to determine the change $d\eta_s$ in the spacetime rapidity of the relative velocity trajectory during a time interval $dt$ at time $t$,
which is given by 
\be\label{eq:d-eta-s}
d\eta_s =\frac{\beta_z dt}{t \gamma \left(1-v(r) \vec{\beta}_\perp \cdot \hat r\right) \ } \,,
\ee
where we have used the fact that   $dz=\tau d\eta_s$  in these transverse coordinates. 
There is an important subtlety here that illustrates why it is important that the
transformation in Section~\ref{step1} is a {\it rapidity-dependent} Lorentz boost.  In order to stay in the transverse coordinates defined in
Section~\ref{step1}, in going from time $t$ to time $t+dt$ we need to change the boost
in Section~\ref{step1} precisely by the $d\eta_s$ defined in Eq.~\eqref{eq:d-eta-s}.
In this way, the relative velocity trajectory stays at $\eta_s=0$ in the (unusual) transverse coordinates defined via this rapidity-dependent boost. 
In the original coordinates, though,
once
$\beta_\perp$ is specified, integrating 
Eq.~\eqref{eq:d-eta-s} from $t=t^d=0$ to $t=t^f=\tau^f$ must yield $\eta_s^f-\eta_s^d=\eta_s^f$.  
This condition determines the value of $\beta_z$.
Because our background flow is boost-invariant, however, we shall use a very simple prescription for 
specifying the $\eta_{s\,{\rm temp}}$ of the point in the template solution whose fluctuations we will map
to a point on the freezeout surface
with $\eta_s=\eta_s^f$: we shall choose $\eta_{s\,{\rm temp}}=\eta_s^f$. This means that we will not actually need to know the value of $\beta_z$ in the following.

We note that, because we are assuming that
the background flow takes the blast-wave form~\eqref{blastwave_flow},
the proper time along a relative velocity trajectory
from the deposition point to the point where this trajectory crosses the freezeout surface, 
as measured by observers in the local fluid rest 
frame at each point on the trajectory, 
takes the form~\eqref{eq:Dtc} that we derived for a flow line, with the only modifications being: (i) we denote this quantity for a relative velocity trajectory by $\Delta t_{\vec\beta_\perp}$; (ii) on the right-hand side of Eq.~\eqref{eq:Dtc}, $r^f_{\vec\beta=0}$ is 
replaced by $r^f_{\vec\beta_\perp}$, the radial position at which the relative velocity trajectory crosses the 
freezeout hypersurface, obtained by 
solving Eq.~\eqref{tofr_flow} with $\vec\beta_\perp$ nonzero; and (iii) the function $r(t')$ is the solution to 
 Eq.~\eqref{eq:relfl}, which must be solved
 together with solving Eq.~\eqref{eq:dthetadt} as these two equations are coupled.

For each deposition point, 
we solve  Eqs.~\eqref{eq:relfl} and~\eqref{eq:dthetadt} for ``all possible'' relative  perpendicular velocities $\vec\beta_\perp$, in actuality for a grid of magnitudes and directions of $\vec\beta_\perp$. 
We choose our grid of $\vec\beta_\perp$'s to correspond to the grid of points $(x_{\rm temp},y_{\rm temp})$ in the transverse plane at which we have tabulated the hydrodynamic perturbations in the template solutions of Section~\ref{sec:NumericalSolution}. In the template solutions, there is no transverse flow. This means that in the template solutions, in transverse coordinates, the transverse component of the relative flow trajectory from the deposition point to the
freezeout point $(\tau^f_{\rm temp},x^f_{\rm temp},y^f_{\rm temp},\eta^f_{s\,{\rm  temp}})$ 
is a straight line in the 
$(x_{\rm temp},y_{\rm temp})$-plane.
For each point $(x^i_{\rm temp},y^j_{\rm temp})$ from the grid of points 
at which we have tabulated the hydrodynamic perturbations in the each of our 34 template solutions, we draw this straight line, which defines a $\vec\beta_\perp$. 
Because the different template solutions have different deposition times $\tau^d_{\rm temp}$, the $\vec\beta_\perp$ for a given grid point $(x^i_{\rm temp},y^j_{\rm temp})$ will be different in each of the templates.  We shall refer to the $\vec\beta_\perp$ that points from the deposition point to the $(i,j)$'th grid point in the transverse  $(x_{\rm temp},y_{\rm temp})$-plane in the $k$'th template solution as
$\vec\beta_\perp^{ijk}$.
This set of possible values of $\vec\beta_\perp$
is our ``grid'' of magnitudes and directions of $\vec\beta_\perp$. 
In the blast-wave background with transverse radial flow that is our actual background of interest,
we then compute the relative flow trajectory originating from a given deposition point for each $\vec\beta_\perp^{ijk}$ by solving Eqs.~\eqref{eq:relfl} and~\eqref{eq:dthetadt}. 
The family of relative flow trajectories originating from a given deposition point that we obtain in this way serves to 
establish a map between the set of  $\vec\beta_\perp^{ijk}$'s and the
points on the freeze-out surface
 $(t^f(r^f),r^f, \phi^f)$ where the relative flow trajectory with a given $\vec\beta_\perp^{ijk}$ intersects the freezeout hypersurface.  
 
 For each deposition point and for each value of $\vec\beta_\perp$, we obtain $\Delta t_{\vec\beta_\perp}$ 
by integrating Eq.~\eqref{eq:Dtc} along 
the corresponding $\vec\beta_\perp^{ijk}$-dependent relative velocity trajectory in the transverse plane. 
In our later analysis, when we want the hydrodynamic perturbations at any given  spacetime point
 $(t^f(r^f),r^f, \phi^f,\eta_s^f)$ on the freezeout hypersurface, we shall choose the $\vec\beta_\perp^{ijk}$ (and the $(i,j)$'th point in the transverse plane in the relevant template solution) 
 that specifies the relative velocity trajectory whose intersection with the freezeout surface is closest in the transverse plane
 to the point on the freezeout surface we are interested in. Because our background flow is boost invariant, it will suffice to choose $\eta_s$ in the relevant template solution to be $\eta_s^f$.

\subsection{\label{step3}Restoring $\eta_s^f$}

We have almost completed the tasks that we need to complete in order to accomplish steps (3) and (4) of our procedure.  We have found $\vec\beta$ for a given point on the freezeout hypersurface and a given deposition point.  And, we have computed $\Delta t_{\vec\beta}$.  However, we have done so entirely using the transverse coordinates introduced in 
Section~\ref{step1} upon absorbing $\eta_s^f$, meaning throughout 
Section~\ref{step2} we have
been working with $\eta_s=0$.  We now need to restore 
$\eta_s^f$.  We do so by performing another Lorentz transformation, corresponding to a boost in the beam direction, as follows:  
\begin{eqnarray}
\label{eq_perpcoord}
\tau^f_{\vec\beta}&=&t^f_{\vec\beta} \cosh\eta^f_s\nonumber \\
z^f_{\vec\beta}&=&t^f_{\vec\beta} \sinh \eta^f_s \,,
\end{eqnarray}
designed such that after this transformation the rapidity at the freezeout point has indeed been restored to its value 
$\eta_s^f$ in the laboratory frame, and
the unperturbed flow fields at the freezeout point
take on the boost-invariant form $u_0^\tau=1$.
Since this transformation is longitudinal, it does not change the direction of $\vec{\beta}_\perp$, and does not change the coordinates in the transverse plane. That is, the logic via which we determined where in the transverse plane the relative velocity trajectory crosses the freezeout surface remains as in the previous Section.
But, does this transformation change the proper time along the relative velocity trajectory?  
Because $\tau^f_{\vec\beta}=\sqrt{\left(t^f_{\vec\beta}\right)^2-\left(z^f_{\vec\beta}\right)^2}=t^f_{\vec\beta}$, 
we see that in fact $\Delta \tau_{\vec\beta}
=\Delta t_{\vec\beta_\perp}$. 
This means that in Section~\ref{step2}, via integrating Eq.~\eqref{eq:Dtc} 
with $\vec\beta_\perp\neq 0$ and without regard for $\beta_z$, we have in fact achieved the  objective in step (4) of our procedure:
we have computed the proper time between the deposition point $(\tau^d,x^d,y^d,0)$ and the 
specified point on the freezeout hypersurface
$(\tau^f,x^f,y^f,\eta_s^f)$.

\subsection{Mapping Hydrodynamic Perturbations from a Template Solution to the Freezeout Hypersurface}
\label{sec:Utility}

We now turn to steps (5) and (6) of our procedure.

With $\vec\beta_\perp$ and $\Delta\tau_{\vec\beta}$ in hand, we are now ready to determine the answer to the following questions. If a unit of energy and momentum with direction $(\cos\phi^d,\sin\phi^d,0)$ in the lab frame is deposited in the fluid at deposition spacetime point $(\tau^d,x^d,y^d,0)$, from
which point, in which template solution, 
will we map the fluctuations in the template solution to the point
$(\tau^f,x^f,y^f,\eta_s^f)$ on the freezeout hypersurface? And, how do we do the mapping?

The first step is to rotate each template solution, by an angle that depends on the
position and time at which the deposition occurs, such that the direction of the deposited momentum in the template solution, which up to this point has been the positive-$x$ direction, points in the
direction of the unit of momentum deposited by the jet parton --- in the local fluid rest frame at the deposition point. 
We denote this direction by $\phi^d_{\vec\beta=0}$ by analogy with our convention (see Section~\ref{step2}) that $t^d_{\vec\beta=0}$ is the deposition time in the local fluid restframe.
We identify $\phi^d_{\vec\beta=0}$
by boosting the lab-frame momentum four-vector  $(1,\cos\phi^d,\sin\phi^d,0)$ to the local fluid rest frame at the deposition point. 
For each choice of deposition point, we rotate each template solution by an angle 
$\phi^d_{\vec\beta=0}$.\footnote{Rotating the template solutions in the transverse plane by an angle $\phi^d_{\vec\beta=0}$
entails two steps: first rotating the $x_{\rm temp}$-axes 
and $y_{\rm temp}$-axes of the template solutions by 
$-\phi^d_{\vec\beta=0}$ while keeping the solutions fixed; and then, since the momentum perturbation
$\delta {\bs g}^\perp_{1{x}}$ (see Eqs.~\eqref{eqn:define_gperp} and \eqref{eq:TemplateSolution})
is itself a transverse vector, rotating $\delta {\bs g}^\perp_{1{x}}$ by
$\phi^d_{\vec\beta=0}$.}

For each deposition point,  in Section~\ref{step3} we have computed $\Delta\tau_{\vec\beta}$ for all of the relative velocity trajectories  
originating from the
deposition point $(\tau^d,x^d,y^d,0)$ 
specified 
by $\vec\beta=(\vec\beta_\perp^{ijk},\beta_z)$
for all the values of
 $\vec\beta_\perp^{ijk}$
in our $(i,j)$ grid for the $k$'th template.
(Recall that $\Delta\tau_{\vec\beta}$ does not depend on $\beta_z$.) 
For each choice $(i,j)$, we choose the value of $k$ (which is to say we choose one rotated template among our 34 rotated template solutions)
for which $\tau^f_{\rm temp}-\Delta\tau_{\vec\beta}$ is closest to (in fact, within 0.2 fm$/c$ of)
 $\tau^d_{\rm temp}$. That is, we choose the $k$'th rotated template such that the proper time between deposition point and freezeout point is as close to the same in the rotated template as along the relativistic velocity trajectory as we can make it, given our 34 templates. 
 For each candidate $(i,j)$, we pick a rotated template in this way.

 For a specified point on the freezeout hypersurface
$(\tau^f,x^f,y^f,\eta_s^f)$, we 
then identify which of the relative velocity trajectories originating from the deposition point $(\tau^d,x^d,y^d,0)$ 
with 
$\vec\beta=(\vec\beta_\perp^{ijk},\beta_z)$ 
(i.e.~for which choice of $i$ and $j$, with $k$ chosen as we have just described, with $\beta_z$ determined by $\eta_s^f$ as we have described) crosses the freezeout hypersurface closest to the desired transverse position $(x^f,y^f)$.
We then choose the $\eta_s$ grid point in the template that is the closest to
the $\eta_s^f$ at the specified point on the hypersurface.  Per the discussion below Eq.~\eqref{eq:d-eta-s}, this is equivalent to fixing $\beta_z$, but we do not need to know the value of $\beta_z$.

For a specified point on the freezeout hypersurface and a specified deposition point, once we have chosen
the $k$'th rotated template solution from among the 34 template solutions from Section~\ref{sec:NumericalSolution}, rotated as we have described, and the grid point $(i,j)$, as described above,
step (5) of the procedure is then straightforward. In the chosen rotated template solution,
we draw a straight-line trajectory that starts from the deposition point, has velocity $\vec\beta$,
and extends for a proper time $\Delta\tau_{\vec\beta}$.
The end of this trajectory identifies a 
point on the freezeout hypersurface in the template solution  $(\tau^f_{\rm temp},x^f_{\rm temp},y^f_{\rm temp},\eta^f_{s\,{\rm temp}})=(\tau^f,x_i+\beta_x\Delta\tau_{\vec\beta},y_i+\beta_y \Delta\tau_{\vec\beta},\eta^f_s)$. 
Note that for a given point
on the freezeout hypersurface, different deposition points will be connected to it by relative velocity
trajectories with different $\vec\beta$ and 
different $\Delta\tau_{\vec\beta}$, which means that for each deposition point in step (5) of our procedure we will choose a different template solution and a different point 
$(\tau^f_{\rm temp},x^f_{\rm temp},y^f_{\rm temp},\eta^f_{s\,{\rm temp}})$
in that template solution.

In step (6) of our procedure, we must map the perturbations in the temperature and velocity at the point 
$(\tau^f_{\rm temp},x^f_{\rm temp},y^f_{\rm temp},\eta^f_{s\,{\rm temp}})$
in the rotated template solution that we have identified to
the perturbations in the temperature and hydrodynamic velocity at the point $(\tau^f,x^f,y^f,\eta_s^f)$ 
on the freezeout hypersurface in the local fluid rest frame at that point.  The perturbations
at the point 
$(\tau^f_{\rm temp},x^f_{\rm temp},y^f_{\rm temp},\eta^f_{s\,{\rm temp}})$
in the rotated template correspond to the wake at that spacetime point in the Bjorken flow caused
by the deposition of a unit of energy and momentum in the $x$-direction
at the deposition point. 
The basis of our procedure is the {\it assumption}
that the hydrodynamic perturbations
at $(\tau^f,x^f,y^f,\eta_s^f)$ 
on the freezeout hypersurface can be modeled
by those at $(\tau^f_{\rm temp},x^f_{\rm temp},y^f_{\rm temp},\eta^f_{s\,{\rm temp}})$,
as both points are a proper time $\Delta\tau_{\vec\beta}$ away from the deposition
point, as measured in the local fluid rest frame along a relative velocity trajectory with relative
velocity $\vec\beta$. We shall test the validity of this assumption in Section~\ref{sect:results}.

Because of all the groundwork that we have laid, step (6) of our procedure, 
mapping the hydrodynamic perturbations from the point in the rotated template solution that we have identified to the point on the freezeout hypersurface, is now straightforward.
We simply map the energy density perturbation $\delta\varepsilon_{1x}$ 
and the longitudinal momentum perturbation $ g^{\eta_s}_{1x}$ from Eqs.~\eqref{eq:TemplateSolution} (equivalently, the temperature perturbation and the longitudinal velocity perturbation
$\delta u^{\eta_s}$, see Eq.~\eqref{eqn:define_geta})
at the point $(\tau^f_{\rm temp},x^f_{\rm temp},y^f_{\rm temp},\eta^f_{s\,{\rm temp}})$ in the rotated template solution that we have identified to $\delta\varepsilon$ and $g^{\eta_s}$ in the local fluid rest frame
at the point $(\tau^f,x^f,y^f,\eta_s^f)$ 
on the freezeout hypersurface.
And, because we have been careful to rotate 
the transverse momentum perturbation
${\bs g}^\perp_{1{x}}$ 
(equivalently, the perturbation $\delta \bs u^\perp$ to the transverse fluid velocity --- see Eq.~\eqref{eqn:define_gperp})
in the template solution by $\phi_{\vec\beta=0}$, this too we can now simply map to 
$ {\bs g}^\perp$ in the local fluid rest frame at the freezeout point 
$(\tau^f,x^f,y^f,\eta_s^f)$.

This concludes our description of how we determine the hydrodynamic perturbations at one point on the freezeout surface from the wake in the QGP fluid that originates from the deposition of a unit of energy and momentum at a specified point in spacetime, with the momentum pointing in a specified transverse direction.

\subsection{\label{seclinsup}Linear Superposition of Perturbations Described by Template 
Solutions}

We can now use the linearity of linearized hydrodynamics to 
determine the hydrodynamic perturbations at one point on the freezeout surface originating from the wake that a high-energy jet parton traversing the expanding, cooling droplet of QGP created in a heavy ion collision along some trajectory with $\eta_s=0$ leaves behind in the fluid --- namely step (7) of our procedure.
In a future Monte Carlo implementation, it will be straightforward to apply the same procedure to determine the perturbations at a point on the freezeout surface originating from the wake of an entire jet, consisting of a branching shower of high-energy partons that each deposit energy and momentum in the fluid.

For the purposes of benchmarking our approach, in this initial study we will consider two samples of 50 different high-energy partons following straight-line trajectories with $\eta_s=0$. In one sample, all the partons have an initial energy $E_{\rm in}^{\rm lab}=10$~GeV in the lab frame and in the other all have an initial energy of $E_{\rm in}^{\rm lab}=50$~GeV in the lab frame. In each case, we sample the transverse positions at which the
partons are produced from a Glauber model, and choose their transverse directions $\phi$ in the laboratory frame
at random from a uniform distribution between 0 and $2\pi$.
As in 
previous Hybrid Model 
studies~\cite{Casalderrey-Solana:2014bpa,Casalderrey-Solana:2015vaa,Casalderrey-Solana:2016jvj,Hulcher:2017cpt,Casalderrey-Solana:2018wrw,Casalderrey-Solana:2019ubu,Hulcher:2022kmn,Bossi:2024qho,Kudinoor:2025ilx,Kudinoor:2025gao,Beraudo:2025nvq,Hulcher:2026dht,Kudinoor:2026wcs},
we will use the holographically derived form for the rate at which an energetic parton traversing strongly coupled gauge theory plasma loses energy $E$~\cite{Chesler:2014jva,Chesler:2015nqz}, 
\begin{equation}
\label{eq:elossrate}
   \frac{d E}{d  \tau_{\vec\beta=0}}= - \frac{4}{\pi} E_{\rm in} \frac{\tau_{\vec\beta=0}^2}{\ell_{\rm stop}^2} \frac{1}{\sqrt{\ell_{\rm stop}^2-\tau_{\vec\beta=0}^2}} \quad ,
\end{equation}
where 
\begin{equation}\label{eq:stopping_length}
\ell_{\rm stop}\equiv  \frac{1}{2\kappa_{\rm sc}}\frac{E_{\rm in}^{1/3}}{{T}^{4/3}}
\end{equation}
is the maximum distance the parton can travel before completely thermalizing, which we evaluate at the local temperature $T$ of the QGP at the position of the high-energy parton at a given time.  
Importantly, the time variable that appears in the energy loss rate \eqref{eq:elossrate}
is the proper time in the local fluid rest frame at the spacetime location of the high-energy parton, which in this paper we  refer to as $\tau_{\vec\beta=0}$, and $E_{\rm in}$ and $E$ are the initial and subsequent parton energies in the local fluid rest frame. 
The parameter $\kappa_{\rm sc}$ 
can be calculated in strongly coupled
${\cal N}=4$ supersymmetric Yang-Mills theory~\cite{Chesler:2014jva} or,
for the strongly coupled QGP liquid formed in heavy ion collisions, can be constrained 
by comparing Hybrid Model calculations of 
jet and high-$p_T$ hadron suppression to experimental data in PbPb collisions~\cite{Casalderrey-Solana:2018wrw,Hulcher:2026dht}.
Such a comparison yields  $\kappa_{\rm sc}\simeq 0.404$
for the Hybrid Model using event-averaged hydrodynamic solutions to describe the droplets of QGP~\cite{Casalderrey-Solana:2018wrw}, 
and $\kappa_{\rm sc}\simeq 0.37$ if event-by-event hydrodynamic solutions are 
employed~\cite{Hulcher:2026dht}.
We shall not be comparing our benchmark calculations for samples of 50 high-energy partons to data in this paper, and shall simply take $\kappa_{\rm sc}=0.4$ throughout.

The expression \eqref{eq:elossrate} 
determines the amount of energy --- and momentum in the high-energy parton direction --- injected into the QGP liquid
during each interval $d\tau_{\vec\beta=0}$.
In any future Monte Carlo analysis of jets, for example via the Hybrid Model, it will be much more convenient to track the trajectories of the jet partons in 
the lab frame, and we shall do so also in this initial study. This means that
we must translate Eq.~\eqref{eq:elossrate} to the lab frame. This was done in Ref.~\cite{Casalderrey-Solana:2015vaa}, and the result turns out to be simple to state: $dE^{\rm lab}/dt^{\rm lab}$ is given by the same expressions as in Eqs.~\eqref{eq:elossrate} and \eqref{eq:stopping_length}. 
This means that we can break the high-energy parton trajectory in the lab frame into segments, each with a duration in $t^{\rm lab}$ of $\delta t^{\rm lab}=0.02$~fm$/c$,
and then for the $n$'th such segment 
we evaluate
\be\label{eq:Delta-En-lab}
\Delta E_{n}^{\rm lab} \equiv \int_{t_i^{\rm lab}+n\, \delta t^{\rm lab}}^{t_i^{\rm lab}+(n+1)\delta t^{\rm lab}} d t_{\rm lab} \frac{\diff E^{\rm lab}}{\diff t_{\rm lab}} \,.
\ee
As in Section 2, we shall assume that each high-energy parton begins to lose
energy at $t^{\rm lab}=t^{\rm lab}_i=0.4~{\rm fm}/c$, before which it is just free streaming. The integer $n$ with which we count segments of the high-energy parton trajectory begins at $n=0$ and ends at the $n_e$ where $t^{\rm lab}$ reaches the $t^{\rm lab}_e$ at which the high-energy parton has either lost all of its energy or crossed the freezeout hypersurface.

For each segment of the high-energy parton trajectory, we identify its position in the locally comoving frame as the corresponding deposition point and 
perform the calculation described in Sections~\ref{step1}-\ref{sec:Utility} 
to obtain the hydrodynamic perturbations 
$\delta\varepsilon$, $\delta g^{\eta_s}$
and $\delta{\bs g}^\perp$ in the local fluid rest frame
at a specified point 
$(\tau^f,x^f,y^f,\eta_s^f)$ on the freezeout hypersurface 
resulting from the wake in the hydrodynamic fluid 
sourced by the deposition of energy and momentum into the fluid by the high-energy parton during the $n$'th segment of its trajectory.  We perform this calculation for a unit deposition of energy and momentum at the  deposition point that we have identified. 
Recalling that our unit deposition is a deposition of 1 GeV of energy in the local fluid rest frame at the deposition point in spacetime, we then multiply the perturbations
at $(\tau^f,x^f,y^f,\eta_s^f)$
by the ratio of $\Delta E_n^{\rm lab}$ boosted to the local fluid rest frame at the deposition point to 1 GeV. 
We then repeat this calculation for
every segment of the high-energy parton trajectory beginning at $n=0$ and ending when the high-energy parton itself loses all of its energy or reaches the freezeout hypersurface, and add the resulting perturbations.
In this way, we obtain the hydrodynamic perturbations at the point
$(\tau^f,x^f,y^f,\eta_s^f)$ on the freezeout hypersurface due to the wake excited in the fluid by the high-energy parton. We see that even though our approach is similar in spirit to the method of Green's functions in the sense that in this step we have used the linearity of linearized hydrodynamics to superpose perturbations at a point on the freezeout hypersurface calculated using different template solutions that, in sum, describe the consequences of the entire source we wish to analyze, it differs in detail because we have employed templates constructed by solving the equations in a Bjorken flow background.

Finally, and this is step (8) of our
procedure, we repeat the entire calculation for all points on the freezeout hypersurface.  

Our task in Section~\ref{sect:cf} will be to translate the hydrodynamic perturbations on the freezeout surface coming from the wake of a high-energy parton into perturbations to the particle spectra (as a function of the transverse momentum, direction in azimuthal angle $\phi$, and rapidity y) of the hadrons produced from the hydrodynamic fluid at freezeout.

Our goal in analyzing jet wakes 
for two samples that each contain only 50 high-energy partons is benchmarking the results of our procedure, not doing phenomenology.
We shall do this by comparing the results of our linearized hydrodynamic calculations to the results of two other calculations for the same samples of 50 high-energy partons, with the same rate of energy loss~\eqref{eq:elossrate}.
First, we shall compare our results to results from the very much simpler
description introduced in Ref.~\cite{Casalderrey-Solana:2016jvj}, 
described there as a
``crude treatment of particles from 
the wake'' of a jet, and employed in 
all subsequent Hybrid Model calculations.
Second, we shall compare our results to the results obtained by doing a full nonlinear hydrodynamics calculation of the wakes of the same 50 high-energy partons using the MUSIC code,
as described in Section~\ref{sect:music}.
If linearized, this full calculation would be equivalent to using Green's functions (for every deposition point in three-dimensional space, every deposition time, and every deposition direction) calculated in a background that includes transverse flow. 
We shall find that although our procedure is {\it much} more involved than the crude approach introduced in Ref.~\cite{Casalderrey-Solana:2016jvj},
as judged by comparing to the MUSIC results it yields a {\it much} better description of the soft particle production at freezeout coming from the wakes of high-energy partons.  And, we shall see in 
Section~\ref{sect:music} that because we have built our approach upon using template solutions calculated in a Bjorken flow background, our approach is computationally {\it much}
faster than solving the hydrodynamics equations for each wake.

\section{Particle Production from the Wake of a High-Energy Parton at Freezeout}
\label{sect:cf}

Once we know how the wake of a high-energy parton has perturbed the hydrodynamic fluid at the freezeout hypersurface, we must compute the resulting perturbation to the momentum distribution of the hadrons produced as the hydrodynamic fluid freezes out.
We describe particle production using the standard Cooper-Frye prescription~\cite{Cooper:1974mv} which assumes that the hadrons formed from each cell in the hydrodynamic fluid at the freezeout hypersurface in spacetime
are produced with a momentum distribution that is in thermal equilibrium with temperature $T$, boosted by the velocity of the fluid in the cell. 
(We shall assume that the hydrodynamic fluid has zero net baryon number and hence zero baryon chemical potential, which is a good approximation for heavy ion collisions at LHC energies.)
Equivalently, the particles produced from a cell at the freezeout surface are in thermal equilibrium with temperature $T$ in the local fluid rest frame
defined by the velocity of the fluid cell. 
In this way, the Cooper-Frye prescription ensures that just after freezeout the hadrons produced from a fluid cell 
have the same energy and momentum that the fluid cell had just before freezeout.

The Cooper-Frye prescription consists of integrating the contributions from all of the fluid cells on the freezeout hypersurface, denoted by $\sigma^\mu$, leading to a momentum distribution for each species of hadron with four-momentum $p_\mu$ that is given by~\cite{Cooper:1974mv}
\be
\label{eq:cfgen}
E\frac{\diff N}{\diff^3 p} = \frac{1}{(2\pi)^3} \int \diff \sigma^\mu p_\mu \, f\Big(\frac{u^\mu p_\mu}{T}\Big)\,,
\ee
where the integral is over the entire freezeout surface and $f(x)$ is the Bose-Einstein distribution $(e^x-1)^{-1}$ for hadron species that are bosons or the Fermi-Dirac distribution $(e^x+1)^{-1}$ for hadron species that are fermions. 
As is typical, we shall assume that the temperature of the unperturbed fluid, $T_0$, is the same at all points on the freezeout surface. If we insist that the freezeout temperature is $T_0$ even in the presence of perturbations arising from the wake of a jet, these perturbations would result in perturbations to the shape of the freezeout surface. We shall instead keep the freezeout surface the same as for the unperturbed fluid, and use the equation of state to compute the perturbation to the temperature $T$ due to the perturbations in the fluid coming from the wake.
The momentum distribution of the particles produced from the wake resulting from the energy and momentum lost by a high-energy parton can be obtained by computing the difference between the momentum distribution  with and without the perturbation
to the hydrodynamic fluid at the freezeout surface originating from the wake, i.e.
\be
\label{eqn:cf_wake}
E\frac{\diff \Delta N}{\diff^3 p} = \frac{1}{(2\pi)^3} \int \diff \sigma^\mu p_\mu \,  f \Big(\frac{u^\mu p_\mu}{T}\Big) - \frac{1}{(2\pi)^3} \int \diff \sigma_0^\mu p_\mu \,  f \Big(\frac{u_0^\mu p_\mu}{T_0}\Big) \,,
\ee
in which $T=T_0+\delta T$ where $\delta T$ is related to $\delta\varepsilon$ by the equation of state as we describe below and $u^\mu = u_0^\mu+\delta u^\mu$ where $\delta u^\mu$ is related to the momentum perturbations in the hydrodynamic fluid by Eqs.~\eqref{eqn:define_gperp} and \eqref{eqn:define_geta}. The freezeout hypersurfaces $\diff\sigma$ and $\diff\sigma_0$ could in general differ between the perturbed and unperturbed fluids but, as noted, we shall take $\diff\sigma=\diff\sigma_0$.

Since both $\sigma^\mu p_\mu$ and $u^\mu p_\mu$ are Lorentz invariant, we can choose any Lorentz frame in which to evaluate each of them. We shall use the
local fluid rest frame at each
point on the freezeout hypersurface (found as described in Sections~\ref{step1} and \ref{step2})
to evaluate $u^\mu p_\mu$, below, but for $\sigma^\mu p_\mu$ this is unnecessary as it is straightforward to use the laboratory frame. In either case, the numerical implementation of the freezeout hypersurface demands some care, as this hypersurface is spacelike close to the center of the fireball (where, loosely speaking, the freezeout temperature is reached at a certain proper time as the droplet of QGP cools) and is timelike at the edge of the fireball (where, again loosely speaking, the freezeout temperature is reached at a certain radius that moves outwards as the droplet of QGP expands).

\begin{figure}[t]
    \centering
    \includegraphics[width=0.8\textwidth]{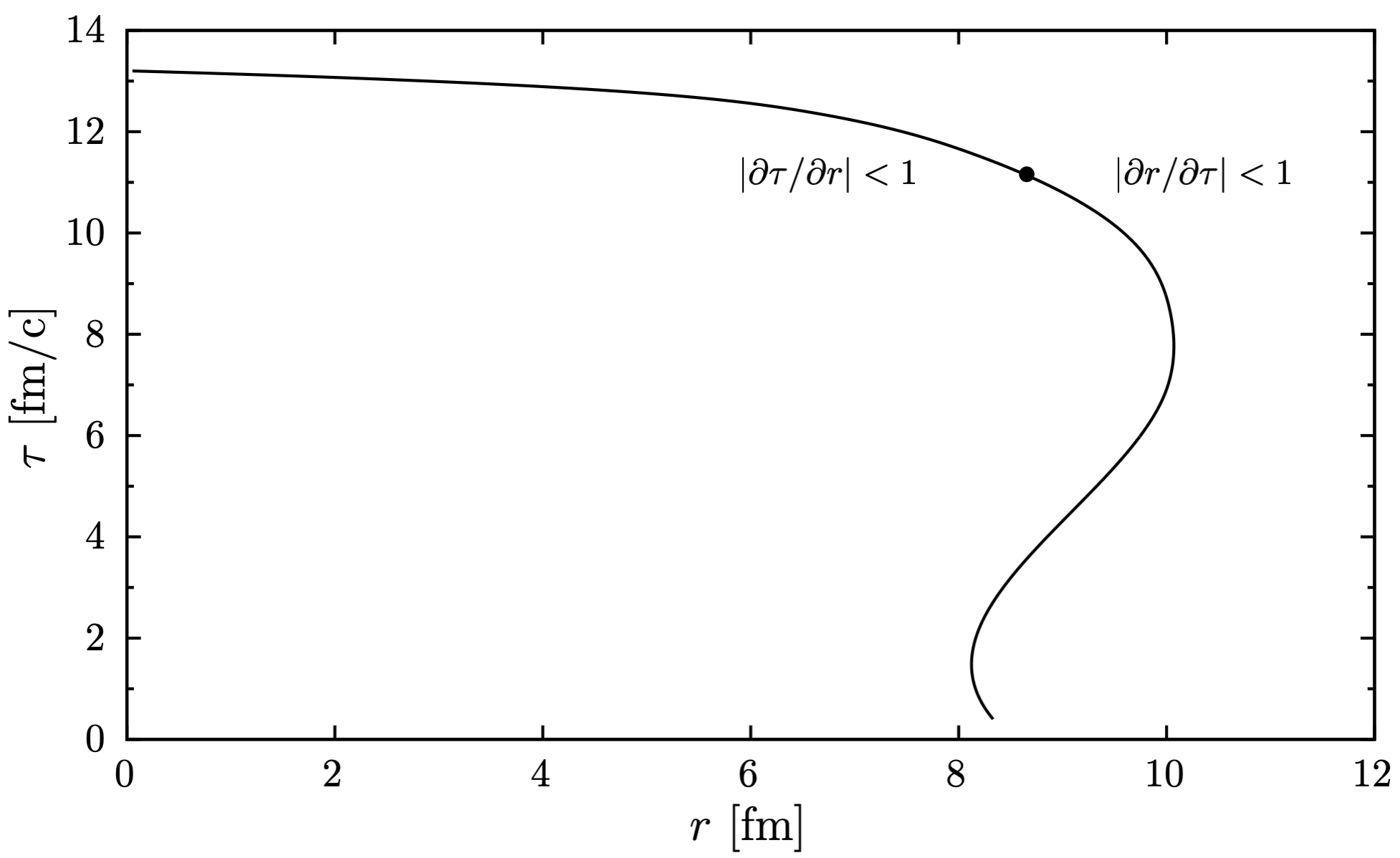}
    \caption{Freezeout hypersurface
    in lab frame Milne coordinates $\tau$
    and transverse radial distance $r$
    for a  PbPb collision with zero impact parameter and collision energy $\sqrt{s_{\rm NN}}=5.02$~TeV from a hydrodynamic simulation done with the MUSIC code~\cite{Schenke:2010nt,Schenke:2010rr,Ryu:2015vwa,Paquet:2015lta}. We use this freezeout hypersurface throughout our calculation, as well as for the MUSIC calculation of Section~\ref{sect:music} against which we shall benchmark our results. 
    The dot corresponds to the point where the hypersurface changes from spacelike to timelike. 
    The hypersurface is boost invariant, meaning that it does not depend on $\eta_s^f$, and is azimuthally symmetric, meaning that it is often referred to as ``mushroom-shaped'' in $(\tau,r,\phi)$. 
     } 
    \label{fig:fzoutpareme}
\end{figure}

In Fig.~\ref{fig:fzoutpareme} we show the
freezeout hypersurface that we employ throughout our calculation. It is obtained from a boost-invariant, azimuthally symmetric, hydrodynamic simulation of the fireball produced in an ultrarelativistic PbPb collision with zero impact parameter and collision energy $\sqrt{s_{\rm NN}}=5.02$~TeV done using the MUSIC code~\cite{Schenke:2010nt,Schenke:2010rr,Ryu:2015vwa,Paquet:2015lta}, and is the hypersurface at which the expanding and cooling droplet of hydrodynamic fluid in this calculation reaches $T^f=145$~MeV.
In any frame, the freezeout hypersurface defined by $\tau^f(x,y)$ has the following normal surface element: 
\be
\diff^3 \sigma_\mu = \Big( \cosh\eta_s, -\frac{\partial\tau^f}{\partial x}, -\frac{\partial\tau^f}{\partial y}, -\sinh\eta_s \Big) \tau^f(x,y) \diff x \diff y \diff\eta_s \,,
\label{eq:NormalSurfaceElement}
\ee
whose derivation is explained in Appendix~\ref{app:dsigma}. 
While this expression is in principle valid anywhere on the hypersurface, 
it is numerically inconvenient to use this expression to describe the 
region where the hypersurface
in Fig.~\ref{fig:fzoutpareme} 
is timelike.  We use Eq.~\eqref{eq:NormalSurfaceElement}
only on the ``top of the mushroom'', to the left of the black dot in Fig.~\ref{fig:fzoutpareme}.
In the region on the ``side of the mushroom'' where the freezeout hypersurface is timelike,
it is better to think of $x^f$ and $y^f$ changing with $\tau$. 
In regions where the freezeout hypersurface is timelike, we find it much more efficient to parametrize the surface by its proper time  coordinate, $r^f(\tau,\phi)$, with $\left(x, y\right)=\left(r\cos\phi,r\sin \phi\right)$. In a central collision, where the surface does not depend on the azimuthal angle $\phi$,  the normal surface element may be equivalently parametrized by (see  Appendix~\ref{app:dsigma}) 
\be\label{eq:NormalTimelikeSurfaceElement}
\diff^3 \sigma_\mu = \Big(-\frac{\partial r^f}{\partial \tau}\cosh\eta_s, \cos\phi, \sin\phi, \frac{\partial r^f}{\partial\tau} \sinh\eta_s \Big) \,
r^f \!\left(\tau\right) \tau \diff\tau \diff\phi \diff \eta_s\ .
\ee
We use the parametrization in Eq.~\eqref{eq:NormalSurfaceElement}
whenever $|\partial \tau^f/\partial r| \leq 1$ and switch to the equivalent
parametrization in Eq.~\eqref{eq:NormalTimelikeSurfaceElement} 
wherever $|\partial r^f/\partial \tau|\leq 1$. In this way, we can cover the full freezeout hypersurface while always avoiding the appearance of large gradients
in our parametrization as these can cause numerical difficulties. 
In practice, with this procedure we find that the contribution of the terms proportional to either $\partial \tau^f/\partial r$ or $\partial r^f/\partial \tau$ is small 
almost everywhere on the freezeout hypersurface.
For this reason, we have dropped the contribution from these terms in the evaluation of our results.
By comparing to the results from the solutions
to full nonlinear $(3+1)$-dimensional viscous hydrodynamics
obtained via the code 
MUSIC~\cite{Schenke:2010nt,Schenke:2010rr,Ryu:2015vwa,Paquet:2015lta}, which we describe in Section~\ref{sect:music}, 
we see that dropping the contribution from these terms does not worsen our description of jet wakes; in fact in some cases it seems to improve it somewhat.

The next element that we need in order to
use the Cooper-Frye expression~\eqref{eqn:cf_wake}
is the relationship between 
the perturbation $\delta T$
to the temperature $T$ of the fluid in the local fluid rest frame at points on the freezeout surface 
and the perturbation $\delta\varepsilon$
to the energy density of the fluid at the freezeout surface caused by the wake of the high-energy parton. 
For simplicity, we will use 
a conformal equation of state with the form
\be
\varepsilon = \frac{3g}{\pi^2}T^4 \,,
\ee
choosing $g=40$ as this is comparable to the result from a lattice QCD calculation for QGP 
with $N_f=2+1$ flavors (2 light and 1 strange quarks) at high temperature~\cite{Bazavov:2014pvz}.
With this equation of state, we can relate the local temperature of the unperturbed fluid at freezeout, namely $T_0=T^f=145$~MeV, and of the perturbed fluid, $T$, to their respective local-fluid-rest-frame energy densities:
\be
\label{eqn:Tperturb}
T_0 = \left(\frac{\pi^2}{3g} \varepsilon_0  \right)^{1/4}\,, \qquad T = T_0 + \delta T = \left(\frac{\pi^2}{3g}\left( \varepsilon_0+\delta\varepsilon \right) \right)^{1/4}  \,,
\ee
where we have 
calculated $\delta\varepsilon$ in the previous Section.

The last element that we need in order to use the Cooper-Frye expression~\eqref{eqn:cf_wake} to compute the perturbation to the momentum distribution of the hadrons after freezeout coming from the wake in the fluid excited by a high-energy parton is $u^\mu p_\mu$. This is in a sense the crucial element, because it links $u^\mu$ --- which describes the hydrodynamic fluid just before freezeout including the wake --- to $p_\mu$ --- the four-momentum of hadrons just after freezeout. 
We will compute $u^\mu p_\mu$ in the 
local rest frame of the fluid, with its background longitudinal and transverse flow,
at each point on the freezeout hypersurface.
We have obtained the fluid velocity perturbations $\delta u^{\eta_s}$ and
$\delta{\bs u}^\perp$ in the local fluid rest frame at each point on the freezeout hypersurface in Section~\ref{sect:lin_flow}
in the form 
\be
\label{eqn:du_depsdp}
\delta u^\mu = \Big( 0, \frac{g^x}{\varepsilon_0+P_0}, \frac{g^y}{\varepsilon_0+P_0}, \tau\frac{g^{\eta_s}}{\varepsilon_0+P_0} \Big) \,,
\ee
with all quantities, including the unperturbed $\varepsilon_0+P_0$ and 
the momentum perturbations, being in the local fluid rest frame expressed in Minkowski coordinates. 
In this frame, $u^\mu=(1,0,0,0)+\delta u^\mu$. To evaluate $u^\mu p_\mu$ in this frame, we need the four-momentum of the hadron immediately after freezeout, $p_\mu$, in the frame that is locally comoving with the fluid at the freezeout point, i.e.~immediately before freezeout. We obtain this by starting from $p_\mu^{\rm lab}$ and performing the same  transformations that we described in Sections~\ref{step1} and \ref{step2}.
At any one point in spacetime, in our case at a point on the freezeout hypersurface, this transformation is a concatenation of Lorentz transformations.
Upon performing this transformation and} using Eqs.~\eqref{eqn:define_gperp} and \eqref{eqn:define_geta}, we find
\be
\label{eqn:udotp}
u^\mu p_\mu &=& m_{T} \cosh \ma{y} - \frac{{\bs g}^\perp \cdot {\bs p}^{\perp}}{\varepsilon_0+P_0} -  \frac{g^{\eta_s} \tau_{\vec\beta=0}^f \, m_{T}\sinh\ma{y}}{\varepsilon_0+P_0} \,,
\ee
where y and $m_T=\sqrt{E^2-p_z^2}$ are the momentum rapidity and transverse mass of the hadron in the local fluid rest frame
at a point 
$(\tau^f,x^f,y^f,\eta_s^f)$ on the freezeout hypersurface, and ${\bs g}^\perp$ and $g^{\eta_s}$ are the momentum perturbations to the fluid at the freezeout point in the local fluid rest frame, calculated in Section~\ref{sect:lin_flow}.
This is the final element that we need in order to use the Cooper-Frye expression to evaluate the distribution of hadron momenta and rapidities after freezeout.
We shall perform the Cooper-Frye integral~\eqref{eqn:cf_wake} over the freezeout hypersurface in the lab frame, noting that the value of $u^\mu p_\mu$ at each point on the hypersurface that
we have evaluated in Eq.~\eqref{eqn:udotp} is the same in any frame.

Substituting Eqs.~\eqref{eqn:Tperturb} and \eqref{eqn:udotp}, together with the parametrizations \eqref{eq:NormalSurfaceElement} and
\eqref{eq:NormalTimelikeSurfaceElement}
of the normal surface elements of the freezeout hypersurface into the Cooper-Frye expression~\eqref{eqn:cf_wake} yields the complete description of particle production from the wake of a high-energy parton. Specifically, it yields the contribution to the momentum distribution of hadrons after freezeout coming from the hydrodynamic wake left behind by a high-energy parton traversing a fluid with a boost invariant longitudinal expansion that is also expanding radially in the transverse plane.

\section{$(3+1)$-Dimensional Hydrodynamics}
\label{sect:music}

We shall benchmark our calculation of the wakes of 50 different high-energy partons with initial energies $E_{\rm in}^{\rm lab}=10$~GeV and 50 GeV by comparing our results to results obtained via the much simpler crude treatment of Ref.~\cite{Casalderrey-Solana:2016jvj} that 
has been employed in Hybrid Model calculations since 2016 (that we shall refer to as ``Old Wake'')
and to results obtained from full nonlinear 
$(3+1)$-dimensional viscous hydrodynamics calculations. 
We shall refer to our results as ``E-Wake'', with the E referring to their computational Efficiency relative to calculations done via full $(3+1)$-dimensional hydrodynamics.  In this way we shall be able to confirm that the E-Wake procedure that we have developed and presented over Sections~\ref{sect:linear}, \ref{sect:lin_flow} and \ref{sect:cf}
represents a substantial improvement over the Old Wake procedure.  We shall at the same time be able to see the degree to which it does not fully reproduce the description of jet wakes by nonlinear 
$(3+1)$-dimensional hydrodynamics.

We have performed full nonlinear relativistic viscous hydrodynamics calculations of the wakes of
the same 50 high-energy partons with each of two initial energies, all with $\eta_s=0$, with different points of origin and directions in the transverse plane, using the 
MUSIC code~\cite{Schenke:2010nt,Schenke:2010rr,Paquet:2015lta}. 
As in a typical MUSIC calculation, we employ the equation of state computed via lattice calculations of QCD thermodynamics by the HotQCD Collaboration~\cite{HotQCD:2014kol}.
As in our E-Wake calculations, we take 
the specific viscosity, namely the ratio of the shear viscosity to the entropy density,
to be
$\eta/s=1/4 \pi$. We do not include bulk viscosity. We use MUSIC to solve the hydrodynamic equations of motion
\begin{equation}
    \nabla_{\mu}T^{\mu \nu}_{\rm hydro} = J^{\nu}
\end{equation}
for a source
term with the same form as the source we have employed in our E-Wake calculations, namely 
\begin{eqnarray}
\label{eq:source}
J^{\nu}(\tau,x,y,\eta_s)&=&\sum_n \frac{P^{{\rm lab}\ \nu}_{\rm parton}(\tau_n)}{E_{\rm parton}^{\rm lab}(\tau_n)}\frac{\Delta E_n^{\rm lab}}{\delta t^{\rm lab}}\frac{1}{ (2 \pi)^{3/2} \sigma_x^2 \sigma_{\eta_s}\tau} \times  \\
&\ &
\exp\left[
-\frac{\left(x-x_{\rm parton}(\tau_n)\right)^2 + \left(y-y_{\rm parton}(\tau_n)\right)^2}{2 \sigma_x^2}  
-\frac{\eta_s^2}
{2 \sigma_{\eta_s}^2} 
\right]\, ,\nonumber
\end{eqnarray}
where $\Delta E_n$ 
is given in Eq.~\eqref{eq:Delta-En-lab} and is specified via Eq.~\eqref{eq:elossrate},
the Gaussian widths of the source are taken to be $\sigma_x=1/\pi T$ fm and $\sigma_{\eta_s}=1/\pi$ as in Eq.~\eqref{eqn:Jmu}, where
$\tau_n\equiv\tau_i+n\,\delta t^{\rm lab}$,
and where we choose to cut away the contributions coming from the tails of the Gaussian that are beyond $5 \sigma$ in any direction.
This is essentially the same setup used in Ref.~\cite{Pablos:2022piv}.
We have done our $(3+1)$-dimensional hydrodynamics calculations with MUSIC 
on 
a $N_x\times N_y\times N_{\eta_s}\times N_\tau = 300\times 300\times 120\times630$
grid with grid spacings of $0.09865$~fm in $x$ and $y$, 0.1 in $\eta_s$, and 0.02 in $\tau$. This ensures that our spatial grid spacings are smaller than the widths of our Gaussian source and that our temporal grid spacing is sufficiently smaller than our spatial grid spacings to ensure numerical stability.

We have also used MUSIC to compute the isothermal freezeout surface at $T^f=145$~MeV, and use this freezeout surface here as we already did in Section~\ref{sect:cf} in our E-Wake calculation.  
We compute the contribution to the 
momentum distributions of particles after freezeout 
coming from the wake of the high-energy parton as described via the full $(3+1)$-dimensional MUSIC calculation using
the same Cooper-Frye formula~\eqref{eqn:cf_wake} as before. 
For simplicity, we shall ignore viscous corrections in the Cooper-Frye formula here, as we have already done in our E-Wake calculation.  We shall show the results of these calculations in the next Section, where we compare results from our new E-Wake calculations for the wakes of 50 partons with each of two different initial energies to these $(3+1)$-dimensional hydrodynamics results and to Old Wake results.

Our $(3+1)$-dimensional nonlinear viscous hydrodynamics calculation of the wake of a single high-energy parton takes about $\sim 3000$ minutes, so $\sim 2\times 10^5$ seconds, on a single CPU core.  
We have not sought to optimize the grid spacings and number of grid points.
Doing so could speed up the MUSIC computation somewhat.
To get a sense of this, we note that the CoLBT collaboration reports~\cite{Wu:2025CoLBTHydroResponse}
being able to do the full hydrodynamics calculation of the wake of a high-energy parton~\cite{Chen:2017zte,Zhang:2018urd,Yang:2021qtl,Zhao:2021vmu,Yang:2022nei}
in 7 to 10 minutes on an A100 
GPU. If this GPU
is around 60 times faster than a single CPU
as in the comparison made in Ref.~\cite{Pang:2018zzo}, 
this would suggest that
further optimization of the MUSIC calculations relative to what we have done could speed them up by a factor $\sim 5$ on the CPU that we have used.

In contrast, with our E-Wake procedure, after the templates have been computed and tabulated  (which must be done only once and which takes $\sim 7000$ minutes on the CPU that we have used) 
and solving Eqs.~\eqref{eq:relfl} and~\eqref{eq:dthetadt} for the grid of magnitudes and directions of $\vec\beta_\perp$ described at the end of Section~\ref{step2} (which also need only be done once and which takes $\sim 500$ minutes)
the time it takes to compute the wake of a high-energy parton is only $\sim 0.8$ seconds.
That is, our E-Wake calculation is  $\sim 2\times 10^5$ times faster than our MUSIC calculation, meaning that it is tens of thousands of times faster than an optimized $(3+1)$-dimensional nonlinear viscous hydrodynamics calculation.  Looking ahead to a future implementation of the E-Wake procedure in the Hybrid Model, we anticipate that it should speed up the calculation of the wake of a single jet shower to the point that it takes 
$\sim 10$ seconds on a single CPU, making it possible to simulate a sample of $10^5$ jets with wakes in less than 
a day on a 16-core cluster.

We have also achieved a modest reduction
(by roughly a factor of 4) in the computation time for the Cooper-Frye calculation  in our E-Wake calculation relative to that in our MUSIC calculation  
 by noting that since 
in Eq.~\eqref{eqn:cf_wake}
we are calculating the difference between
the momentum distribution of the hadrons after freezeout with and without the perturbation originating from the wake of the high-energy parton, we only need to compute the Cooper-Frye integral over those regions of the freezeout hypersurface where $\delta\varepsilon/\varepsilon_0$ is greater than a threshold that we take to be $10^{-7}$.

\section{Results and Discussion}
\label{sect:results}

In this Section, we present our results for the momentum distribution of the hadrons originating from the wakes of high-energy partons, comparing the results from the computationally efficient procedure that we have introduced to results obtained via full $(3+1)$-dimensional hydrodynamical calculations of the wakes using MUSIC and results obtained via the ``Old Wake'' procedure~\cite{Casalderrey-Solana:2016jvj} that has been used in Hybrid Model calculations since 2016 and that neglects all effects of transverse flow. As noted earlier, we have analyzed the wakes of high-energy partons produced at 50 different points in the transverse plane sampled from a Glauber model weighted by the number of collisions, all moving at constant spacetime rapidity $\eta_s=0$, with their directions (in azimuthal angle in the transverse plane) sampled from a uniform distribution.
For each of the 50 initial configurations, we analyze the wakes of partons with initial energies $E_{\rm in}^{\rm lab}=10$~GeV and 50 GeV.

\begin{figure*}[t]
    \centering
    \vspace{-0.4in}
    \begin{subfigure}[!htbp]{1\textwidth}
        \centering
        \includegraphics[width=0.38\textwidth]{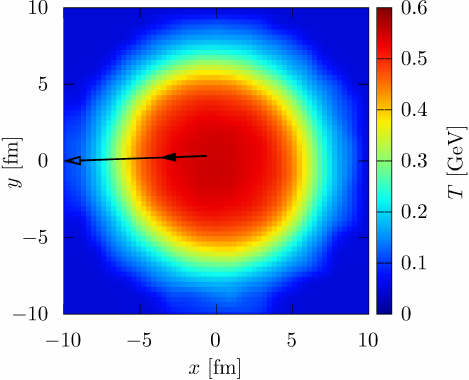}
        \caption{Energy-momentum deposition for $E_{\rm in}^{\rm lab}=10$ GeV (50 GeV), see the filled (open) arrow.}
        \label{fig:comb_contour_20}
    \end{subfigure}%
    
    \begin{subfigure}[!htbp]{0.43\textwidth}
        \centering
        \includegraphics[width=\textwidth]{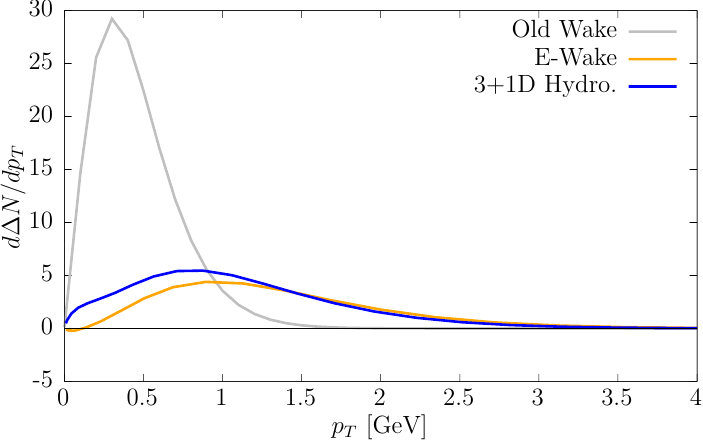}
        \caption{$p_T$ distribution for $E_{\rm in}^{\rm lab}=10$ GeV.}
        \label{fig:pt_10GeV_conf_20}
    \end{subfigure}%
    ~
    \begin{subfigure}[!htbp]{0.43\textwidth}
        \centering
        \includegraphics[width=\textwidth]{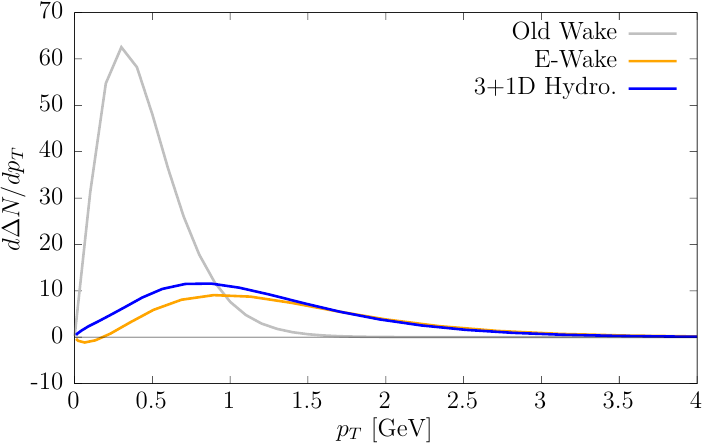}
        \caption{$p_T$ distribution for $E_{\rm in}^{\rm lab}=50$ GeV.}
          \label{fig:pt_50GeV_conf_20}
    \end{subfigure}%
    
    \begin{subfigure}[!htbp]{0.43\textwidth}
        \centering
        \includegraphics[width=\textwidth]{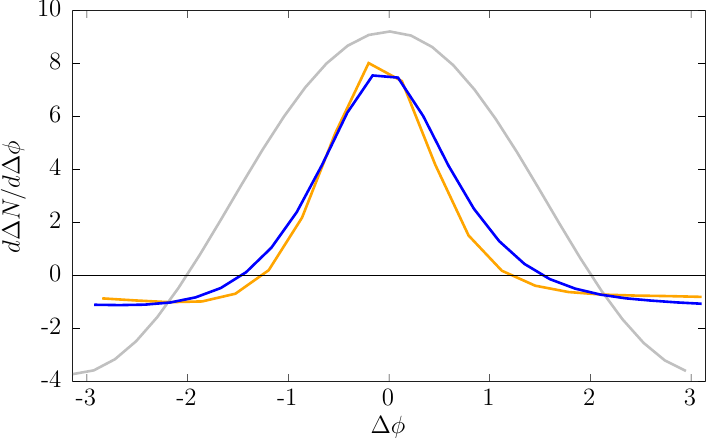}
        \caption{$\phi$ distribution for $E_{\rm in}^{\rm lab}=10$ GeV.}
        \label{fig:phi_10GeV_conf_20}
    \end{subfigure}%
    ~
    \begin{subfigure}[!htbp]{0.43\textwidth}
        \centering
        \includegraphics[width=\textwidth]{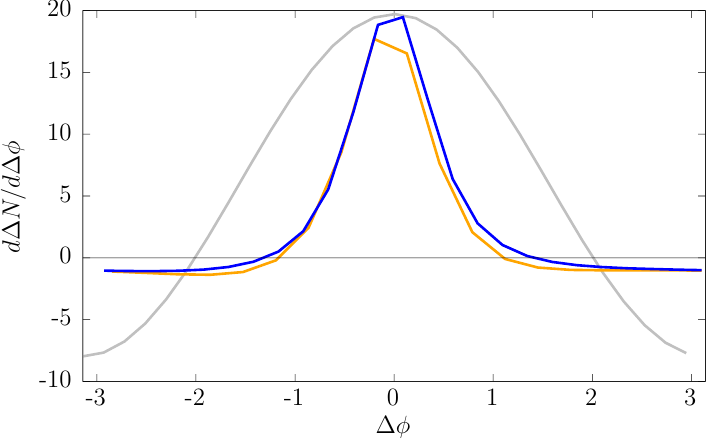}
        \caption{$\phi$ distribution for $E_{\rm in}^{\rm lab}=50$ GeV.}
        \label{fig:phi_50GeV_conf_20}
    \end{subfigure}%
    
    \begin{subfigure}[!htbp]{0.43\textwidth}
        \centering
        \includegraphics[width=\textwidth]{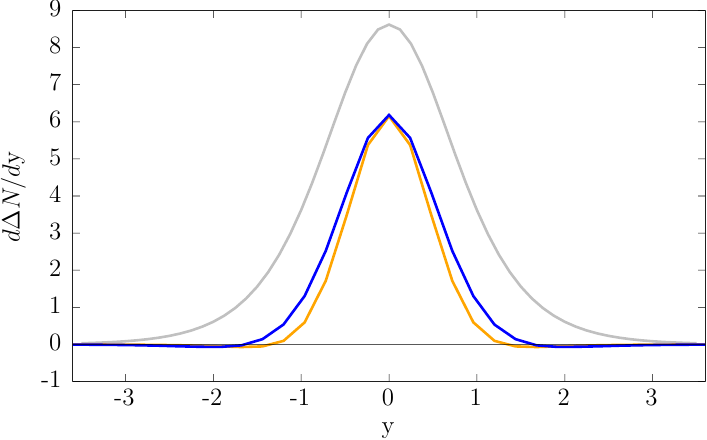}
        \caption{y distribution for $E_{\rm in}^{\rm lab}=10$ GeV.}
        \label{fig:y_10GeV_conf_20}
    \end{subfigure}%
    ~
    \begin{subfigure}[!htbp]{0.43\textwidth}
        \centering
        \includegraphics[width=\textwidth]{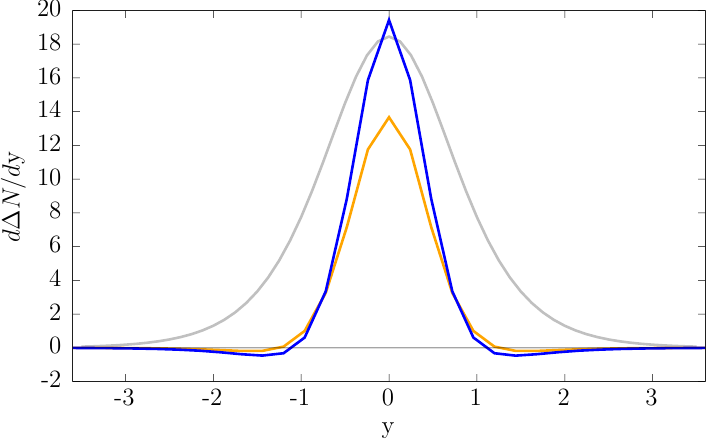}
        \caption{y distribution for $E_{\rm in}^{\rm lab}=50$ GeV.}
        \label{fig:y_50GeV_conf_20}
    \end{subfigure}%
\caption{Comparison between momentum
distributions (in $p_T$, $\phi$ and $\ma{y}$)
of hadrons from the wake of a particular high-energy parton with two different initial energies (see top panel) computed with our new E(fficient)-Wake procedure compared to 
results from the Old Wake calculation 
used in Hybrid Model studies
which excludes all effects of transverse flow 
and from a full $(3+1)$-dimensional hydrodynamical analysis (MUSIC).
}
\label{fig:comp_conf_20}
\end{figure*}

\begin{figure*}[t]
\vspace{-0.4in}
    \centering
    \begin{subfigure}[!htbp]{1\textwidth}
        \centering
        \includegraphics[width=0.38\textwidth]{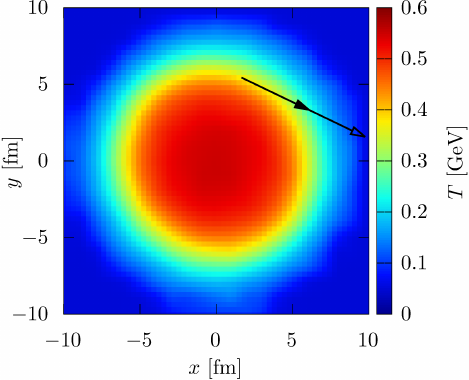}
        \caption{Energy-momentum deposition for $E_{\rm in}^{\rm lab}=10$ GeV (50 GeV), see the filled (open) arrow.}
        \label{fig:comb_contour_17}
    \end{subfigure}%
    
    \begin{subfigure}[!htbp]{0.43\textwidth}
        \centering
        \includegraphics[width=\textwidth]{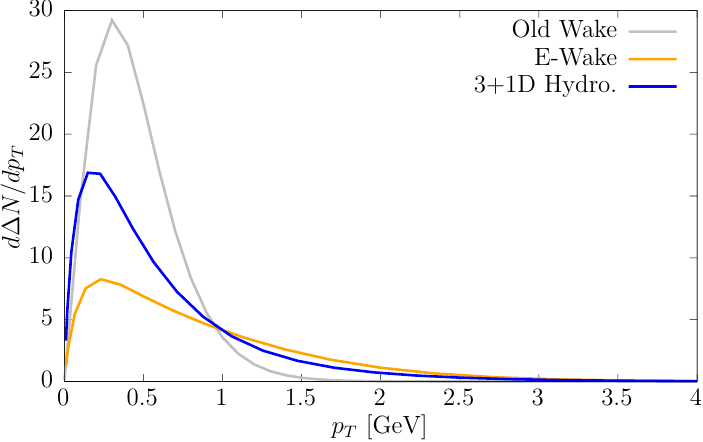}
        \caption{$p_T$ distribution for $E_{\rm in}^{\rm lab}=10$ GeV.}
        \label{fig:pt_10GeV_conf_17}
    \end{subfigure}%
    ~
    \begin{subfigure}[!htbp]{0.43\textwidth}
        \centering
        \includegraphics[width=\textwidth]{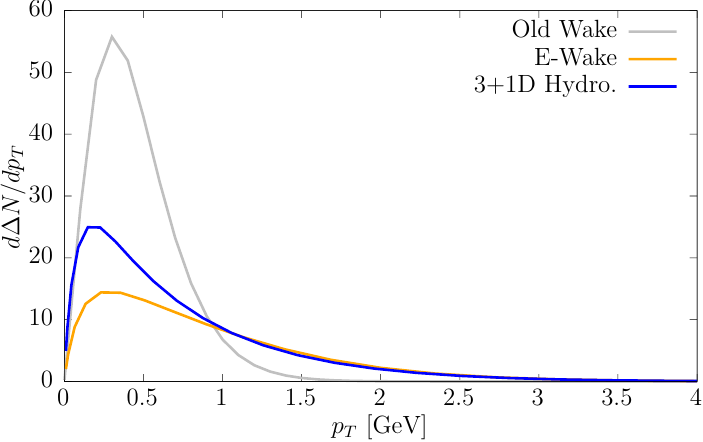}
        \caption{$p_T$ distribution for $E_{\rm in}^{\rm lab}=50$ GeV.}
          \label{fig:pt_50GeV_conf_17}
    \end{subfigure}%
    
    \begin{subfigure}[!htbp]{0.43\textwidth}
        \centering
        \includegraphics[width=\textwidth]{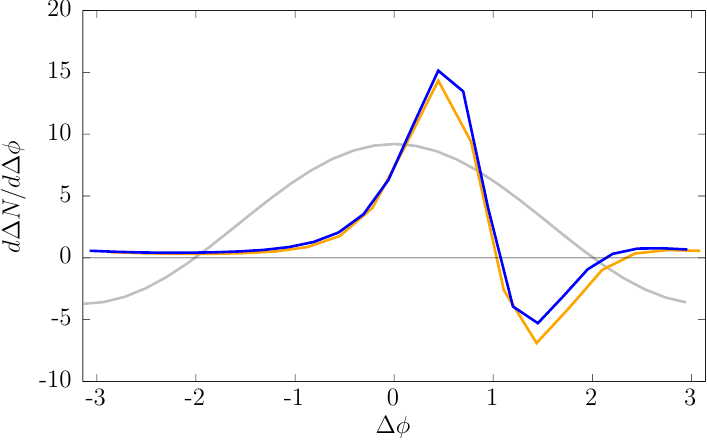}
        \caption{$\phi$ distribution for $E_{\rm in}^{\rm lab}=10$ GeV.}
        \label{fig:phi_10GeV_conf_17}
    \end{subfigure}%
    ~
    \begin{subfigure}[!htbp]{0.43\textwidth}
        \centering
        \includegraphics[width=\textwidth]{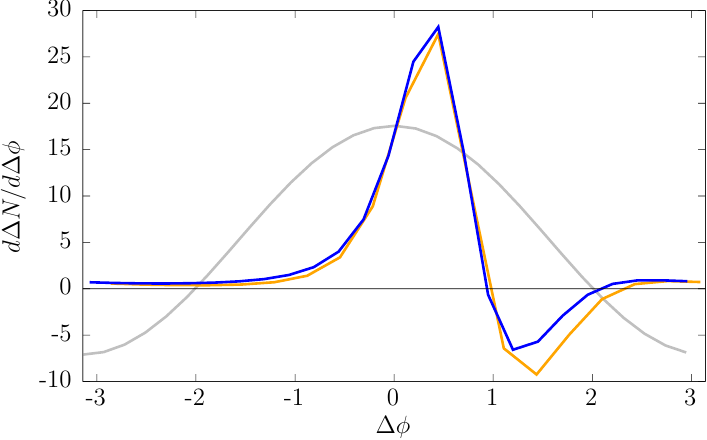}
        \caption{$\phi$ distribution for $E_{\rm in}^{\rm lab}=50$ GeV.}
        \label{fig:phi_50GeV_conf_17}
    \end{subfigure}%
    
    \begin{subfigure}[!htbp]{0.43\textwidth}
        \centering
        \includegraphics[width=\textwidth]{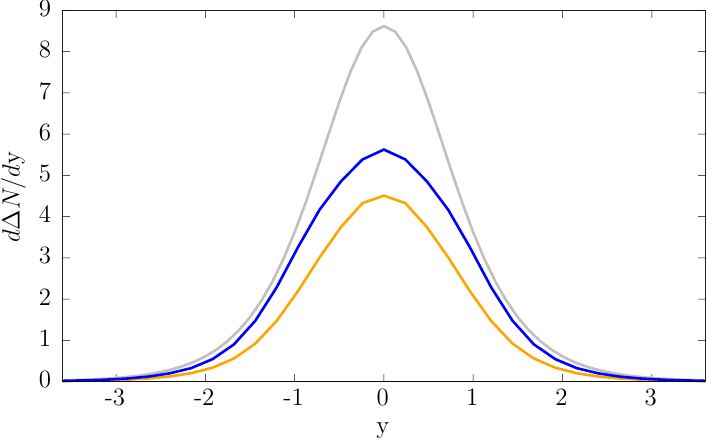}
        \caption{y distribution for $E_{\rm in}^{\rm lab}=10$ GeV.}
        \label{fig:y_10GeV_conf_17}
    \end{subfigure}%
    ~
    \begin{subfigure}[!htbp]{0.43\textwidth}
        \centering
        \includegraphics[width=\textwidth]{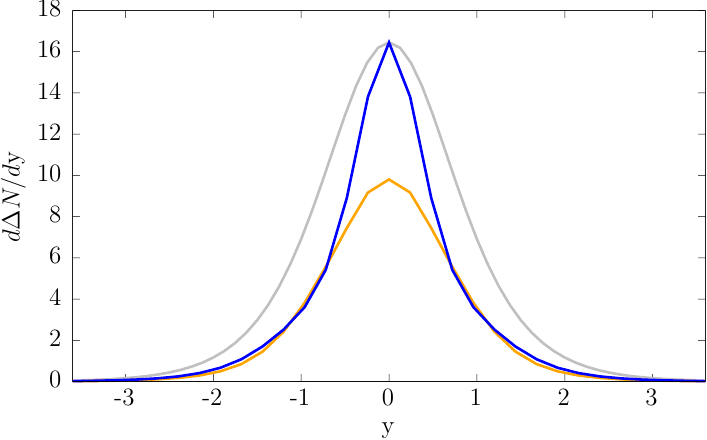}
        \caption{y distribution for $E_{\rm in}^{\rm lab}=50$ GeV.}
        \label{fig:y_50GeV_conf_17}
    \end{subfigure}%
\caption{Comparison between momentum
distributions (in $p_T$, $\phi$ and $\ma{y}$)
of hadrons from the wake of a particular high-energy parton with two different initial energies (see top panel) computed with our new E(fficient)-Wake procedure compared to 
results from the Old Wake calculation 
used in Hybrid Model studies
which excludes all effects of transverse flow 
and from a full $(3+1)$-dimensional hydrodynamical analysis (MUSIC).
}
\label{fig:comp_conf_17}
\end{figure*}

\begin{figure*}[t]
\vspace{-0.4in}
    \centering
    \begin{subfigure}[!htbp]{1\textwidth}
        \centering
        \includegraphics[width=0.38\textwidth]{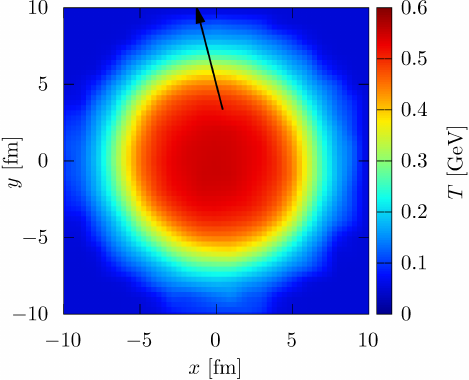}
        \caption{Energy-momentum deposition for $E_{\rm in}^{\rm lab}=10$ GeV (50 GeV), see the filled (open) arrow.}
        \label{fig:comb_contour_18}
    \end{subfigure}%
    
    \begin{subfigure}[!htbp]{0.43\textwidth}
        \centering
        \includegraphics[width=\textwidth]{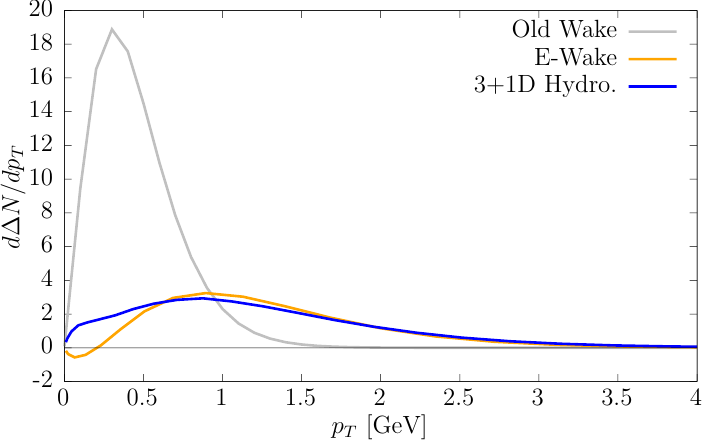}
        \caption{$p_T$ distribution for $E_{\rm in}^{\rm lab}=10$ GeV.}
        \label{fig:pt_10GeV_conf_18}
    \end{subfigure}%
    ~
    \begin{subfigure}[!htbp]{0.43\textwidth}
        \centering
        \includegraphics[width=\textwidth]{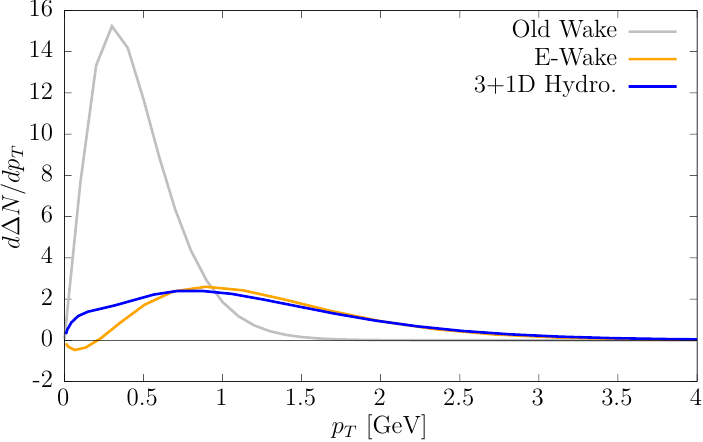}
        \caption{$p_T$ distribution for $E_{\rm in}^{\rm lab}=50$ GeV.}
          \label{fig:pt_50GeV_conf_18}
    \end{subfigure}%
    
    \begin{subfigure}[!htbp]{0.43\textwidth}
        \centering
        \includegraphics[width=\textwidth]{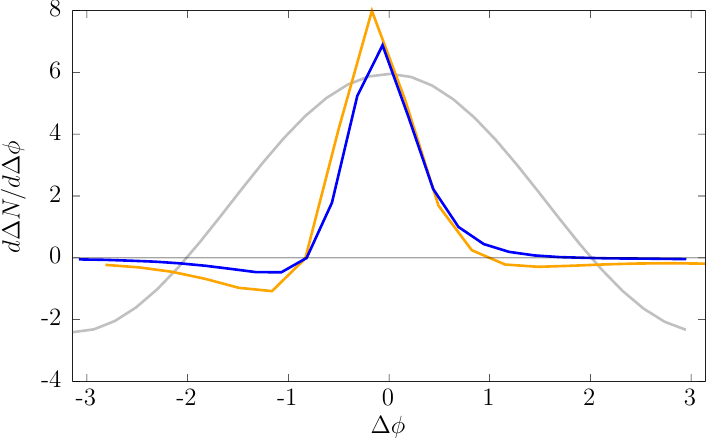}
        \caption{$\phi$ distribution for $E_{\rm in}^{\rm lab}=10$ GeV.}
        \label{fig:phi_10GeV_conf_18}
    \end{subfigure}%
    ~
    \begin{subfigure}[!htbp]{0.43\textwidth}
        \centering
        \includegraphics[width=\textwidth]{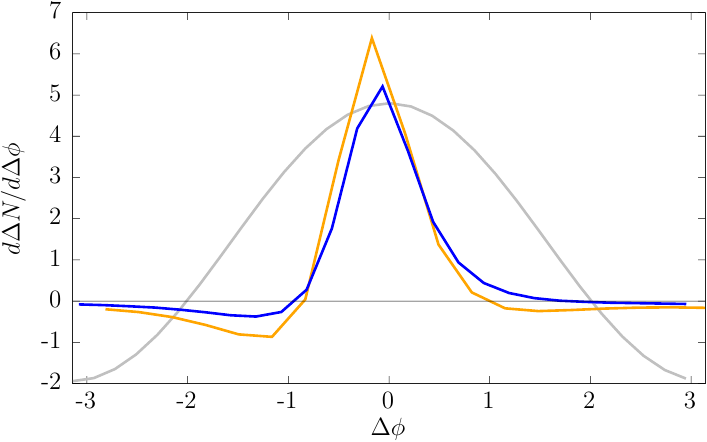}
        \caption{$\phi$ distribution for $E_{\rm in}^{\rm lab}=50$ GeV.}
        \label{fig:phi_50GeV_conf_18}
    \end{subfigure}%
    
    \begin{subfigure}[!htbp]{0.43\textwidth}
        \centering
        \includegraphics[width=\textwidth]{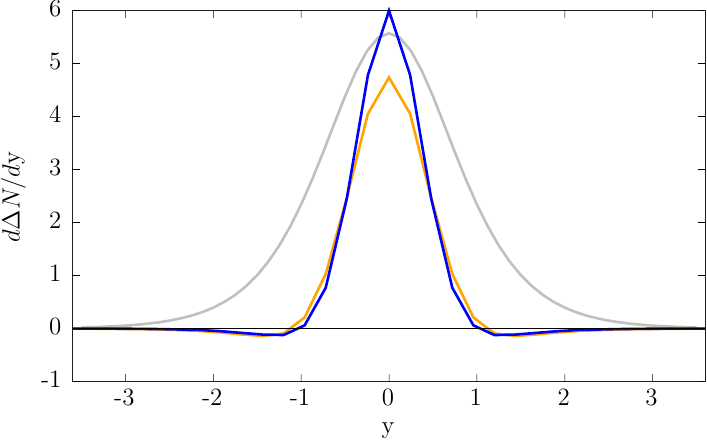}
        \caption{y distribution for $E_{\rm in}^{\rm lab}=10$ GeV.}
        \label{fig:y_10GeV_conf_18}
    \end{subfigure}%
    ~
    \begin{subfigure}[!htbp]{0.43\textwidth}
        \centering
        \includegraphics[width=\textwidth]{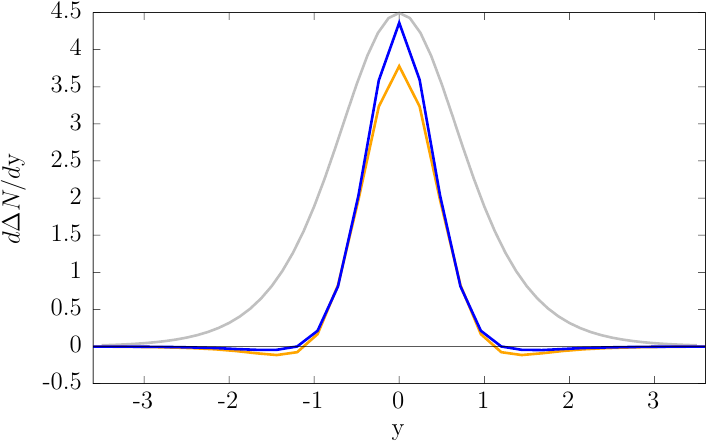}
        \caption{y distribution for $E_{\rm in}^{\rm lab}=50$ GeV.}
        \label{fig:y_50GeV_conf_18}
    \end{subfigure}%
\caption{Comparison between momentum
distributions (in $p_T$, $\phi$ and $\ma{y}$)
of hadrons from the wake of a particular high-energy parton with two different initial energies (see top panel) computed with our new E(fficient)-Wake procedure compared to 
results from the Old Wake calculation 
used in Hybrid Model studies
which excludes all effects of transverse flow 
and from a full $(3+1)$-dimensional hydrodynamical analysis (MUSIC).
}
\label{fig:comp_conf_18}
\end{figure*}

\begin{figure*}[t]
\vspace{-0.4in}
    \centering
    \begin{subfigure}[!htbp]{1\textwidth}
        \centering
        \includegraphics[width=0.38\textwidth]{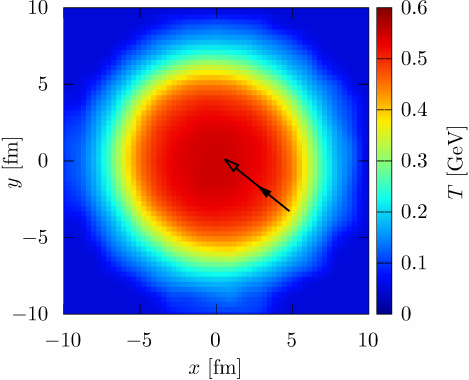}
        \caption{Energy-momentum deposition for $E_{\rm in}^{\rm lab}=10$ GeV (50 GeV), see the filled (open) arrow.}
        \label{fig:comb_contour_42}
    \end{subfigure}%
    
    \begin{subfigure}[!htbp]{0.43\textwidth}
        \centering
        \includegraphics[width=\textwidth]{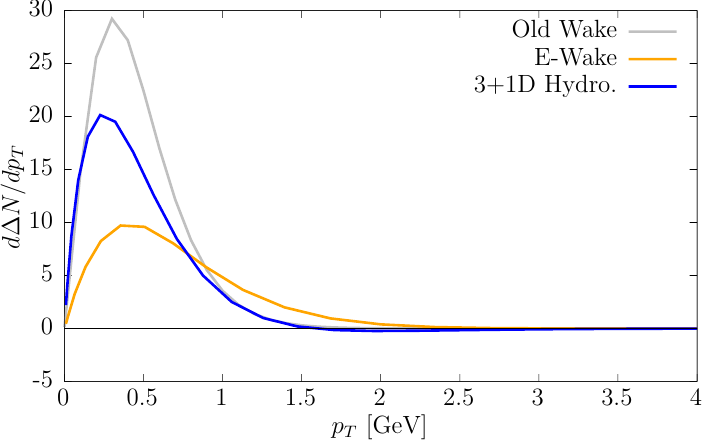}
        \caption{$p_T$ distribution for $E_{\rm in}^{\rm lab}=10$ GeV.}
        \label{fig:pt_10GeV_conf_5}
    \end{subfigure}%
    ~
    \begin{subfigure}[!htbp]{0.43\textwidth}
        \centering
        \includegraphics[width=\textwidth]{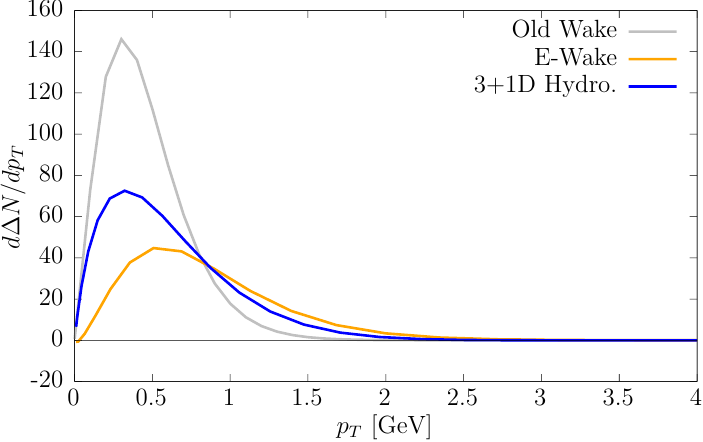}
        \caption{$p_T$ distribution for $E_{\rm in}^{\rm lab}=50$ GeV.}
          \label{fig:pt_50GeV_conf_42}
    \end{subfigure}%
    
    \begin{subfigure}[!htbp]{0.43\textwidth}
        \centering
        \includegraphics[width=\textwidth]{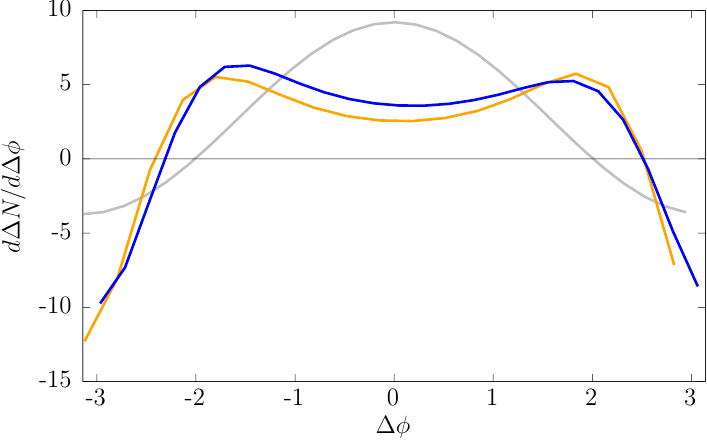}
        \caption{$\phi$ distribution for $E_{\rm in}^{\rm lab}=10$ GeV.}
        \label{fig:phi_10GeV_conf_42}
    \end{subfigure}%
    ~
    \begin{subfigure}[!htbp]{0.43\textwidth}
        \centering
        \includegraphics[width=\textwidth]{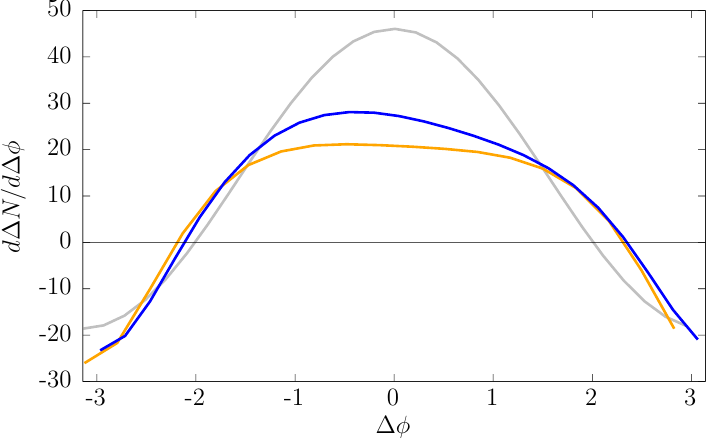}
        \caption{$\phi$ distribution for $E_{\rm in}^{\rm lab}=50$ GeV.}
        \label{fig:phi_50GeV_conf_42}
    \end{subfigure}%
    
    \begin{subfigure}[!htbp]{0.43\textwidth}
        \centering
        \includegraphics[width=\textwidth]{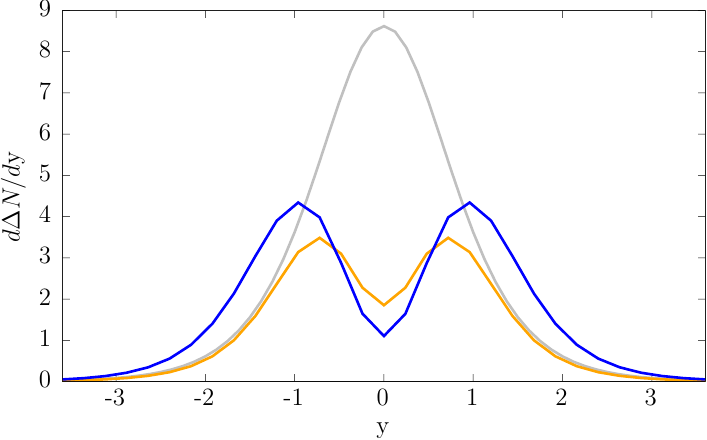}
        \caption{y distribution for $E_{\rm in}^{\rm lab}=10$ GeV.}
        \label{fig:y_10GeV_conf_42}
    \end{subfigure}%
    ~
    \begin{subfigure}[!htbp]{0.43\textwidth}
        \centering
        \includegraphics[width=\textwidth]{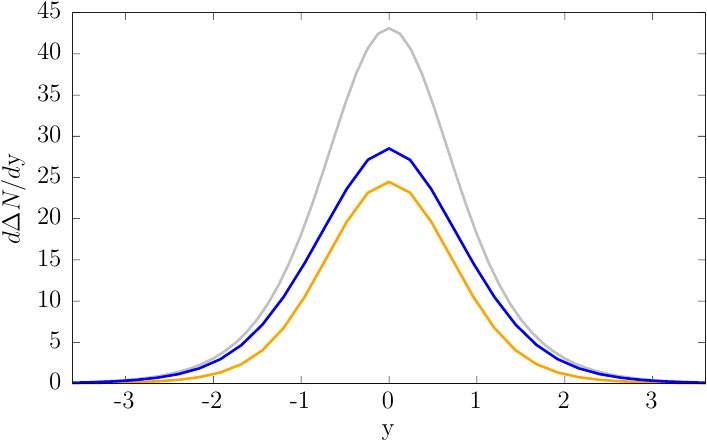}
        \caption{y distribution for $E_{\rm in}^{\rm lab}=50$ GeV.}
        \label{fig:y_50GeV_conf_42}
    \end{subfigure}%
\caption{Comparison between momentum
distributions (in $p_T$, $\phi$ and $\ma{y}$)
of hadrons from the wake of a particular high-energy parton with two different initial energies (see top panel) computed with our new E(fficient)-Wake procedure compared to 
results from the Old Wake calculation 
used in Hybrid Model studies
which excludes all effects of transverse flow 
and from a full $(3+1)$-dimensional hydrodynamical analysis (MUSIC).
}
\label{fig:comp_conf_42}
\end{figure*}

Rather than displaying the results for all 50 of the randomly chosen initial configurations, in Figs.~\ref{fig:comp_conf_20}, \ref{fig:comp_conf_17}, \ref{fig:comp_conf_18} and \ref{fig:comp_conf_42}
we have selected four of them in which the high-energy partons with 
$E_{\rm in}^{\rm lab}=10$~GeV and 50 GeV
are produced near the center traveling roughly outwards (Fig.~\ref{fig:comp_conf_20}) and at a radius of order 4-6~fm traveling roughly tangentially, roughly outwards, and roughly inwards (Figs.~\ref{fig:comp_conf_17}, \ref{fig:comp_conf_18} and \ref{fig:comp_conf_42} respectively).
In the top panels of each of these Figures, 
the color indicates the temperature of the background fluid at $\tau=\tau_i=0.4~{\rm fm}/c$,
the base of the black arrow denotes the point in the transverse plane where the high-energy partons 
begin depositing energy, with the tip of the solid (open) black arrowhead denoting the point in the transverse plane where the parton with initial energy 
$E_{\rm in}^{\rm lab}=10$~GeV (50 GeV) 
finds itself at the later time when it has
either lost all of its energy to the expanding cooling droplet of QGP or has escaped.  (The 10 GeV partons lose all of their energy to the hydrodynamic droplet in Figs.~\ref{fig:comp_conf_20}, \ref{fig:comp_conf_17} and \ref{fig:comp_conf_42}; the 50 GeV parton loses all of its energy only in Fig~\ref{fig:comp_conf_42}.)

In each of these four Figures, the six lower panels display the momentum distributions (in transverse momentum $p_T$, azimuthal angle $\phi$ in the transverse plane, and momentum rapidity $\ma{y}$) of the
hadrons originating from the wakes of the two high-energy partons with initial energies $E_{\rm in}^{\rm lab}=10$~GeV and 50 GeV.  
(Note that $\Delta\phi\equiv \phi-\phi_{\rm parton}^{\rm lab}$ is the azimuthal angle between the momentum of a hadron originating from the wake at freezeout and the momentum of the high-energy parton that sourced the wake.)
In the present benchmark study, for simplicity we have computed results upon assuming that all hadrons formed at freezeout are massless bosons.
In a future phenomenological study in which the method that we have introduced and benchmarked here is implemented in a 
jet Monte Carlo, it will be straightforward to incorporate pions, kaons, protons, $\ldots$ at freezeout. In the results we present here, we require that the total energy of the massless ``hadrons'' produced at freezeout is the same as the total energy lost by the high-energy parton.  In a future phenomenological study this could be ensured via including all hadron species consistent with the equation of state in the freezeout calculation.

In each of Figs.~\ref{fig:comp_conf_20}, \ref{fig:comp_conf_17}, \ref{fig:comp_conf_18} and \ref{fig:comp_conf_42}, the orange curves (``E-Wake'') show the results obtained via our new computationally efficient procedure, namely from the Cooper-Frye freezeout (described in Section~\ref{sect:cf}) of the hydrodynamic perturbations computed as described in Section~\ref{sect:lin_flow} built upon the Bjorken flow template solutions constructed in Section~\ref{sect:linear}.
We benchmark our results by comparing the orange curves to the blue curves, which 
show the results obtained from the full nonlinear $(3+1)$-dimensional viscous hydrodynamics calculations obtained using the MUSIC code~\cite{Schenke:2010nt,Schenke:2010rr,Paquet:2015lta} as described in Section~\ref{sect:music}.
By comparing the orange curves to the grey curves, we can also compare our results to 
results obtained via the ``Old Wake'' procedure, namely the early description of the soft hadrons originating from jet wakes introduced in Ref.~\cite{Casalderrey-Solana:2016jvj} and used since then in the Hybrid Model that is crude in a number of respects, including because it completely neglects
the effects of transverse radial flow on jet wakes.

The overall message of Figs.~\ref{fig:comp_conf_20}, \ref{fig:comp_conf_17}, \ref{fig:comp_conf_18} and \ref{fig:comp_conf_42} together is
that our new Efficient Wake procedure does a {\it much} better job of describing soft hadron production originating from the wakes of high-energy partons than the Old Wake procedure did: the orange curves come {\it much} closer to reproducing the features of the blue curves than the grey curves did. With respect to the distribution of the hadrons in azimuthal angle $\phi$, the orange curves reproduce the features of the blue curves remarkably well in all eight cases (four configurations; two initial parton energies) shown in the Figures. With respect to the $p_T$ and $\ma{y}$ distributions of the hadrons from the wake, for two out of the four configurations shown the orange curves are slightly narrower in $\ma{y}$ than the blue curves and --- related --- the orange $p_T$-distributions are slightly harder than the blue $p_T$-distributions.
In all cases, though, the orange curves  provide a much better characterization of the blue curves than the grey curves do, meaning that our Efficient Wake procedure does a much better job of reproducing the distribution of the hadrons coming from the wakes of high energy partons than does the Old Wake procedure.

The effects of radial flow on the
distribution of soft hadrons originating from the wake of a high-energy parton is most straightforward to see and understand in Figs.~\ref{fig:comp_conf_20} and \ref{fig:comp_conf_18}.  
In both these cases, the high energy parton is produced with a position and direction such that it is traveling outwards, meaning that it is moving in a direction that is roughly parallel to the outward radial flow of the hydrodynamic fluid in which it finds itself.
In this case,
the outward radial flow boosts the wake in the direction of the deposited momentum.  This hardens the $p_T$ distribution of the hadrons originating from the wake, and collimates the distribution of hadrons in both $\phi$ and $\ma{y}$. All three of these effects can easily be seen by comparing the orange or blue curves in both Figs.~\ref{fig:comp_conf_20} and \ref{fig:comp_conf_18} to the grey curves, since the Old Wake calculation whose results are shown in grey includes no  effects of radial flow.

In the results plotted in both Figs.~\ref{fig:comp_conf_17} and \ref{fig:comp_conf_42} we see qualitative features that are not present in 
Figs.~\ref{fig:comp_conf_20} and \ref{fig:comp_conf_18}.  
In Fig.~\ref{fig:comp_conf_17},
the azimuthal distribution of hadrons
originating from the wake of the high-energy parton is not centered at $\Delta\phi=0$. This is easily understood: the momentum deposited in the fluid by the high-energy parton points in the roughly tangential direction indicated by the black arrow, but the resulting wake in the fluid
is boosted outwards by the radial flow.  This results in a $\Delta\phi$ distribution peaked somewhere between 0 (where it would be peaked if the radial flow boost were along the high energy parton direction) and $\sim \pi/2$ (where it would be peaked if the momenta of the particles originating from the wake were  determined entirely by the radial flow). Both here and in
Figs.~\ref{fig:comp_conf_20} and \ref{fig:comp_conf_18}, the momentum distribution of the hadrons originating from the wake in the fluid depends significantly on the boost that the wake receives from the transverse radial flow of the background fluid, but here this boost shifts the peak of the $\Delta\phi$ distribution since the radial flow boost is far from aligned with the momentum deposition.

In Fig.~\ref{fig:comp_conf_42}, for the 10 GeV parton we see a double-humped momentum distribution, most clearly in $\ma{y}$ but also in $\Delta\phi$.  What is unique about this case among the eight cases that we have shown in these four Figures (although we do see instances of the same phenomenon in others among the 50 configurations that we have analyzed where the deposition point is peripheral and the direction of the high-energy parton is roughly inwards) is that this is the case where the wake travels through the droplet of QGP for the longest time before freezeout.  Relative to the cases illustrated in the other three Figures, this is so because 
wake created near the deposition point has a chance to travel across much more of the droplet before freezeout.  This is much more so for the 10 GeV parton than for the 50 GeV parton because, with $dE/d\tau$ taking the form~\eqref{eq:elossrate}, most of the energy lost by a parton that loses all of its momentum as happens in this case is lost ``near the arrowhead'', as its distance traveled approaches the stopping distance $\ell_{\rm stop}$.  Hence, the 50 GeV parton loses most of its momentum later than, and much closer to the center of the droplet than, the 10 GeV parton. This means that the 10 GeV parton in Fig.~\ref{fig:comp_conf_42} sources a wake whose sound waves can travel the farthest in both spacetime rapidity $\eta_s$ and azimuthal angle in position space.
As a consequence of the longitudinal expansion and the transverse radial flow, this
results in particle production at freezeout that is substantially displaced in both $\ma{y}$  and $\Delta\phi$, yielding the double-humped features in the left column of Fig.~\ref{fig:comp_conf_42}.  In agreement with what we found in our earlier work~\cite{Casalderrey-Solana:2020rsj} where we added radial flow by hand, this double-humped structure in the $\ma{y}$- and $\phi$-distributions of the hadrons produced at freezeout develops only for those high-energy partons that deposit their energy and momentum into the fluid long before freezeout, giving the wake a long time to evolve hydrodynamically in the expanding background fluid.

\begin{figure*}[t]
    \centering
    \begin{subfigure}[!htbp]{0.44\textwidth}
        \centering
        \includegraphics[width=\textwidth]{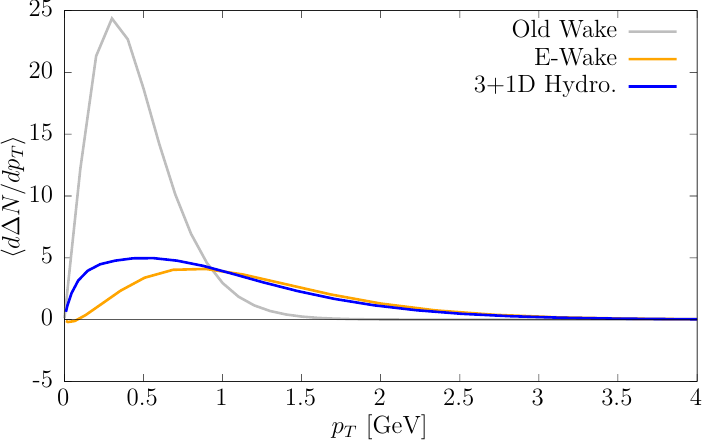}
        \caption{$p_T$ distribution for $E_{\rm in}^{\rm lab}=10$ GeV.}
        \label{fig:ave_pt_10GeV}
    \end{subfigure}%
    ~
    \begin{subfigure}[!htbp]{0.44\textwidth}
        \centering
        \includegraphics[width=\textwidth]{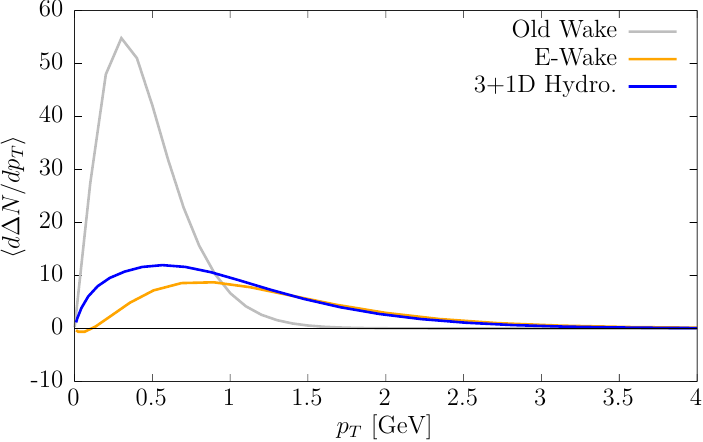}
        \caption{$p_T$ distribution for $E_{\rm in}^{\rm lab}=50$ GeV.}
          \label{fig:ave_pt_50GeV}
    \end{subfigure}%
    
    \begin{subfigure}[!htbp]{0.44\textwidth}
        \centering
        \includegraphics[width=\textwidth]{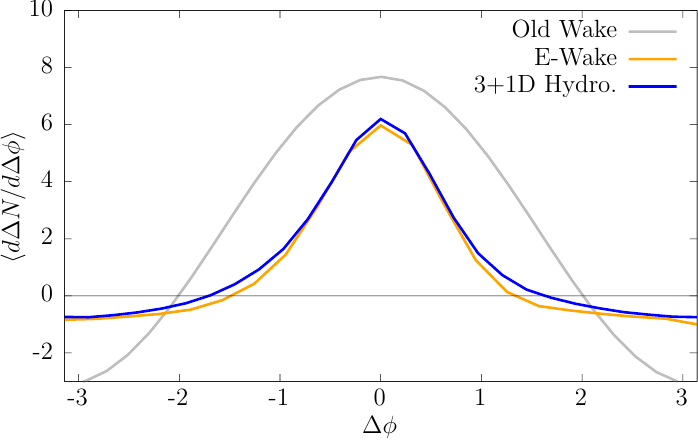}
        \caption{$\phi$ distribution for $E_{\rm in}^{\rm lab}=10$ GeV.}
        \label{fig:ave_phi_10GeV}
    \end{subfigure}%
    ~
    \begin{subfigure}[!htbp]{0.44\textwidth}
        \centering
        \includegraphics[width=\textwidth]{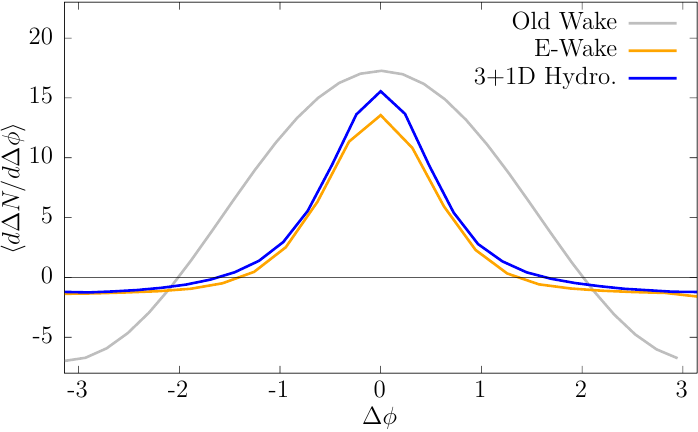}
        \caption{$\phi$ distribution for $E_{\rm in}^{\rm lab}=50$ GeV.}
        \label{fig:ave_phi_50GeV}
    \end{subfigure}%
    
    \begin{subfigure}[!htbp]{0.44\textwidth}
        \centering
        \includegraphics[width=\textwidth]{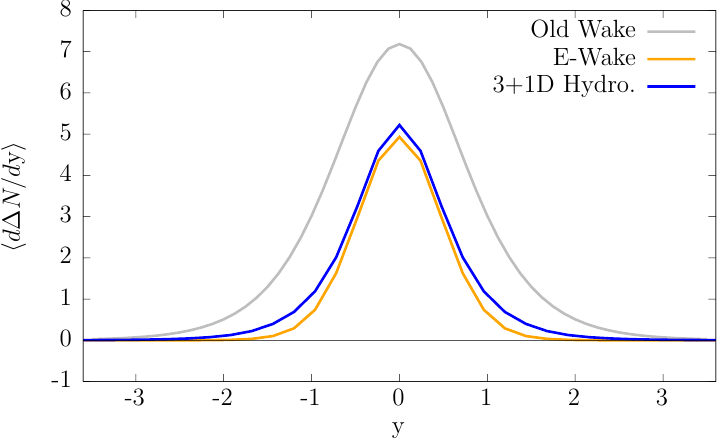}
        \caption{y distribution for $E_{\rm in}^{\rm lab}=10$ GeV.}
        \label{fig:ave_y_10GeV}
    \end{subfigure}%
    ~
    \begin{subfigure}[!htbp]{0.44\textwidth}
        \centering
        \includegraphics[width=\textwidth]{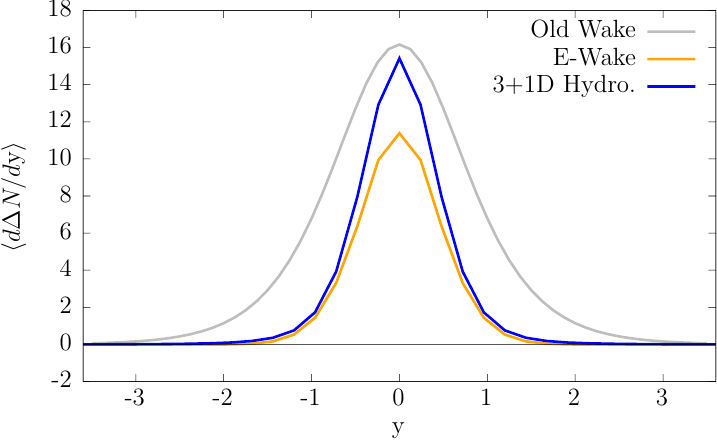}
        \caption{y distribution for $E_{\rm in}^{\rm lab}=50$ GeV.}
        \label{fig:ave_y_50GeV}
    \end{subfigure}%
\caption{Comparison between the averaged (over 50 high-energy parton configurations) momentum
distributions (in $p_T$, $\phi$ and $\ma{y}$)
of hadrons computed with our new E(fficient)-Wake procedure compared to 
results from the Old Wake calculation 
used in Hybrid Model studies
which excludes all effects of transverse flow 
and from a full $(3+1)$-dimensional hydrodynamical analysis (MUSIC).
}
\label{fig:ave_comp}
\end{figure*}

Finally, we discuss the averaged distributions of the particles produced from the wakes in the $50$ configurations that we have analyzed,
shown in Fig.~\ref{fig:ave_comp}.
As can be seen from the plots, 
many of the interesting features that originate from the details of the interplay between the transverse radial flow and the jet wake momentum that we have seen earlier in our results for individual high-energy partons with particular points of origin and orientations disappears after averaging. 
That said, it would have been impossible to obtain the averaged distributions here without having first obtained a good description of the distribution of hadrons originating from the wakes in individual events, with all their intricacies. Although we shall make no attempt to do phenomenology with only 50 high-energy partons with each of two initial energies rather than with a sample of (hundreds of thousands of) parton showers, we expect that it is the averaged distributions in Fig.~\ref{fig:ave_comp}
that provide the best guide to anticipating the experimentally observable consequences of our improved description of jet wakes.

We see from Fig.~\ref{fig:ave_comp}
that our new Efficient Wake procedure (orange curves) 
does a very much better job of describing the averaged momentum distributions of the hadrons originating from the wakes of our sample of 50 high-energy partons than the Old Wake calculation (grey curves), coming very much closer to the results obtained from a full nonlinear hydrodynamics calculation (blue curves).  Relative to the Old Wake calculation, our results show that the transverse radial flow results in a narrower distribution in both $\phi$ and $\ma{y}$  and a harder distribution in $p_T$.
This is pleasing since already in the paper in which the Old Wake calculation was
introduced~\cite{Casalderrey-Solana:2016jvj} the first comparisons between Hybrid Model results obtained via this approach and experimental data that was then available suggested that the Old Wake resulted in a distribution of hadrons that was too soft in $p_T$ and too wide in angle. 
Interestingly, we see in panels (c) and (d) of Fig.~\ref{fig:ave_comp} that in both our Efficient Wake calculation and the full nonlinear hydrodynamics calculation
the ``negative wake'' feature, namely the negative region in the $\phi$-distribution centered on $\Delta\phi=\pi$ corresponding to the direction opposite to that of the jet itself, remains just as broad as it is in the Old Wake calculation, although it is smaller in magnitude.
We very much look forward to future comparisons between results obtained by implementing our new Efficient Wake procedure in the Hybrid Model and contemporary measurements in  heavy ion collisions, including measurements in which
clear evidence of jet wakes has been reported 
via the detection of the negative wake~\cite{CMS:2025dua,CMS:2026mur}.
It will be just as important to
analyze such comparisons for 
many other classes of observables with significant sensitivity to jet wakes,
including the measurements from Ref.~\cite{CMS:2015hkr}
analyzed in Ref.~\cite{Casalderrey-Solana:2016jvj}, many more recent
measurements including those in Refs.~\cite{ATLAS:2020wmg,ALICE:2023qve,PHENIX:2024twd,CMS:2025ydi,STAR:2025yhg}, as well as future measurements of proposed observables~\cite{Bossi:2024qho,Kudinoor:2025ilx,Yang:2025nhy,Wu:2026nef}. In all these cases, 
we expect that implementing our new Efficient Wake procedure in the Hybrid model will yield a better description of experimental data than can be obtained with the Old Wake calculation.

\section{Concluding Remarks and a Look Ahead}
\label{sect:conclusions}

In this paper, we have developed a computationally efficient description of the 
distributions of hadrons produced at freezeout from jet wakes left behind by high-energy partons traversing the droplets of QGP in heavy ion collisions. Our approach relies upon the linearity of linearized hydrodynamics, as we compose the hydrodynamic perturbations at each point on the freezeout hypersurface by superposing perturbations sourced by a high-energy parton during each segment of its trajectory through the droplet of QGP.
We have introduced a procedure that allows us to approximate the perturbation at a freezeout point caused by a unit deposition of energy and momentum at a specified deposition point in space and time with a specified momentum direction by taking the hydrodynamic perturbations from a suitably chosen point in a suitably chosen template solution and applying suitably chosen rotations and boosts. Crucially from the perspective of computational efficiency, the template solutions we employ are constructed by solving linearized hydrodynamics in a Bjorken flow background with no transverse flow, and they need only be computed once.  Equally crucially, the procedure that we introduce takes account for the (very significant) effects of the transverse radial flow of the background fluid on the wake of a high-energy parton. After having determined the 
hydrodynamic perturbations to the temperature and velocity of the fluid at any point on the freezeout hypersurface, we then describe the momentum distributions (in transverse momentum $p_T$, azimuthal angle $\phi$, and momentum rapidity $\ma{y}$) of the hadrons originating from the wakes of high-energy partons
via the conventional Cooper-Frye freezeout of a --- in this case perturbed --- hydrodynamic fluid.
The procedure that we introduce to describe the hydrodynamic perturbations at freezeout in a radially expanding droplet of QGP using templates computed in a Bjorken flow background
is not a controlled approximation and for this reason a central goal of this paper has been to benchmark its validity quantitatively.

We have benchmarked our procedure by
comparing the momentum distributions of the hadrons originating from the wakes of high-energy partons
calculated via our Efficient Wake procedure
with those obtained from a full $(3+1)$-dimensional nonlinear hydrodynamic calculation, done using the MUSIC code, for partons with two different initial energies, $10$ GeV and $50$ GeV, in 50 different configurations. The high-energy partons in these 
configurations differ in their creation points and 
in the directions of their trajectories.
We find that the effects of the transverse radial flow in the expanding droplet of QGP on the wakes, and hence on the resulting distribution of the hadrons originating from the wakes at freezeout,
are both substantial in magnitude and sensitively dependent on ``details'' like where the high-energy parton was produced and in what direction it was moving relative to the direction of the transverse flow of the fluid in which it finds itself.
We have found that our Efficient Wake procedure reproduces the hadron momentum distributions obtained from a full $(3+1)$-dimensional nonlinear hydrodynamics calculation reasonably well, both for the distributions averaged over all 50 events (see Fig.~\ref{fig:ave_comp}) and for the individual distributions originating from the wakes of individual high-energy partons (see Figs.~\ref{fig:comp_conf_20}, \ref{fig:comp_conf_17}, \ref{fig:comp_conf_18} and \ref{fig:comp_conf_42} for four examples), including reproducing the large and striking differences between the wakes of high-energy partons in different configurations.
And, the Efficient Wake procedure does so via  calculations that are faster than our $(3+1)$-dimensional nonlinear hydrodynamics calculations by about a factor of 200,000, and that are tens of thousands of times faster than the best such calculations.

We have shown that our Efficient Wake procedure yields a {\it much} better characterization of the distributions of the hadrons originating from the wakes of high-energy partons than does the oversimplified procedure introduced in Ref.~\cite{Casalderrey-Solana:2016jvj} and used in Hybrid Model studies since then. That ``Old Wake'' procedure is oversimplified in many ways including, crucially, that it ignores transverse radial flow in the background fluid.  We have found 
that many interesting features in the distributions of hadrons originating from wakes in individual events can only be explained by the interplay between the local transverse flow along the trajectory of the high-energy parton and the direction of the momentum of, and hence the direction of the momentum deposited in the fluid by, the high-energy parton.
After averaging, the more intricate  features are averaged out, yielding averaged distributions that are harder in $p_T$ and narrower in $\phi$ and $\ma{y}$ 
than those obtained via the Old Wake procedure.
However, to correctly describe the averaged distributions it is imperative that one correctly describes the intricate and interesting effects of transverse radial flow on wakes in individual events. 
Our Efficient Wake procedure does so, as we can see
via comparing its results to those obtained via full $(3+1)$-dimensional nonlinear hydrodynamics calculations of the 
(unique)
wakes of the high-energy parton in each of the 50 different configurations in our sample. In this context it is interesting to note that
machine-learning techniques have recently been developed as an alternative means of accelerating the calculation of jet wakes and the resulting particle production~\cite{Wu:2026pdi}, where the algorithm is taught to statistically 
predict the event-averaged particle spectra
from the fully nonlinear response of the medium to an entire jet. Although 
we have not made a direct comparison, the increase in computational speed obtained in this way reported in Ref.~\cite{Wu:2026pdi}
is at least as impressive as the factor of 200,000 speed-up that we have obtained with our Efficient Wake procedure.  
Our approach, based upon a linearized hydrodynamics analysis of the unique wake of each individual high-energy parton via the superposition of perturbations in suitably chosen templates, yields a good description on a jet-by-jet basis, which is a significant advantage.

The next step in the road ahead is obvious.
In future work, we plan to implement our Efficient Wake procedure in the Hybrid Model so as to have a computationally efficient description of the hadrons originating from the wakes of real jets, which are parton showers not individual high-energy partons, in a sample of $10^5-10^6$ jets produced in, and selected as in, a heavy ion collision experiment. 
The implementation of our Efficient Wake procedure in the Hybrid Model or in any other Monte Carlo treatment of jets in heavy ion collisions will open the door to rapid and efficient, quantitatively reliable, 
phenomenological studies of the observable consequences of jet wakes in heavy ion collision experiments. 

Reliable phenomenological conclusions must await this future work, but we can make some immediate remarks based upon what we have found in our results averaged over 50 high-energy parton configurations in Fig.~\ref{fig:ave_comp}. By comparing the Old Wake and Efficient Wake $p_T$-distributions in Fig.~\ref{fig:ave_comp}, we can confirm that excluding hadrons with $p_T\gtrsim 1.5$~GeV from the analysis of some experimental observable serves to eliminate effects of jet wakes in a Hybrid Model calculation that employs the Old Wake and observe that 
the results from the full $(3+1)$-dimensional hydrodynamics calculation, correctly captured by our Efficient Wake results, 
 indicate that doing so in experimental data would require excluding 
hadrons with $p_T\gtrsim 3$~GeV. 
We can also speculate about how the narrowing of the
distribution of hadrons originating from jet wakes in both $\phi$ and $\ma{y}$, and the reduction in magnitude of the negative contribution to the hadron distribution around $\Delta\phi\sim\pi$,
both of which are apparent in Fig.~\ref{fig:ave_comp}, may translate into an improved description of the recent experimental measurements in Refs.~\cite{CMS:2025dua} and \cite{CMS:2026mur} 
and of many prior measurements~\cite{CMS:2015hkr,ATLAS:2020wmg,ALICE:2023qve,PHENIX:2024twd,CMS:2025ydi,STAR:2025yhg}
when the Efficient Wake procedure has been implemented in the Hybrid Model.

Our results for individual configurations, as shown in Figs.~\ref{fig:comp_conf_20}, \ref{fig:comp_conf_17}, \ref{fig:comp_conf_18} and \ref{fig:comp_conf_42}
are also thought-provoking. For example, if it is possible to select a sample of events in an analysis of experimental data where the jet was produced away from the center of the collision in a direction that is roughly perpendicular to the direction of the radial flow at its production point, as in Fig.~\ref{fig:comp_conf_17}, then the direction in $\phi$ defined by the distribution of hadrons originating from the wake of the jet should be misaligned with the direction of the jet 
itself. Observational signatures of this effect may be possible in events with a nonzero impact parameter~\cite{Yang:2025nhy,Wu:2026nef}, 
which motivates extending our Efficient Wake procedure to such events.  This too we leave for future work.  
We also observe that this idea can be turned on its head, with the effect
seen in Fig.~\ref{fig:comp_conf_17} 
used as the criterion for selecting events. Selecting events in which the distribution of soft hadrons (e.g.~hadrons originating from the wake of a jet) are misaligned with the direction of the hard core of a jet (identified via the winner-take-all algorithm) could be a means to select a sample of jets produced far from the center of the collision pointing in a tangential direction,
as in Fig.~\ref{fig:comp_conf_17}, even in head-on collisions. Although likely more challenging in practice, one can also speculate that if it were possible to select events in which the soft hadrons around a jet have a distribution that is double-humped in both $y$ and $\Delta\phi$ as in Fig.~\ref{fig:comp_conf_42}, doing so could be a way to select a sample of highly quenched jets produced near the periphery of the collision pointing inwards. Extending this speculation even further, to its logical extreme, perhaps Fig.~\ref{fig:comp_conf_42} may point the way to a method of selecting a sample of completely quenched jets that are entirely wake, with no remaining hard core at all.

\acknowledgments
KR is grateful to the CERN Theory Department for hospitality during the completion of this work.
JCS acknowledges financial support from the “Center of Excellence Maria de Maeztu 2025–2029” award to the ICCUB, grant CEX2024-001451-M, funded by AEI/ 10.13039/ 501100011033, as well as Grant No. PID2022-136224NB-C21 from the Spanish Ministry of Science, Innovation and Universities, and from Grant No. 2021-SGR-872 funded by the Catalan Government.
JGM is supported by European Research Council (ERC) under the European Union’s Horizon 2020 research and innovation programme (Grant agreement No. 835105, YoctoLHC) and Funda\c c\~ao para a Ci\^encia e a Tecnologia (FCT I.P.), under ERC-PT A-Projects ‘Unveiling’, financed by PRR, NextGenerationEU, and he gratefully acknowledges the hospitality of the CERN theory group. 
DP is supported by the Spanish Ram\'on y
Cajal fellowship RYC2023-044989-I. The work of KR was supported by the U.S. Department of Energy, Office of Science, Office of Nuclear Physics grant DE-SC0011090. 
The work of XY is supported by the U.S. Department of Energy, Office of Science, Office of Nuclear Physics, InQubator for Quantum Simulation (IQuS)\footnote{\url{https://iqus.uw.edu/}} under Award Number DOE (NP) Award DE-SC0020970 via the program on Quantum Horizons:\footnote{\url{https://science.osti.gov/np/Research/Quantum-Information-Science}}~QIS Research and Innovation for Nuclear Science.

\appendix

\section{Blast-wave Approximation}
\label{app:model}

\begin{figure}[t]
    \centering
    \begin{subfigure}[!htbp]{0.49\textwidth}
        \centering
        \includegraphics[width=\textwidth]{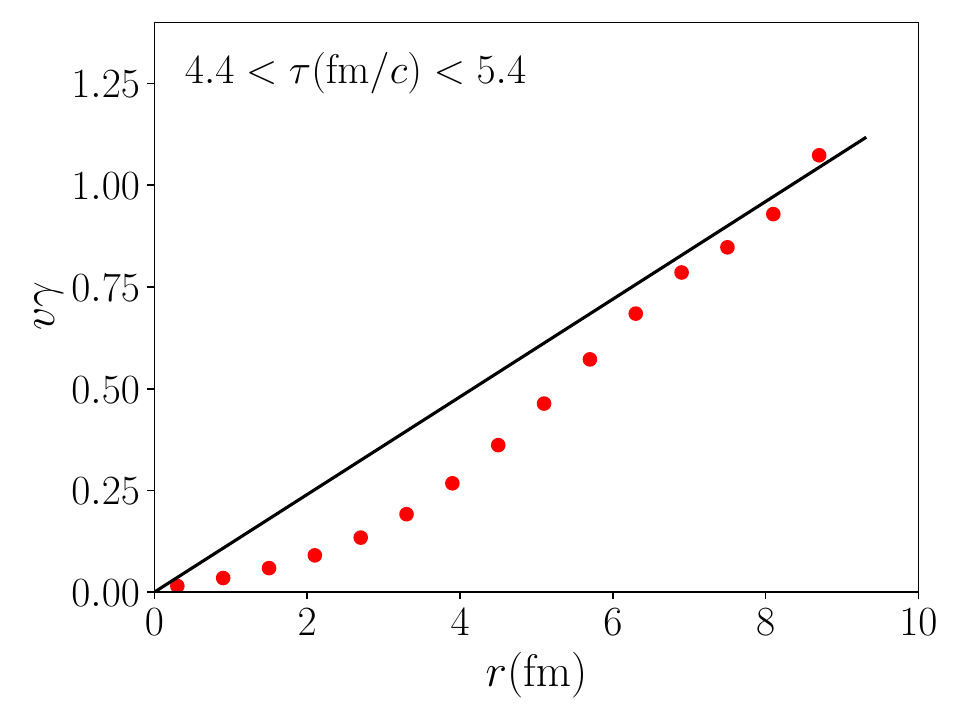}
        \caption{}
        \label{fig:}
    \end{subfigure}%
    ~
    \begin{subfigure}[!htbp]{0.49\textwidth}
        \centering
        \includegraphics[width=\textwidth]{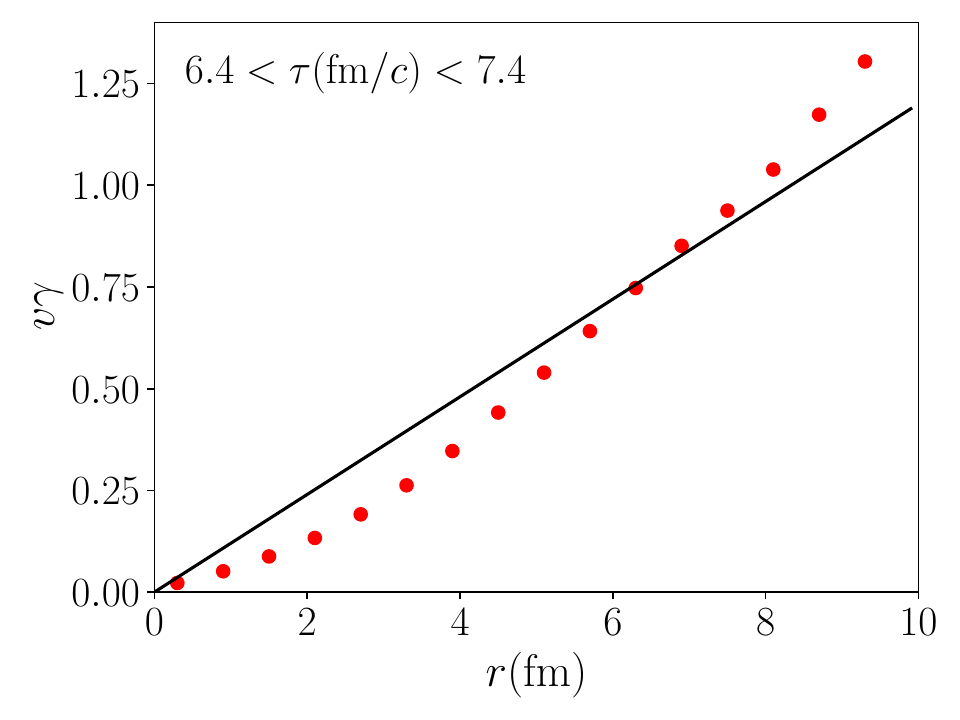}
        \caption{}
        \label{fig:}
    \end{subfigure}%

    \begin{subfigure}[!htbp]{0.49\textwidth}
        \centering
        \includegraphics[width=\textwidth]{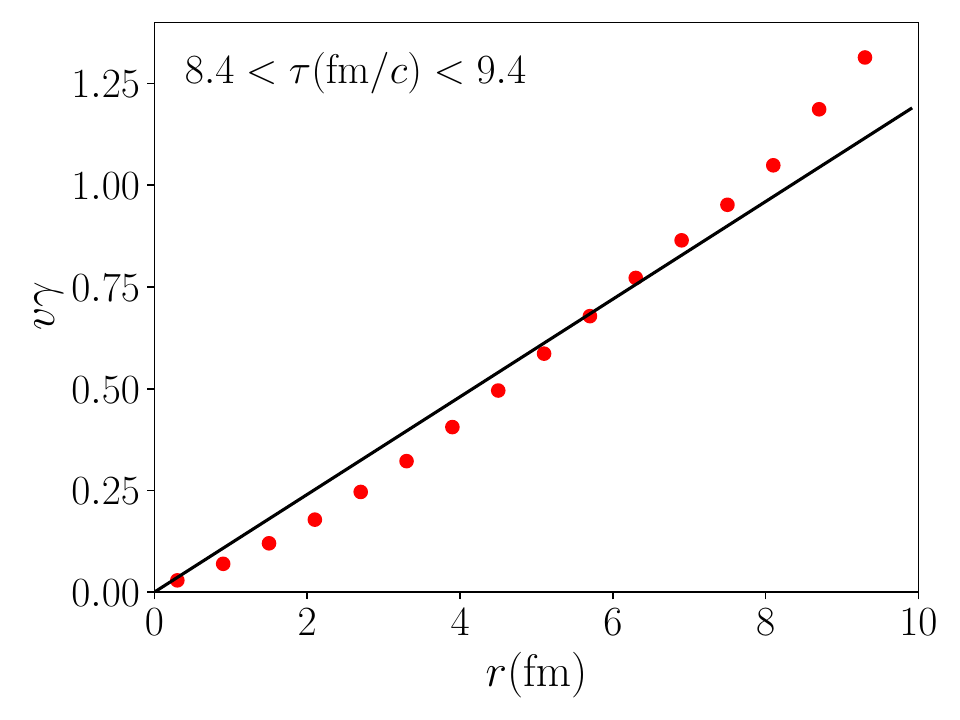}
        \caption{}
        \label{fig:}
    \end{subfigure}%
    ~
    \begin{subfigure}[!htbp]{0.49\textwidth}
        \centering
        \includegraphics[width=\textwidth]{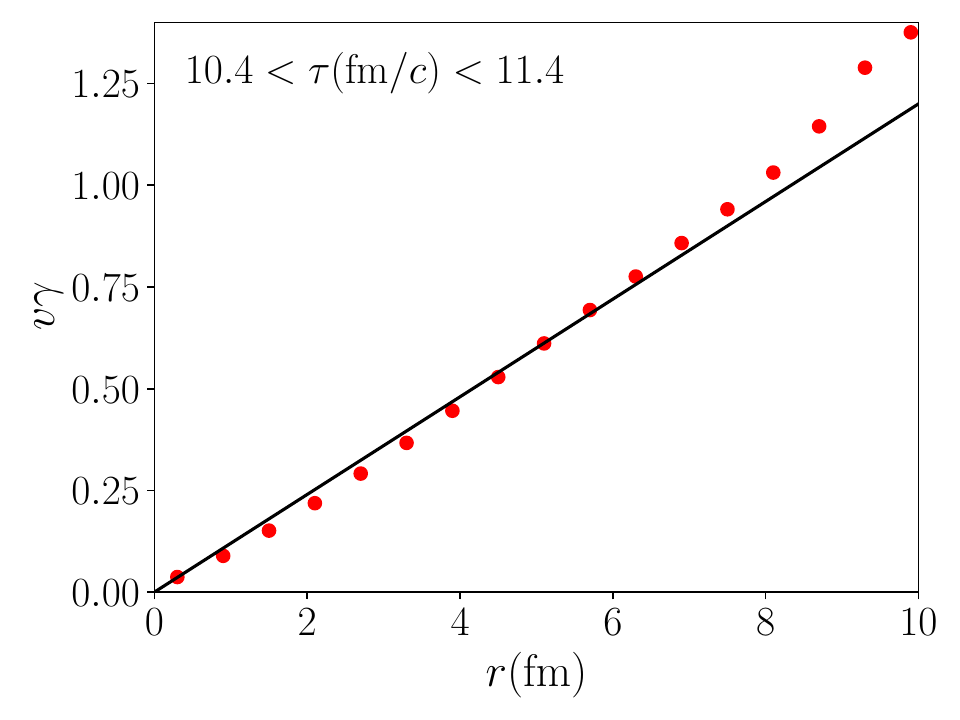}
        \caption{}
        \label{fig:}
    \end{subfigure}%

    \begin{subfigure}[!htbp]{0.49\textwidth}
        \centering
        \includegraphics[width=\textwidth]{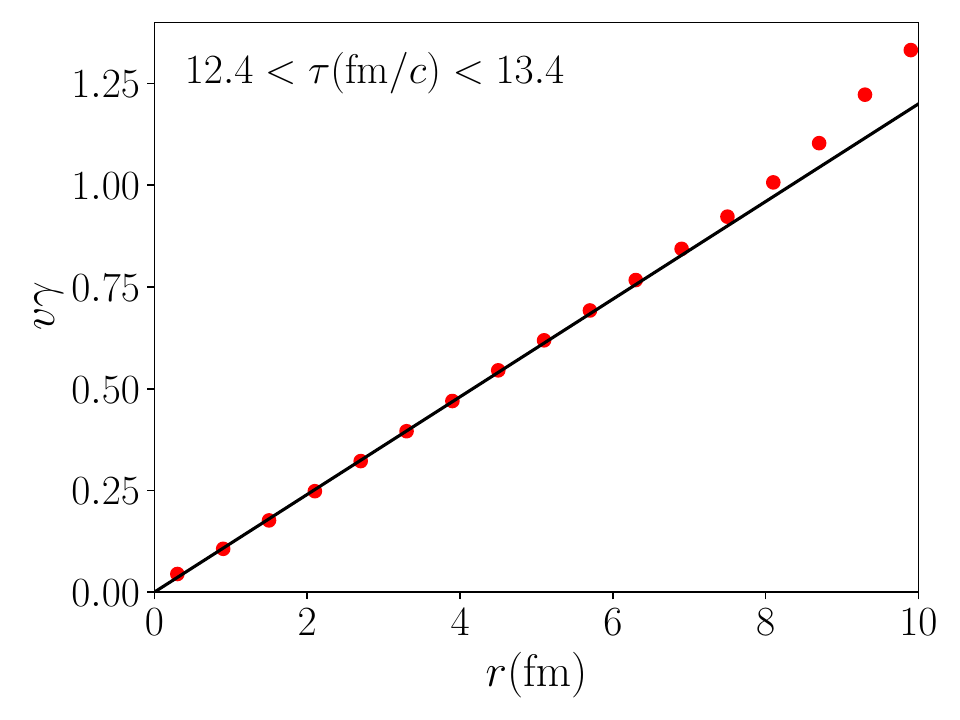}
        \caption{}
        \label{fig:}
    \end{subfigure}%
\caption{Blast-wave approximation 
\eqref{blastwave_flow} with parameter $\sigma=0.12~{\rm fm}^{-1}$ (black lines) for $v\gamma$ (with $v$ the radial flow velocity) as a function of $r$ at five different proper times compared to the same quantity obtained via hydrodynamic simulations (red dots) done with the MUSIC code~\cite{Schenke:2010nt,Schenke:2010rr,Ryu:2015vwa,Paquet:2015lta} for  heavy ion collisions with zero impact parameter and $\sqrt{s_{\rm NN}}=5.02$~TeV. 
}
\label{fig:app_model}
\end{figure}

In Section~\ref{step2}, in order to obtain an analytical description of the 
relative velocity trajectories that we have used in Section~\ref{sec:Utility} to determine from which point in which template solution we will map the hydrodynamic fluctuations to a given point on the freezeout hypersurface
we restricted our attention to collisions 
with zero impact parameter in which the transverse flow velocity is purely radial and introduced an analytical form with which to describe the spatial dependence of the transverse flow velocity, called the blast-wave approximation:
\be
\gamma v = \sigma r \,,
\ee
where $v=\sqrt{v_x^2+v_y^2}$ is the magnitude of the radial transverse fluid velocity, $\gamma \equiv \sqrt{1-v^2}$, and $\sigma$ is a parameter that we must choose. 
In this Appendix, in Fig.~\ref{fig:app_model} we show that this
simplified model provides a reasonable description of the radial flow in central heavy ion collisions after an early period of time which is roughly
$4~{\rm fm}/c$. In the figure, the black lines show the blast-wave form for $v\gamma$ as a function of $r$ at different times, where we have chosen the value of the parameter $\sigma=0.12~{\rm fm}^{-1}$, while the red dots show the same quantity obtained from a full  viscous hydrodynamic simulation of central heavy ion collisions at LHC collision energies $\sqrt{s_{\rm NN}}=5$~TeV performed with the  MUSIC code~\cite{Schenke:2010nt,Schenke:2010rr,Ryu:2015vwa,Paquet:2015lta}. The description provided by the blast-wave approximation becomes better and better at later and later times.

\section{Isothermal Hypersurface in Laboratory Frame}
\label{app:dsigma}

In this Appendix, we review the two parameterizations of the isothermal hypersurface that we use in Section~\ref{sect:cf}. In general, any three-dimensional differentiable hypersurface may be described by any intrinsic set of coordinates $\{\xi^a \}$, with $a=1\dots 3$, such that the embedding of the surface is given by $x^\mu \left(\xi^a\right)$.  In those coordinates, the surface element is given by 
\be
\diff^3\sigma_\mu = n_\mu \sqrt{\mp h} \diff \xi^1\diff \xi^2 \diff \xi^3 \,,
\ee
where the sign $\mp$ depends on whether the hypersurface is spacelike $(-)$ or timelike $(+)$, $h$ is the determinant of the induced metric on the surface,
\be
h_{a b}=g_{\mu \nu} \frac{\partial x^\mu}{\partial \xi^a} \frac{\partial x^\nu}{\partial\xi^b }
\ee
with $g_{\mu \nu} $ the metric in $(3+1)$-dimensional space-time, and $n_\mu$ is the outward unit normal (co)vector  to the hypersurface. This vector may be constructed from the cross product of the three tangent vectors
\be
e_a^\mu \equiv \frac{\partial x^\mu}{\partial \xi^a} \, 
\ee
as $n_\mu \equiv \pm \sqrt{|g|/|h|}  \varepsilon_{\mu \nu \rho \sigma} e^\nu_1 e^\rho_2 e^\sigma_3$ where $g$ is the determinant of the spacetime metric and where the sign is to be chosen such that the vector points outwards.

We may now choose a specific parameterization for the freezeout surface.  In a central collision, for the spacelike regions of the hypersurface we choose the intrinsic coordinates  $\{\xi^1,\xi^2,\xi^3\}=\{x,y,\eta_s\}$. 
As a consequence of boost invariance and azimuthal symmetry in the transverse plane, the freezeout hypersurface is given by a function $\tau^f(r)$ with $r=\sqrt{x^2+y^2}$. With this choice, the normal, future-pointing covector is 

\be
n_\mu = \frac{\tau^f}{\sqrt{-h}} \Big( \cosh\eta_s, -\frac{\partial\tau^f}{\partial x}, -\frac{\partial\tau^f}{\partial y}, -\sinh\eta_s \Big) \,,
\ee
where the determinant of the induced metric is given by 
\be
h=-\left(\tau^f\right)^2 \left( 1 - \left(\frac{\partial\tau^f}{\partial x}\right)^2
- \left(\frac{\partial\tau^f}{\partial y}\right)^2 \right)\ .  
\ee
Putting everything together, the surface element on the freezeout hypersurface becomes 
\be
\diff^3\sigma_\mu = \Big( \cosh\eta_s, -\frac{\partial\tau^f}{\partial x}, -\frac{\partial\tau^f}{\partial y}, -\sinh\eta_s \Big) \tau^f \diff x \diff y \diff \eta_s \,.
\ee
This is the expression \eqref{eq:NormalSurfaceElement} we use in the regions where the freezeout hypersurface is spacelike.

For the timelike regions of the freezeout hypersurface, a more convenient choice is to set $\{\xi^1,\xi^2,\xi^3\}=\{\tau,\phi,\eta_s\}$. 
Then the freezeout hypersurface is given by a function $r^f(\tau)$. With this choice, the normal, outward-pointing covector becomes 
\be
n_\mu =  \frac{ \tau r^f}{\sqrt{h}}\Big(- \frac{\partial r^f}{\partial \tau}\cosh\eta_s, \cos\phi, \sin\phi, \frac{\partial r^f}{\partial\tau} \sinh\eta_s \Big) \,.
\ee
where here 
\be
h=  \tau^2 \left(r^f\right)^2 \left(1-\left(\frac{\partial r^f}{\partial\tau}\right)^2 \right)\ . 
\ee
Here, $n_\mu$ is the outward-pointing normal on the ``side of the mushroom'' in Fig.~\ref{fig:fzoutpareme}, meaning that it points upward in $\tau$ (downward in $\tau$) in those regions of the side of the mushroom where $\partial r^f/\partial \tau$ is negative (positive).
In this case, the surface element becomes 
\be\label{eq:NormalTimelikeSurfaceElement_app}
\diff^3 \sigma_\mu = \Big(-\frac{\partial r^f}{\partial \tau}\cosh\eta_s, \cos\phi, \sin\phi, \frac{\partial r^f}{\partial\tau} \sinh\eta_s \Big) \,
r^f \!\left(\tau\right) \tau \diff\tau \diff\phi \diff \eta_s\, ,
\ee
which is Eq.~\eqref{eq:NormalTimelikeSurfaceElement}.

\bibliographystyle{jhep}
\bibliography{main.bib}

\end{document}